\documentclass{aa}  

\usepackage{graphicx}
\usepackage{txfonts}
\usepackage{placeins}
\usepackage{bm}
\usepackage{stfloats}
\usepackage[pdfpagelabels=false]{hyperref}
\hypersetup{colorlinks=true,linkcolor=blue,citecolor=blue,filecolor=blue,urlcolor=blue}

\newcommand{\kpc}{\hbox{$\mathrm{kpc}$}}
\newcommand{\kms}{\hbox{$\mathrm{km \, s^{-1}}$}}
\newcommand{\kmskpc}{\hbox{$\mathrm{km \, s^{-1} \, kpc^{-1}}$}}

\newcommand{\Rg}{\hbox{$R_{\mathrm{g}}$}}

\begin{document}
\title{Ripples spreading across the Galactic disc}
\titlerunning {Ripples across the Galactic disc}
\subtitle{Interplay of direct and indirect effects of the Sagittarius dwarf impact}

\author{Tetsuro Asano
	\inst{1,2,3,4},
	Michiko S. Fujii\inst{4},
	Junichi Baba\inst{5,6},
	Simon Portegies Zwart\inst{7},
	\and
	Jeroen B\'edorf\inst{7,8}
}
\authorrunning{T. Asano et al.}

\institute{Departament de F\'isica Qu\`antica i Astrof\'isica (FQA), Universitat de Barcelona (UB), c. Mart\'i i Franqu\`es, 1, 08028 Barcelona, Spain\\
	\email{asano@fqa.ub.edu}
	\and
	Institut de Ci\`encies del Cosmos (ICCUB), Universitat de Barcelona (UB), c. Mart\'i i Franqu\`es, 1, 08028 Barcelona, Spain
	\and
	Institut d’Estudis Espacials de Catalunya (IEEC), c. Gran Capit\`a, 2-4, 08034 Barcelona, Spain
	\and
	Department of Astronomy, Graduate School of Science, The University of Tokyo, 7-3-1 Hongo, Bunkyo-ku, Tokyo 113-0033, Japan
	\and
	Amanogawa Galaxy Astronomy Research Center, Kagoshima University, 1-21-35 Korimoto, Kagoshima 890-0065, Japan
	\and
	National Astronomical Observatory of Japan, Mitaka-shi, Tokyo 181-8588, Japan
	\and
	Leiden Observatory, Leiden University, NL-2300RA Leiden, the Netherlands
	\and
	Minds.ai, Inc., Santa Cruz, CA 95060, USA
}

\date{Received 20 January 2025; Accepted: 12 June 2025}

\abstract
{\textit{Gaia} data have revealed vertically asymmetric phase-space structures in the Milky Way (MW) disc, such as phase spirals, indicating vertical oscillations. These oscillations exhibit two distinct modes, the bending mode and the breathing mode, associated with one-arm and two-arm phase spirals, respectively. The mechanisms driving these modes remain debated, with both external and internal origins proposed.}
{With this study, we aim to explore the excitation mechanisms of the bending and breathing modes and their subsequent evolution in the MW disc, focusing on the interplay between direct perturbations from the Sagittarius dwarf galaxy and indirect contributions from tidally induced spiral arms.}
{We performed high-resolution $N$-body simulations with five billion particles to model the interaction between an MW-like disc galaxy and a Sagittarius dwarf-like satellite. 
These simulations resolve fine phase-space structures, enabling analysis of the bending and breathing modes at both macroscopic (global bending and breathing waves) and microscopic (local phase spirals) scales.
}
{Our simulations demonstrate that the satellite's perturbation directly excites the bending mode and induces spiral arms in the Galactic disc. 
These spiral arms, in turn, excite the breathing mode, making it an indirect consequence of the satellite interaction. Initially, the bending mode dominates, but it rapidly decays due to horizontal mixing. In contrast, the breathing mode persists for a longer duration, sustained by the spiral arms, leading to a transition from a bending-dominated to a breathing-dominated state. This transition progresses faster in the inner galaxy than in the outer galaxy. 
The simulations successfully reproduce the one-arm phase spiral observed in the solar neighbourhood and reveal two-arm phase spirals, particularly in the inner galaxy, associated with spiral arm-induced breathing modes. The two-arm phase spirals emerge approximately 200–250 Myr after the bending-to-breathing transition.
}
{
Our findings highlight the combined effects of direct satellite perturbations and indirect spiral arm dynamics in shaping the vertical structure of the MW disc. The emergence of the two-arm phase spiral after the bending-to-breathing transition suggests that the MW disc experienced a significant perturbation more than $\sim 400$ Myr ago, likely caused by the Sagittarius dwarf galaxy. This study underscores the importance of considering the dynamic interplay between direct and indirect mechanisms in understanding the vertical dynamics of the MW disc.
}


\maketitle

\section{Introduction}
In the \textit{Gaia} Data Release 2 \citep[DR2;][]{2016A&A...595A...1G, 2018A&A...616A...1G} data, \citet{2018Natur.561..360A} discovered a `one-arm' spiral pattern within the stellar distribution in the vertical position ($z$) versus vertical velocity ($v_z$) space.
This discovery provides evidence of incomplete phase mixing, indicating that the Milky Way (MW) was perturbed relatively recently (\( \lesssim 1 \)~Gyr) and has not yet reached equilibrium.
The phase spiral has been studied extensively because it contains two kinds of important information about the Galactic dynamics: the nature of past perturbation event(s) \citep[e.g.][]{2018MNRAS.481.1501B, 2019A&A...622L...6K, 2019MNRAS.485.3134L, 2019MNRAS.486.1167B} and the Galactic potential \citep[e.g.][]{2021A&A...650A.124W,2021A&A...653A..86W, 2022A&A...663A..16W, 2022A&A...663A..15W, 2024ApJ...960..133G}.

While various mechanisms have been proposed to explain the origin of the one-arm phase spiral, the external perturbation by the Sagittarius dwarf (Sgr) is considered a plausible scenario and has been widely studied.
Theoretical studies using analytical models \citep[e.g.][]{2018MNRAS.481.1501B,2022ApJ...935..135B, 2023ApJ...952...65B} and controlled $N$-body simulations \citep[e.g.][]{2018MNRAS.481..286L, 2019MNRAS.485.3134L, 2021MNRAS.504.3168B, 2022MNRAS.516L...7H, 2022ApJ...927..131B} have confirmed the emergence of the phase spiral in galactic discs perturbed by Sgr-like satellites.
Not only present in such controlled simulations, phase spirals have also been discovered in cosmological simulations \citep{2022MNRAS.510..154G, 2023MNRAS.524..801G}.
\citet{2023MNRAS.524..801G} demonstrated that the dark-matter (DM) wake generated by a satellite, rather than its direct impact, can excite the phase spiral.
Similarly, \citet{2018MNRAS.481..286L, 2019MNRAS.485.3134L} showed that the torque from the DM wake is larger than that from the main body of the Sgr during the early stage of the MW-Sgr interaction.
Theoretical studies further suggest that Sgr induces other phase-space substructures and the Galactic warp \citep[e.g.][]{2013MNRAS.429..159G, 2019MNRAS.489.4962K, 2022MNRAS.515.5951T, 2024MNRAS.535.1898B}.
Apart from the Sgr scenario, other phase spiral formation mechanisms have been proposed, including bar buckling \citep{2019A&A...622L...6K} and the combined effects of many small perturbations \citep{2023MNRAS.521..114T}, for instance, due to DM sub-halos \citep{2025ApJ...980...24G}. However, their contributions to other distorted structures such as the warp are unclear.

\citet{2022MNRAS.516L...7H} discovered another type of phase spiral in the \textit{Gaia} DR3 data \citep{2023A&A...674A...1G}, finding that stars with small guiding radii exhibit the `two-arm' phase spiral.
The one-arm and two-arm phase spirals correspond to different global vertical oscillation modes in the Galactic disc.
The one-arm phase spiral is associated with the `bending' mode, where the regions above and below the mid-plane oscillate in the same direction.
In contrast, the two-arm phase spiral is associated with the `breathing' mode, where the regions above and below the mid-plane oscillate in opposite directions.
Although the one-arm phase spiral is observed in a wide area of the Galactic disc \citep[e.g.][]{2019ApJ...877L...7W, 2020ApJ...905....6X,  2023ApJ...956...13X, 2023A&A...673A.115A}, the two-arm phase spiral is detected only in specific radial and angular momentum ranges \citep{2024A&A...690A..15A}; therefore, two-arm phase spirals are generally considered to arise from local internal processes rather than global external perturbations \citep{2014MNRAS.437.3702C, 2022MNRAS.516L...7H, 2023MNRAS.524.6331L, 2025arXiv250320869C}.
Among potential internal drivers, spiral arms are plausible origins of the two-arm phase spiral, as they naturally excite the breathing mode \citep{2014MNRAS.443L...1D, 2014MNRAS.440.2564F, 2016MNRAS.457.2569M, 2016MNRAS.461.3835M, 2022MNRAS.517L..55K, 2022MNRAS.511..784G, 2022MNRAS.516.1114K, 2024MNRAS.529L...7A}.

Beyond the phase spirals, \textit{Gaia} and other surveys have provided further evidence of bending and breathing oscillations on both local and global scales \citep[e.g.][]{2012ApJ...750L..41W, 2013MNRAS.436..101W, 2015ApJ...801..105X, 2018A&A...616A..11G, 2018MNRAS.479L.108K, 2018MNRAS.478.3809S, 2018MNRAS.477.2858W, 2018MNRAS.478.3367W, 2020MNRAS.491.2104W, 2019MNRAS.490..797C, 2019MNRAS.482.1417B, 2019MNRAS.489.4962K, 2020A&A...634A..66L, 2020ApJ...905...49C, 2022MNRAS.516.4988M, 2022MNRAS.511..784G, 2022A&A...668A..95W, 2022MNRAS.510L..13L, 2023A&A...678A.111A, 2024MNRAS.529L...7A, 2024MNRAS.533L..31W, 2024MNRAS.528.3281L, 2024arXiv240718659P}.  
Many studies have investigated internal and external processes separately, but in reality the vertical oscillations in the MW disc should be driven by a combination of both. \citet{2021MNRAS.503..376B} suggested that the vertical asymmetry in the solar neighbourhood cannot be fully explained by the Sgr impact alone.  
Moreover, internal and external processes are not entirely independent, as external perturbations influence the internal structures of the disc. For instance, tidal interactions with satellite galaxies induce spiral arms in host discs \citep{2022A&A...668A..61A}, which can further perturb the motion of disc stars.  
To fully understand the excitation and evolution of the bending mode and the breathing mode, it is crucial to consider the interplay between internal and external processes.

$N$-body simulations are useful tools to study the self-consistent evolution of galactic discs.
In the context of the Sgr impact, most of the previous studies used dynamically hot discs, where spontaneous bar and spiral formations are suppressed to isolate the effects of external perturbation \citep[but see][]{2021MNRAS.504.3168B}.
However, such hot disc models underestimate the influence of self-gravity, which affects the amplitude and winding rate of the phase spiral \citep{2019MNRAS.490..114D, 2019MNRAS.484.1050D, 2023MNRAS.522..477W}.
To address these limitations and bridge the gap between theory and observations, we employ a `cold' disc model that accounts for the influence of self-gravity and better replicates the dynamical conditions of the MW. Our high-resolution simulations, powered by a graphics processing unit (GPU)-based code \citep{2012JCoPh.231.2825B, 2014hpcn.conf...54B}, utilise five billion particles to resolve fine phase-space structures and directly compare the results with \textit{Gaia} data. Unlike previous billion-particle $N$-body models \citep{2021MNRAS.508.1459H, 2022ApJ...927..131B, 2024ApJ...977..252S}, our disc model is colder and incorporates disc and bulge properties more consistent with the real MW. This approach allows us to investigate the disc’s dynamical evolution under external perturbations caused by a Sgr-like satellite galaxy while also capturing the potential role of internal processes such as spiral arms.

This paper is organised as follows.
In Section~\ref{sec:simulation}, we describe the details of the $N$-body simulations.
In Section~\ref{sec:repeated_impact}, we analyse the global vertical oscillations of the MW disc induced by repeated impacts from the satellite, focusing on the bending and breathing modes and their time evolution.
In Section~\ref{sec:long_term_evolution}, we isolate the effects of a single impact to clarify the role of the dwarf galaxy’s perturbation in driving the global vertical oscillations, by examining the long-term evolution of the bending and breathing modes after a single satellite passage.
In Section~\ref{sec:phase_spiral}, we shift to the local scale, investigating the spatial variation and time evolution of phase spirals, which link global vertical oscillations to local phase-space structures.
In Section~\ref{sec:discussion}, we discuss what we can infer about the Galactic disc structure and its evolution history from the real observation results and our simulation. Finally, Section~\ref{sec:conclusion} provides a summary of our findings.

\section{\texorpdfstring{$N$}{N}-body simulation}\label{sec:simulation}
\subsection{Initial condition and simulation}\label{sec:initial_condition}
We performed $N$-body simulations of a MW-like disc galaxy perturbed by a Sgr-like satellite galaxy.
The MW model is identical to MWa of \citet{2019MNRAS.482.1983F}.  We generated the initial condition from the same parameters as MWa using \texttt{Galactics} \citep{1995MNRAS.277.1341K,2005ApJ...631..838W,2008ApJ...679.1239W}.  It comprises a DM halo, a classical bulge, and a disc.

The DM halo follows the Navarro-Frenk-White \citep[NFW;][]{1997ApJ...490..493N} profile:
\begin{align}
  \rho_{\mathrm{h}}(r) = \frac{\rho_{\mathrm{h}0}}{(r/a_{\mathrm{h}}) (1+r/a_{\mathrm{h}})^2},
\end{align}
where $a_{\mathrm{h}}$ and $\rho_{\mathrm{h}0}$ are the scale radius and the characteristic density, respectively.
Here, $a_{\mathrm{h}}$ is set to 10~kpc, and $\rho_{\mathrm{h},0}$ is determined through the characteristic velocity dispersion $\sigma_{\mathrm{h}} = (4\pi G a_{\mathrm{h}}^2\rho_{\mathrm{h}0})^{1/2}=420\,\kms$, where $G$ is the gravitational constant. The truncation parameter \citep{2005ApJ...631..838W} is set to $\epsilon_{\mathrm{h}}=0.85$. 
The rotation fraction, $\alpha_{\mathrm{h}}$, dictates the spin of the halo. This parameter represents the fraction of particles rotating in the same direction as the disc. Thus, if $\alpha_{\mathrm{h}}$ equals 0.5, the halo exhibits no rotation. 
The chosen value of $\alpha_{\mathrm{h}}=0.8$ results in a pattern speed and bar length consistent with the observed values \citep{2016ARA&A..54..529B}.

The classical bulge follows the Hernquist profile \citep{1990ApJ...356..359H}:
\begin{align}
  \rho_{\mathrm{b}}(r) = \frac{\rho_{\mathrm{b}0}}{(r/a_{\mathrm{b}}) (1+r/a_{\mathrm{b}})^3}.
\end{align}
The scale radius, $a_{\mathrm{b}}$, the characteristic velocity dispersion, $\sigma_{\mathrm{b}} = (4\pi G a_{\mathrm{b}}^2 \rho_{\mathrm{b}0})^{1/2}$, and the truncation parameter, $\epsilon_{\mathrm{b}}$, are set to 0.75~kpc, 330~\kms, and 0.99, respectively. The rotation fraction, $\alpha_{\mathrm{b}}$, is set to 0.5, indicating that the classical bulge does not initially rotate.

The density distribution of the disc is given by a radially exponential and vertically isothermal profile:
\begin{align}
  \rho_{\mathrm{d}}(R, z) = \rho_{\mathrm{d}0}
  \exp \left(-\frac{R}{R_{\mathrm{d}}} \right)
  \mathrm{sech}^2 \left(\frac{z}{z_{\mathrm{d}}} \right).
\end{align}
The five parameters: scale radius, $R_{\mathrm{d}}$,  the scale height, $z_{\mathrm{d}}$, the total mass, $M_{\mathrm{d}}$, the disc truncation radius, $R_{\mathrm{out}}$, and the sharpness of truncation, $\delta R_{\mathrm{out}}$, characterise the model.
They are set to $R_{\mathrm{d}}=2.3\,\kpc$, $z_{\mathrm{d}}=0.2\,\kpc$, $M_{\mathrm{d}}=3.61\times 10^{10}M_{\sun}$, $R_{\mathrm{out}}=30\,\kpc$, and $\delta R_{\mathrm{out}}=0.8\,\kpc$, respectively.
The radial velocity dispersion follows $\sigma_R(R)^2=\sigma_{R0}^2\exp(-R/R_{\mathrm{d}})$, with $\sigma_{R0}=94\,\kms$.

The particle numbers for the DM halo, bulge, and disc components are $N_{\mathrm{h}}=4.9$B, $N_{\mathrm{b}}=3.1$M, and $N_{\mathrm{d}}=213$M, respectively.
We evolved the MW model for 8 Gyr and used the final snapshot for the initial condition of the host galaxy in the perturbed simulation.
The initial condition parameters for the host galaxy are summarised in Table~\ref{table:model_mw}.
\begin{table}
		\caption{Initial condition parameters for the host.}
		\label{table:model_mw}
\begin{center}
\begin{tabular}{l c}
 \hline\hline
 DM Halo (NFW profile)& \\
 \hline
 Scale radius $a_{\mathrm{h}}$ &  10 kpc  \\
 Characteristic velocity dispersion $\sigma_{\mathrm{h}}$ &  $420\,\mathrm{km \, s^{-1}}$  \\
 Truncation parameter $\epsilon_{\mathrm{h}}$ &  $0.85$  \\
 Spin fraction $\alpha_{\mathrm{h}}$ &  $0.8$  \\
 Total mass $M_{\mathrm{h}}$ & $8.7\times10^{11} M_{\sun}$\\
 Particle number $N_{\mathrm{h}}$ &  4.9B  \\
 \hline
 Classical bulge (Hernquist profile)& \\
 \hline
 Scale radius $a_{\mathrm{b}}$ &  0.75 kpc  \\
 Characteristic velocity dispersion $\sigma_{\mathrm{b}}$ &  $330\,\mathrm{km \, s^{-1}}$  \\
 Truncation parameter $\epsilon_{\mathrm{b}}$ &  $0.99$  \\
 Spin fraction $\alpha_{\mathrm{b}}$ &  $0.5$  \\
 Total mass $M_{\mathrm{b}}$ & $5.4\times10^9 M_{\sun}$\\
 Particle number $N_{\mathrm{b}}$ &  3.1M \\
 \hline
 Stellar disc & \\
 \hline
 Scale radius $R_{\mathrm{d}}$ &  2.3 kpc  \\
 Scale height $z_{\mathrm{d}}$ &  0.2 kpc  \\
 Total mass $M_{\mathrm{d}}$ &  $3.61\times10^{10} M_{\sun}$  \\
 Truncation radius $ R_{\mathrm{out}}$ &  30 kpc \\
 Sharpness of the truncation $\delta R_{\mathrm{out}}$ &  0.8 kpc  \\
 Central velocity dispersion $\sigma_{R0}$ &  $94\,\kms$  \\
 Particle number $N_{\mathrm{d}}$ &  213M \\
 \hline

\end{tabular}
\end{center}
\end{table}

The dwarf model comprises both DM and stellar components. We generated the initial condition with \texttt{Agama} \citep{2019MNRAS.482.1525V}. The distribution of the DM particles follows the NFW profile. The total mass and the scale radius of the DM halo are $5\times10^{10}M_{\sun}$ and $7.5\,\kpc$, respectively. The outer cut-off radius is set to $33\,\kpc$, which corresponds to the tidal radius of the dwarf at the initial position. 
The stellar component follows the Hernquist profile. The total mass and the scale radius  are $1\times10^9M_{\sun}$ and $2\,\kpc$, respectively.  Neither the DM component nor the stellar component has a spin.
The numbers of the DM and stellar particles are $N_{\mathrm{DM}}=$297M and $N_{\mathrm{star}}=$5.8M, respectively.
The initial condition parameters for the dwarf galaxy are summarised in Table~\ref{table:model_dwarf}.

\begin{table}
		\caption{Initial condition parameters for the dwarf.}
		\label{table:model_dwarf}
\begin{center}
\begin{tabular}{l c}
 \hline\hline
 DM (NFW profile)& \\
 \hline
 Scale radius $a_{\mathrm{DM}}$ &  7.5 kpc  \\
 Total mass $M_{\mathrm{DM}}$ &  $5\times10^{10}M_{\sun}$  \\
 Particle number $N_{\mathrm{DM}}$ &  297M  \\
 \hline
 Star (Hernquist profile)& \\
 \hline
 Scale radius $a_{\mathrm{star}}$ &  2 kpc  \\
 Total mass $M_{\mathrm{star}}$ &  $1\times10^9M_{\sun}$  \\
 Particle number $N_{\mathrm{star}}$ &  5.8M \\
 \hline
\end{tabular}
\end{center}
\end{table}

Combining the MW-like host with the Sgr-like satellite, we simulated their interaction for $\sim3$ Gyr.
We used the parallel GPU tree-code \texttt{Bonsai}\footnote{\url{https://github.com/treecode/Bonsai}} \citep{2012JCoPh.231.2825B, 2014hpcn.conf...54B} on Pegasus at Center for Computational Sciences, University of Tsukuba. 
We adopted a softening length of $0.01\,\kpc$, an opening angle of $0.4$, and a shared integration timestep of $0.61$ Myr.
Snapshots were output at intervals of 9.78~Myr.
The simulation took about 10 hours with 43 NVIDIA H100 GPUs.
The simulation data is available at \url{http://galaxies.astron.s.u-tokyo.ac.jp}.

\subsection{The time evolution of the dwarf}
\begin{figure}[ht!]
	\begin{center}
	\includegraphics[width=0.8\hsize]{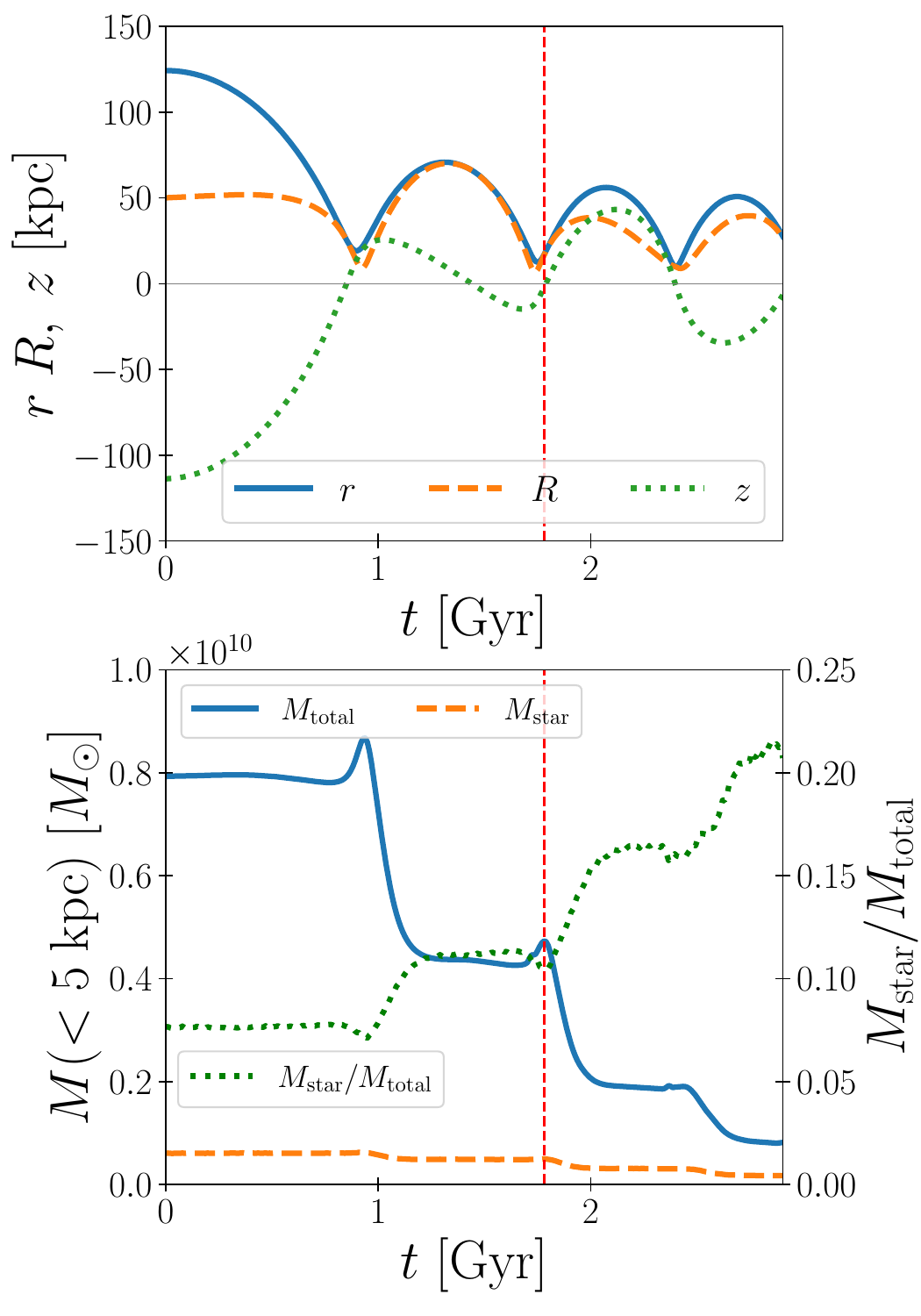}
	\caption{\textit{Upper panel:} Position of the dwarf as a function of time. Blue solid, orange dashed, and green dotted lines show $r=\sqrt{x^2 + y^2 + z^2}$, $R=\sqrt{x^2 + y^2}$, and $z$, respectively.
		\textit{Lower panel:} Mass of the dwarf as a function of time. Blue solid and orange dashed lines show the total mass and stellar mass enclosed within 5~kpc. The Green dotted line shows the stellar mass fraction.
			The red vertical line indicates the time of $t=1.78$~Gyr when the dwarf's position is close to the present-day position of the Sgr.
			The final total mass is consistent with the present-day Sgr mass.
		}\label{fig:dwarf_pos_mass}
	\end{center}
\end{figure}
\begin{figure}[ht!]
	\begin{center}
	\includegraphics[width=0.75\hsize]{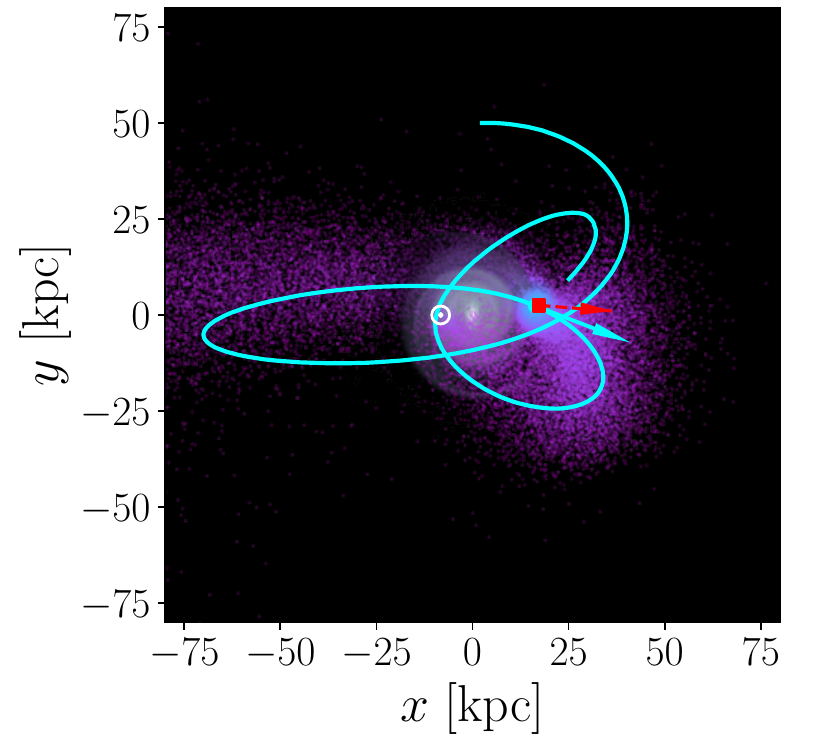}
	\includegraphics[width=0.75\hsize]{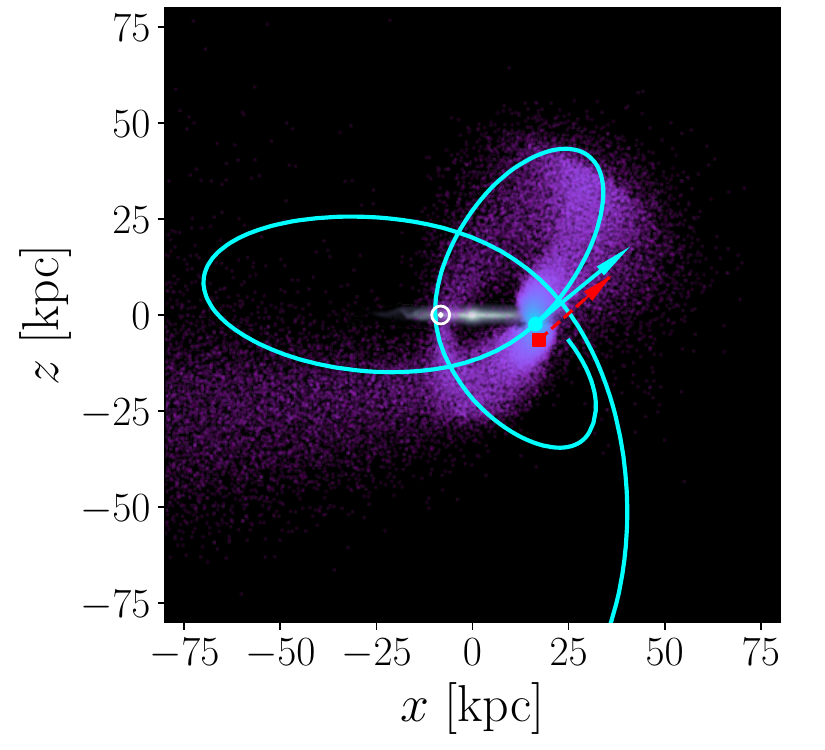}
	\caption{Projection of the dwarf's orbit in the $x$-$y$ plane (\textit{upper panel}) and in the $x$-$z$ plane (\textit{lower panel}). The cyan line indicates the orbital trajectory of the dwarf. The cyan point and cyan solid arrow indicate the dwarf's position and velocity at $t=1.78$~Gyr, respectively. The red square and red dashed arrow indicate those for the Sgr at the present day estimated by \citet{2020MNRAS.497.4162V}. Purple dots show the distribution of the stellar particles stripped away from the dwarf. The surface density of the host disc and the solar position are displayed on the background.
		The dwarf follows a nearly polar orbit similar to the Sgr, leaving a stellar stream of stripped particles along its orbital path.
	}\label{fig:dwarf_orbit}
	\end{center}
\end{figure}
Fig.~\ref{fig:dwarf_pos_mass} shows the orbit and mass evolution of the dwarf.
The dwarf is initially located at the apocentre of $r\approx 130\,\kpc$ and completes three orbit loops by the end of the simulation, with its orbit decaying due to dynamical friction. 
The dwarf's position at $t=$1.78 Gyr is closest to the present-day position of the Sgr \citep{2020MNRAS.497.4162V}.
The vertical red line in the figure indicates this time.  
The lower panel shows the time evolution of the dwarf's mass.
The solid and dashed lines indicate the total mass ($M_{\mathrm{total}}=M_{\mathrm{DM}} + M_{\mathrm{star}}$) and the stellar mass ($M_{\mathrm{star}}$) within 5 kpc from the centre of the dwarf, respectively.
The centre is defined as the density peak of the stellar component.
The dotted line represents the stellar mass fraction with respect to the total mass ($M_{\mathrm{star}}/M_{\mathrm{total}}$).
It exhibits staircase patterns, reflecting the efficient tidal stripping at the pericentre.
In the final epoch of the simulation, the total mass is $8\times10^8 M_{\sun}$, which is comparable with the observationally estimated Sgr mass, $\sim5\times10^8M_{\sun}$ \citep{2020MNRAS.497.4162V}.
Fig.~\ref{fig:dwarf_orbit} presents the projections of the dwarf's orbital trajectory into the $x$-$y$ and $x$-$z$ planes. The dwarf follows a nearly polar orbit similar to that of the Sgr.
The cyan point and the cyan arrow indicate the position and velocity of the dwarf at $t=1.78$~Gyr, respectively.
The red square and the red dashed arrow indicate those of the Sgr \citep{2020MNRAS.497.4162V}.
The purple dots represent stellar particles stripped from the dwarf, forming a stellar stream which roughly follows the orbit of the dwarf.
This feature is qualitatively consistent with the observed Sgr stream  \citep[e.g.][]{2003ApJ...599.1082M} as well as results of other simulations \citep[e.g.][]{2022ApJ...940L...3W}.

\section{Global vertical oscillations from repeated impacts}
\label{sec:repeated_impact}

In this section, we investigate the global vertical oscillations of the MW disc induced by repeated impacts from the Sgr-like dwarf galaxy. We first provide a qualitative description of the face-on evolution of the disc structure and then perform a quantitative analysis of the bending and breathing modes, focusing on their time evolution.

\subsection{Face-on maps of the galactic disc}\label{sec:face_on}
\begin{figure*}[ht!]
	\begin{center}
	\includegraphics[width=\hsize]{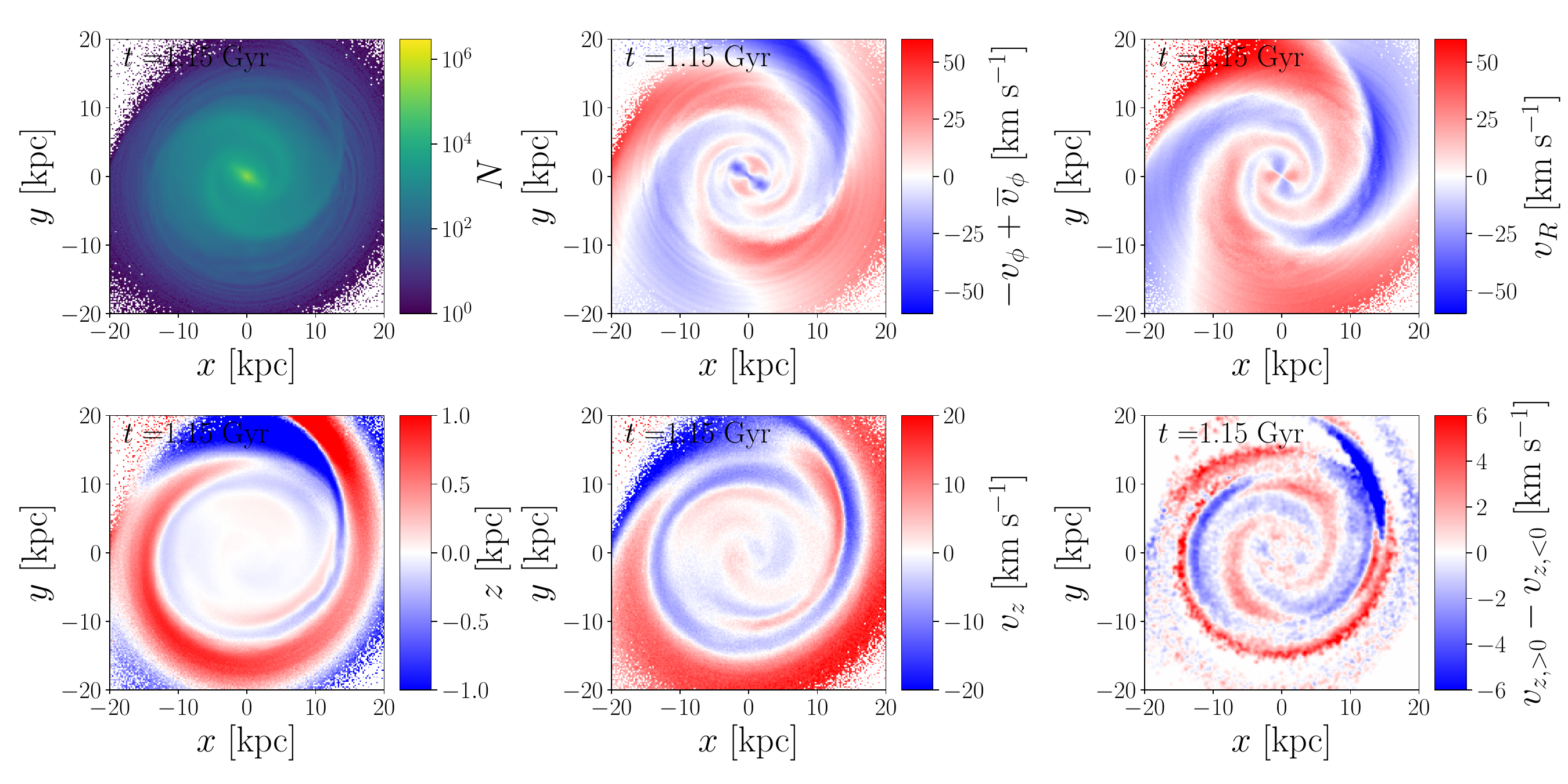}
	\caption{Face-on maps of the disc at $t=1.15$ Gyr, which is $\sim250$~Myr after the first pericentre passage of the dwarf. 
		Tidal forces from the dwarf induce prominent two-arm spiral arms and cause vertical corrugations (bending) and a breathing pattern in the galactic disc. The breathing pattern is aligned with the spiral arms.
	}\label{fig:faceon}
	\end{center}
\end{figure*}
The external perturbation by the dwarf causes a shift in the centre of mass and a tilt of the disc with respect to the initial galactic plane.
To focus on the internal structure of the disc, we redefined the coordinate system by setting $(x, y, z) = (0, 0, 0)$ at the centre of mass of the host’s stellar components (bulge and disc) and aligning the $z$-axis with their total angular momentum vector.

Fig.~\ref{fig:faceon} shows face-on maps of the galactic disc at $t=1.15$~Gyr, which is $\sim250$~Myr after the first pericentre passage of the dwarf.\footnote{The animated version is available at \url{http://galaxies.astron.s.u-tokyo.ac.jp}.}
The upper left panel is colour-coded by the particle numbers in $0.2\times0.2\, \kpc ^2$ bins, which represent the surface density of the disc.
The other panels show the mean azimuthal velocity ($v_{\phi}$) with respect to the azimuthally averaged velocity ($\bar{v}_{\phi}$), the mean radial velocity ($v_{R}$), the mean vertical coordinate ($z$), the mean vertical velocity ($v_z$), and the breathing velocity which is defined as the difference in the mean vertical velocities above and below the mid-plane ($v_{z,>0}-v_{z, <0}$). 
To reduce noise, the breathing velocity map is smoothed with a Gaussian filter of 0.2~kpc.

The surface density map (upper left panel) exhibits a bar and spiral arms.
While the bar exists before the dwarf’s interaction, the prominent two-arm spiral pattern is induced by the perturbation from the dwarf.
Additionally, we can identify faint arms.
They are transient dynamic arms, which result from the disc's self-instability not from the external perturbation.

The azimuthal and radial velocity maps (upper middle and upper right panels) exhibit quadrupole patterns in the central region of the galaxy, which are signatures of the bar.
Beyond the bar radius, the two-arm patterns align with the tidally induced spiral arms.
As studied by \citet{2022A&A...668A..61A}, a satellite fly-by initially induces a global quadrupole velocity pattern in the disc, which corresponds to the initial conditions for forming a two-arm kinematic density wave \citep{1973PASA....2..174K}.
This wave develops into the two-arm spiral pattern due to the differential rotation of the disc.

The vertical coordinate and the vertical velocity maps (lower left and lower middle panels) exhibit a corrugation of the galactic disc.
The amplitude of the corrugation increases with the galactocentric radius.
During the early stage of interaction, the galactic disc exhibits an S-shaped warp.
It starts to emerge in the outer galaxy $\sim200$~Myr before the dwarf's pericentre passage.
The amplitude of the warp grows until the pericentre passage, after which it winds up rapidly, evolving into a corrugated structure whose amplitude decreases over time.
This damping is faster in the inner galaxy ($R\lesssim10$~kpc) than in the outer galaxy ($R\gtrsim 10$~kpc).

In the breathing map (lower right panel), a two-arm pattern is observed along the tidally induced spiral arms.
As demonstrated by previous theoretical studies \citep[e.g.][]{2014MNRAS.443L...1D}, spiral arms naturally excite the breathing mode.
\citet{2021MNRAS.506...98K, 2022MNRAS.516.1114K} also performed $N$-body simulations of disc galaxies interacting with fly-by satellites and identified a prominent large-scale breathing pattern excited by tidally influenced spiral arms.
Figure~1 of \citet{2019ApJ...879L..15H} shows similar face-on breathing maps from \citet{2018MNRAS.481..286L}'s $N$-body model, where we can identify two-arm breathing pattern possibly excited by tidally induced arms.
In the central region, a quadrupole breathing pattern arises from the bar, while in the outermost region ($R \gtrsim 15$~kpc), a more complex breathing pattern emerges.
This complexity likely results from the combination of the directly excited breathing mode due to the satellite impact and the spiral-induced breathing mode, as the former cannot be neglected in the outer galaxy \citep{2023ApJ...952...65B}.

\subsection{Bending and breathing modes}\label{sec:bend_breath_ana}
We quantitatively evaluate the amplitudes of the bending and breathing modes and investigate their time evolution. 
Based on the symmetry of these two modes, we can evaluate their amplitude by comparing the vertical velocities above and below the mid-plane \citep{2014MNRAS.440.1971W, 2022MNRAS.517L..55K,2024A&A...683A..47G}.
We defined the bending and breathing velocities as
\begin{align}
	V_{\mathrm{bend}}(R,\phi,z,t)=\frac{1}{2}[\overline{v}_z(R,\phi,z,t)+\overline{v}_z(R,\phi,-z,t)],\\
	V_{\mathrm{breath}}(R,\phi,z,t)=\frac{1}{2}[\overline{v}_z(R,\phi,z,t)-\overline{v}_z(R,\phi,-z,t)],
\end{align}
where $\overline{v}_z(R,\phi,z,t)$ is the mean vertical velocity at  position $(R,\phi, z)$ at time $t$.
To obtain a vertically averaged value for each mode, $V_{n}$ ($n\in$~\{bend, breath\}), we calculated
\begin{align}
	\langle V_{n} \rangle(R,\phi,t)=\frac{1}{Z_{\mathrm{max}}}\int_0^{Z_{\mathrm{max}}}V_{n}(R,\phi,z,t)\mathrm{d}z,
	\label{eq:bend_breath_vel}
\end{align}
where $Z_{\mathrm{max}}=1.5\,\kpc$. 
We then computed the azimuthally averaged squared amplitudes of the modes:
\begin{align}
	A_n^2(R,t)=\frac{1}{2\pi}\int_0^{2\pi}{\langle V_{n}\rangle}^2(R,\phi,t)\mathrm{d} \phi.
\label{eq:bend_breath_amp}
\end{align}
For numerical evaluation, we applied binning with radial, azimuthal, and vertical resolutions of 0.5 kpc, 1$^{\circ}$, and 0.1 kpc, respectively.

\begin{figure}[ht!]
	\includegraphics[width=\hsize]{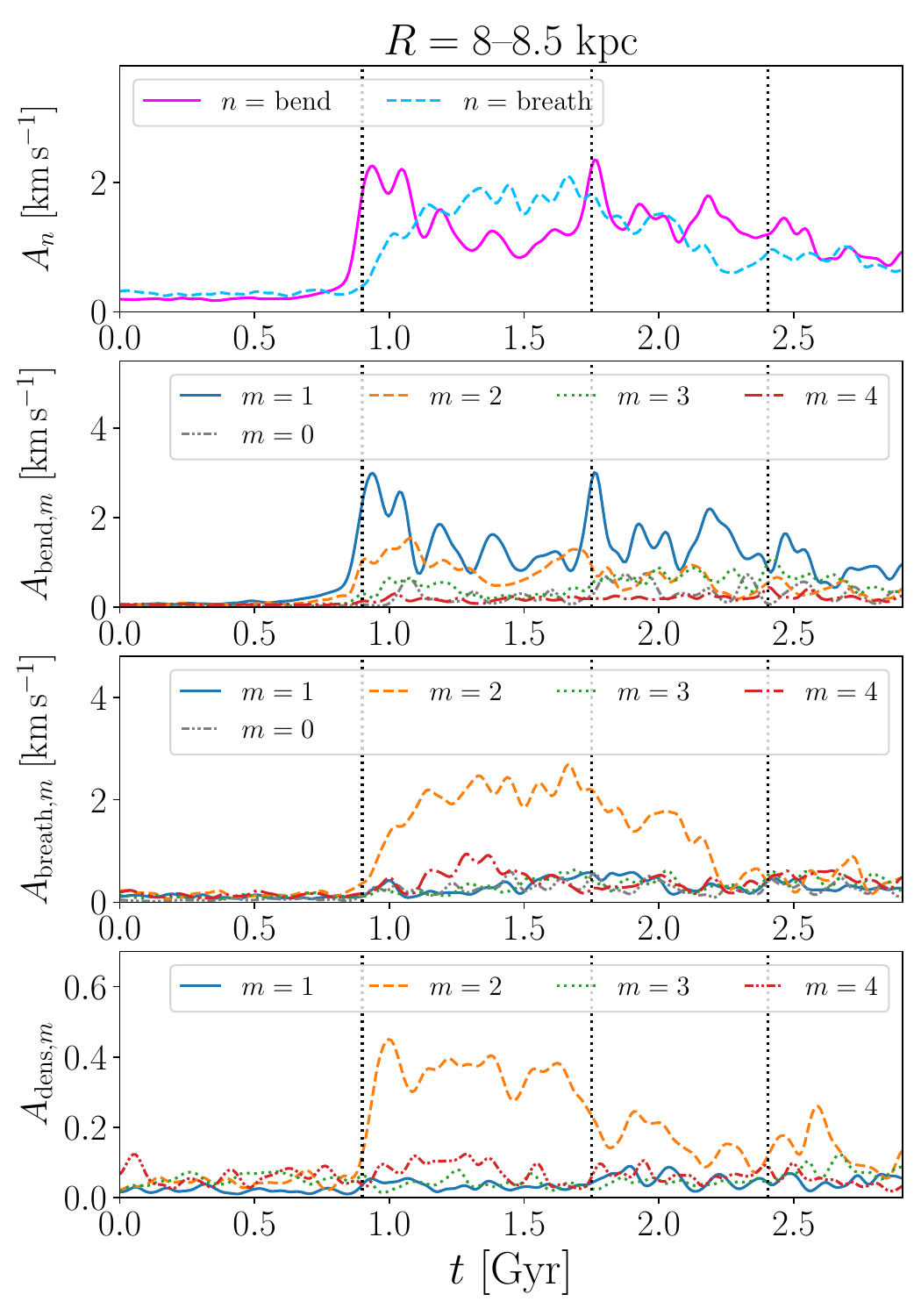}
	\caption{Time evolution of the bending mode, breathing mode, and surface density in the ring of $R=8$--8.5~kpc. \textit{First panel:} Total amplitudes of the bending (solid line) and breathing (dashed line) as functions of time. \textit{Second panel:}  Fourier amplitudes of the bending mode for $m=0$, 1, 2, 3, and 4. Different line styles correspond to different Fourier components as indicated in the legend. \textit{Third panel:} Same as the second panel but for the breathing mode. \textit{Fourth panel:} Time evolution of the Fourier amplitudes of the surface density. In all panels, vertical lines indicate the timing of the pericentre passages of the dwarf.
		The bending mode responds sharply to pericentre passages and decays quickly, while the breathing mode increases more slowly and persists longer, leading to a transition from bending-dominated to breathing-dominated state. The $m=1$ and $m=2$ are the dominant Fourier components for the bending and breathing modes, respectively.
	}\label{fig:bend_breath_dens_amp}
\end{figure}

The first panel of Figure~\ref{fig:bend_breath_dens_amp} shows the time evolution of the bending and breathing amplitudes ($A_n$) in the radial range $R = 8$–8.5 kpc.
The vertical lines indicate the dwarf's pericentre passages.
Both the bending and breathing amplitudes begin to increase around the time of the first pericentre passage ($t\sim 0.9$ Gyr), indicating that external perturbation triggers both modes.
However, their responses differ in detail.
The bending mode (solid line) increases sharply at $t\sim 0.85$~Gyr ($\sim 50$~Myr before the pericentre passage) and decays shortly after the pericentre passage.
In contrast, the breathing mode (dashed line) starts to increase after the pericentre passage, with its amplitude reaching saturation around $t=1.2$~Gyr.
Initially, the bending mode dominates, but approximately 300 Myr after the pericentre passage, the breathing mode becomes dominant.
This transition of the dominant mode reflects the differences in the growth and decay timescales of the two modes, as discussed in more detail in the next section.

After the second pericentre passage ($t \sim 1.75$~Gyr), the bending mode again shows a sharp increase in amplitude, whereas the breathing mode shows no clear response.
During this epoch, the breathing amplitude gradually decreases. 
Between the second and third pericentres, $1.75 \lesssim t \lesssim 2.4$ Gyr, the bending amplitude remains higher or comparable to the breathing amplitude.
The third pericentre passage does not significantly impact either the bending or breathing modes.

\subsection{Fourier decomposition}
To gain deeper insight into the temporal evolution of the bending and breathing modes, we performed the Fourier decomposition of the bending and breathing modes.
Similar Fourier analyses were employed in previous studies \citep[e.g.][]{2017MNRAS.472.2751C,2018MNRAS.480.4244C, 2021MNRAS.508..541P, 2021MNRAS.507.2825G, 2024A&A...683A..47G}.
The vertically averaged bending and breathing velocities of Equation~\eqref{eq:bend_breath_vel}
were decomposed as
\begin{align}
	\langle V_{n}\rangle(R,\phi,t) &= 
	\sum_{m=-\infty}^{\infty}Q_{n,m}(R,t) \exp(im\phi) \\
	&= \sum_{m=0}^{\infty}A_{n,m}(R,t) \cos\{m[\phi-\phi_{n,m}(R,t)]\},
\end{align}
where $Q_{n,m}$, $A_{n,m}$, and $\phi_{n,m}$ are the complex coefficient, amplitude, and phase of the $m$-th Fourier component, respectively.
Since $\langle V_n \rangle$ is a real function, the complex Fourier coefficients are related to the amplitude and the phase as
\begin{align}
	Q_{n, m}(R,t)=
	\begin{cases}
		A_{n,0}(R,t) & \text{for }  m= 0,\\
		\frac{1}{2}A_{n,m}(R,t) e^{-im\phi_{n, m}(R,t)} & \text{for }  m= 1,\,2,\,\ldots\\
	\frac{1}{2}A_{n,-m}(R,t) e^{im\phi_{n, -m}(R,t)} & \text{for }  m=-1, \, -2, \, \ldots
	\end{cases}
\end{align}

Similarly, we performed the Fourier decomposition of the normalised surface density,
\begin{align}
	\frac{\Sigma(R,\phi,t)}{\Sigma_0(R,t)} &=
	\sum_{m=-\infty}^{\infty}Q_{\mathrm{dens},m}(R,t) \exp(im\phi) \\
	&= 1 + \sum_{m=1}^{\infty} A_{\mathrm{dens}, m}(R,t)\cos \{m[\phi-\phi_{\mathrm{dens}, m}(R,t)]\},
\end{align}
where $\Sigma_0(R, t)$ is the azimuthally averaged surface density.

The second panel of Fig.~\ref{fig:bend_breath_dens_amp} shows the Fourier amplitudes of the bending mode for  $m=0$, 1, 2, 3, and 4.
The $m=1$ and $m=2$ components correspond to the S-shaped and U-shaped warps, respectively. The $m=0$ component represents the ring-shaped vertical oscillation. It is sometimes referred to as the bell mode \citep{1994ApJ...425..551M, 1994ApJ...425..530S}.
Among these, the $m=1$  component is the most dominant, growing at $t\sim0.85$ Gyr, reaching its peak at $t\sim0.95$ Gyr, and subsequently undergoing a damping oscillation.

The  $m=2$  component, the second most dominant, also starts growing at $t\sim0.85$ Gyr. However, it lags behind the  $m=1$  component, reaching its maximum amplitude at $t\sim1.2$ Gyr.
Interestingly, the $m=2$ component experiences a secondary increase in amplitude after its initial decay just before the second pericentre passage of the dwarf, although the cause of this behaviour is unclear.
In other similar simulations, the $m=2$ component dominates over the $m=1$ component.
In the $N$-body simulation of \citet{2018MNRAS.481..286L}, $m=2$ component becomes dominant across a wide radial range 200--300~Myr after the first disc crossing \citep{2021MNRAS.508..541P}.
Similarly, \citet{2021MNRAS.504.3168B} show that the $m=2$ component is the dominant bending component in the inner galaxy ($R\lesssim10$~kpc) in their simulation.
The difference in the dominant Fourier component may be due to the differences in satellite orbits, since the relative strength of the $m=1$ and $m=2$ components depends on the satellite's velocity \citep{2023ApJ...952...65B}.

After the second pericentre passage, the $m=1$ component responds sensitively to external perturbation, while the amplitudes of the other components remain largely unchanged and comparable to each other.
The third pericentre passage does not significantly affect any of the Fourier components. 

The third panel of Fig.~\ref{fig:bend_breath_dens_amp} illustrates the Fourier amplitudes of the breathing mode over time. The $m=2$  component dominates, consistent with the total amplitude shown in the first panel. While the first pericentre passage triggers the $m=2$ component, the second and third passages have little impact.
The $m=2$ breathing amplitude slowly decreases from $t\sim1.7$~Gyr and drops rapidly at  $t\sim2.3$~Gyr.

The fourth panel of Fig.~\ref{fig:bend_breath_dens_amp} shows the Fourier amplitudes of the surface density. The $m=2$ component grows after the first pericentre passage, indicating the formation of tidally induced spiral arms, but its amplitude decreases around the second pericentre passage. It remains unclear whether this decline is directly caused by the second passage, as the trend begins earlier. A slight resurgence is observed after the third pericentre passage. While single fly-by interactions are known to generate two-arm spiral arms in smooth discs \citep{2011MNRAS.414.2498S, 2016MNRAS.458.3990P, 2021MNRAS.504.3168B, 2022A&A...668A..61A}, it is uncertain whether such perturbations strengthen or weaken pre-existing arms. Notably, the similarity in the Fourier amplitudes of the breathing mode and the surface density suggests that the breathing mode is excited by spiral arms.

\subsection{Bending and breathing amplitudes as functions of time and radius}
\begin{figure}[ht!]
	\begin{center}
	\includegraphics[width=\hsize]{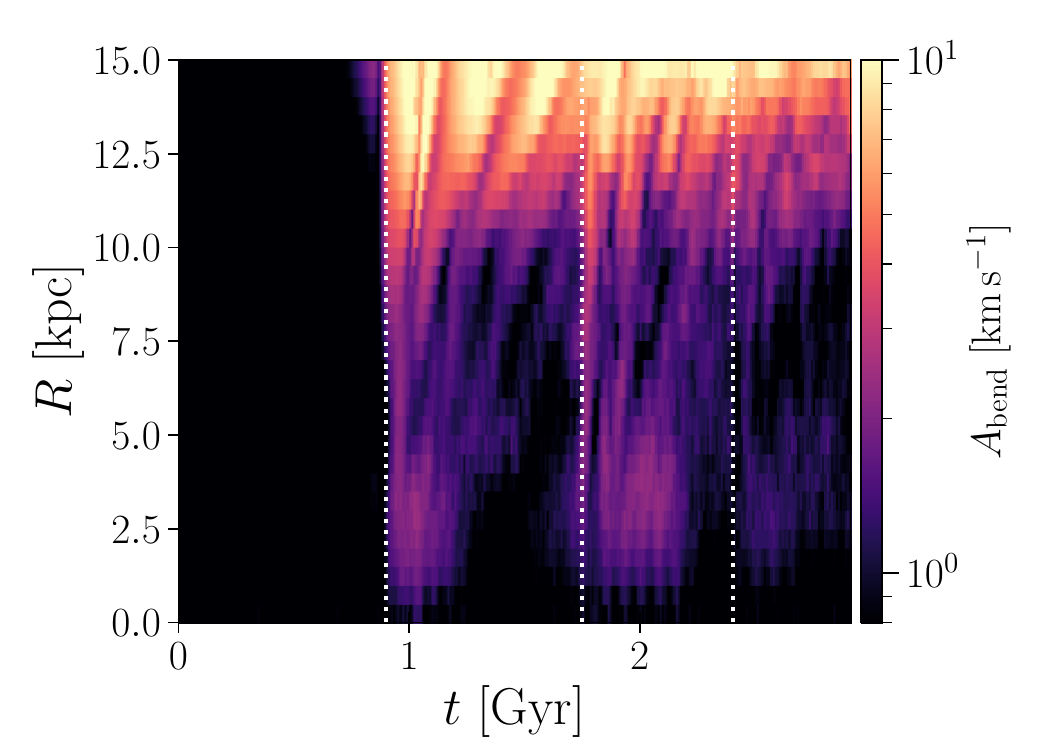}
	\includegraphics[width=\hsize]{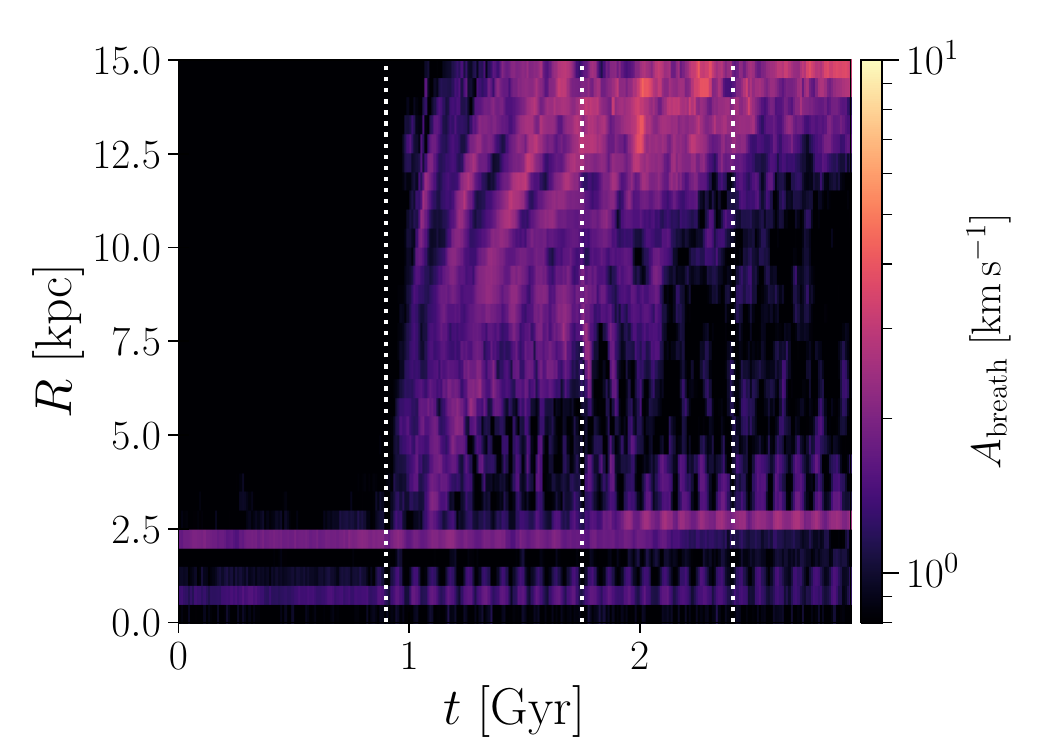}
	\includegraphics[width=\hsize]{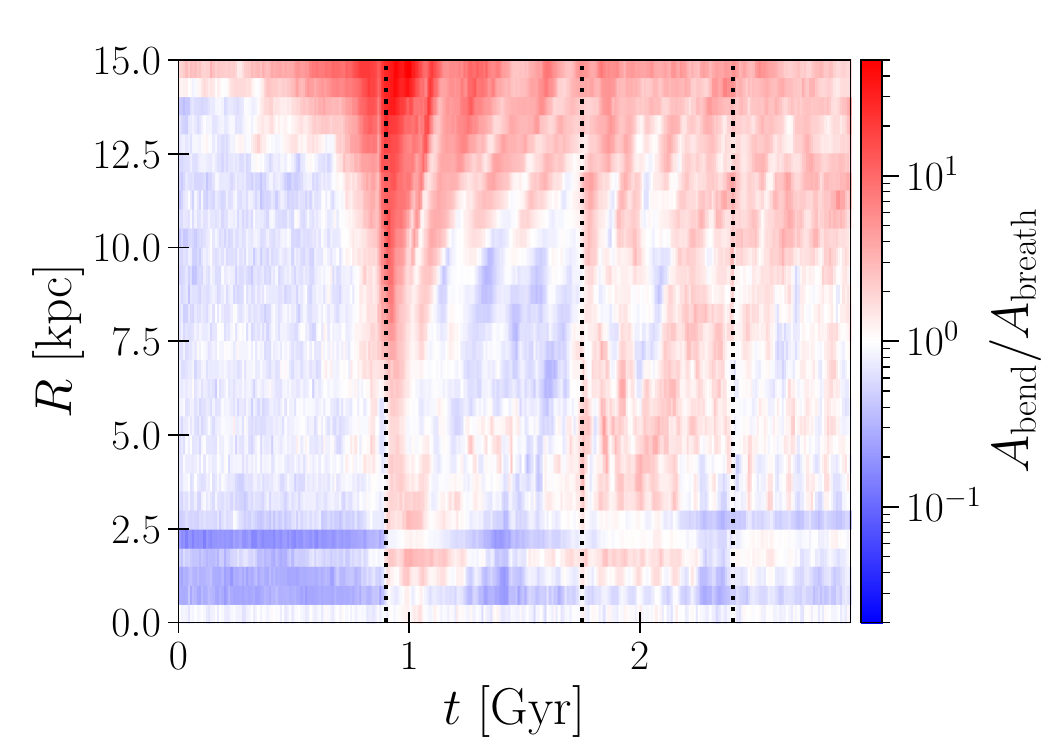}
	\caption{Bending amplitude, breathing amplitude, and their ratio as functions of time and galactocentric radius. \textit{First panel:} Total amplitude of the bending mode. \textit{Second panel:} Total amplitude of the breathing mode. \textit{Third panel:} Amplitude ratio between the bending mode and the breathing mode.  The vertical lines indicate the times of the dwarf’s pericentre passages.
	The dominant mode changes from bending to breathing. This transition occurs faster in the inner galaxy.
	}\label{fig:t_R_bend_breath}
	\end{center}
\end{figure}

In the previous subsection, we examined the temporal evolution of the bending and breathing modes at a fixed radius. Here, we extend our analysis to explore how these modes vary as functions of both time and radius. This allows us to investigate the spatial dependence of their excitation and decay processes.

The first panel of Fig.~\ref{fig:t_R_bend_breath} shows the bending amplitudes, as defined in Equation~\eqref{eq:bend_breath_amp}, as functions of time and the galactocentric radius. 
The first close encounter excites the bending mode across a wide radial range.
The amplitude increases with $R$, as the dwarf crosses the outer disc ($R=25.5$~kpc).
The maps reveal ridge structures, indicating that the bending amplitude oscillates with time.
These ridges gradually fade over time, suggesting that the bending oscillation behaves as a damping wave rather than a stationary wave.
The second pericentre passage excites the bending mode again, but the ridge pattern becomes less regular compared to that after the first passage, reflecting increased complexity from repeated interactions.
While the third pericentre passage causes a slight increase in the bending amplitude, its impact is notably weaker than the earlier passages.

The second panel of Fig.~\ref{fig:t_R_bend_breath} shows the same plot but for the breathing mode.
Before the dwarf’s first pericentre passage, a bar-induced breathing mode is visible as double peaks at $R\sim1$~kpc and $R\sim2.5$~kpc.
These features are linked to the boxy-peanut shape of the bar (Asano et al.\ in prep.).
Outside the bar region, the breathing mode grows after the first pericentre passage, driven by tidally induced spiral arms, as observed in Figs.~\ref{fig:faceon} and \ref{fig:bend_breath_dens_amp}.
Unlike the bending mode, the breathing mode exhibits slower growth and decay, persisting over longer timescales.
For example, at $R=8$~kpc, the breathing mode remains significant from $t\sim1$~Gyr to $\sim2$~Gyr.
The second and third pericentre passages, however, do not strongly influence the breathing mode.

The third panel of Fig.~\ref{fig:t_R_bend_breath} is colour-coded by the amplitude ratio between the bending and the breathing modes.
In this map, red and blue regions represent dominance by the bending and breathing modes, respectively.
Before the dwarf's first pericentre passage, the breathing amplitude exceeds the bending amplitude across most parts of the disc, except in the outermost region.
At the first pericentre passage, the bending mode becomes dominant.
Over time, however, the dominant mode transitions back to the breathing mode, with this transition occurring faster in the inner galaxy than in the outer galaxy.
By the time of the second pericentre passage, the breathing mode dominates within $R\lesssim10$~kpc.

This temporal transition reflects the distinct excitation and decay behaviours of the two modes. While the bending mode is excited instantaneously by the satellite’s perturbation and decays rapidly, the breathing mode is continuously driven by the spiral arms and persists as long as the arms remain sufficiently strong.
This difference in temporal behaviour between the two modes drives the transition of the dominant mode.

Between the second and third pericentre passages, no clear transition is observed, since the breathing mode also begins to decay during this period. After the third pericentre passage, the outer galaxy ($R \gtrsim 10$~kpc) is dominated by the bending mode, while the inner galaxy ($R \lesssim 3$~kpc) is dominated by the bar-induced breathing mode.
At intermediate radii, the amplitudes of the two modes are comparable, reflecting a balance between the decaying bending and breathing modes.

\section{Long-term evolution of the bending mode and the breathing mode after a single impact}\label{sec:long_term_evolution}

In the previous section, we investigated how repeated interactions with the dwarf galaxy introduce complexities into the MW disc. To isolate the impact of a single perturbation, we now investigate the long-term evolution of global vertical oscillations in the disc after a single impact.
To this end, we performed the same $N$-body simulation of the MW-Sgr system but stopped the run at $t=1.2$~Gyr, when the dwarf reaches its first apocentre.
At this point, we removed the dwarf's stellar and DM particles within 30~kpc from its centre.
After removing the main body of the dwarf, we continued the simulation for an additional 1.5~Gyr. 

\subsection{Damping of the bending and breathing modes}
\begin{figure}
	\includegraphics[width=\hsize]{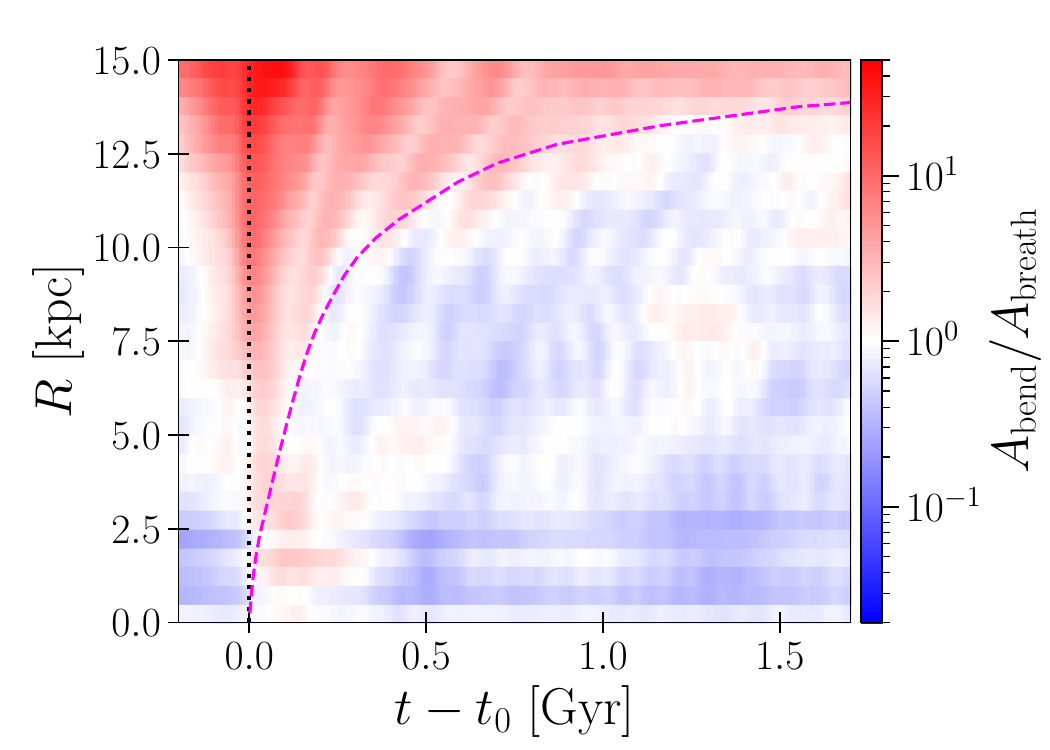}
	\includegraphics[width=\hsize]{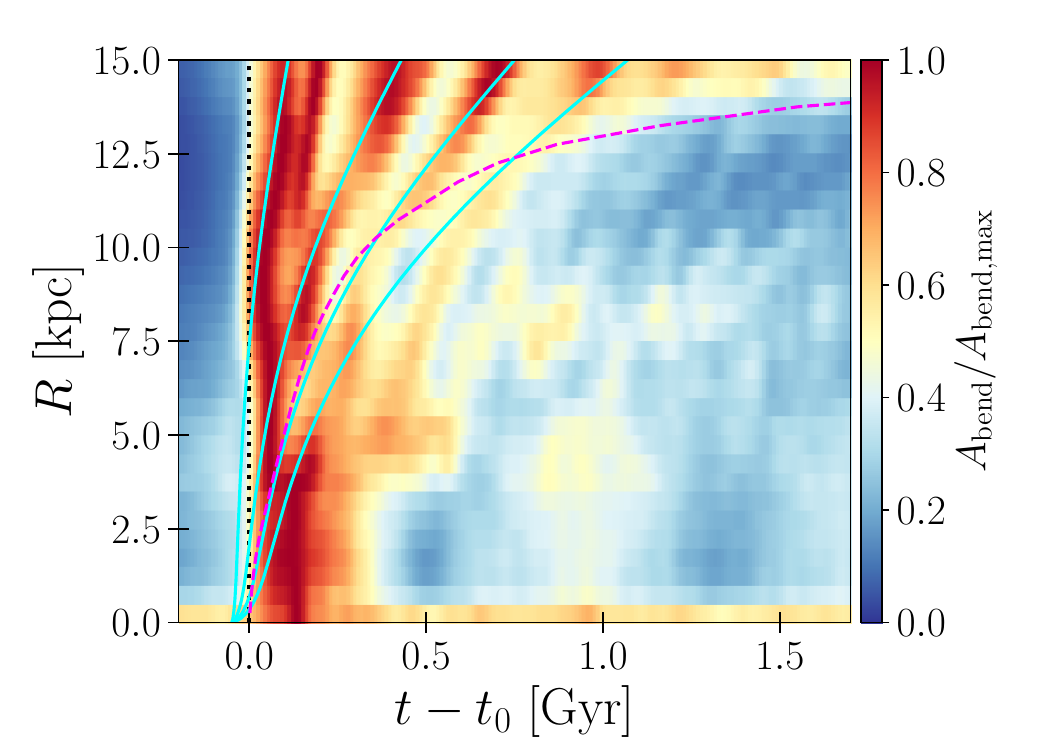}
	\includegraphics[width=\hsize]{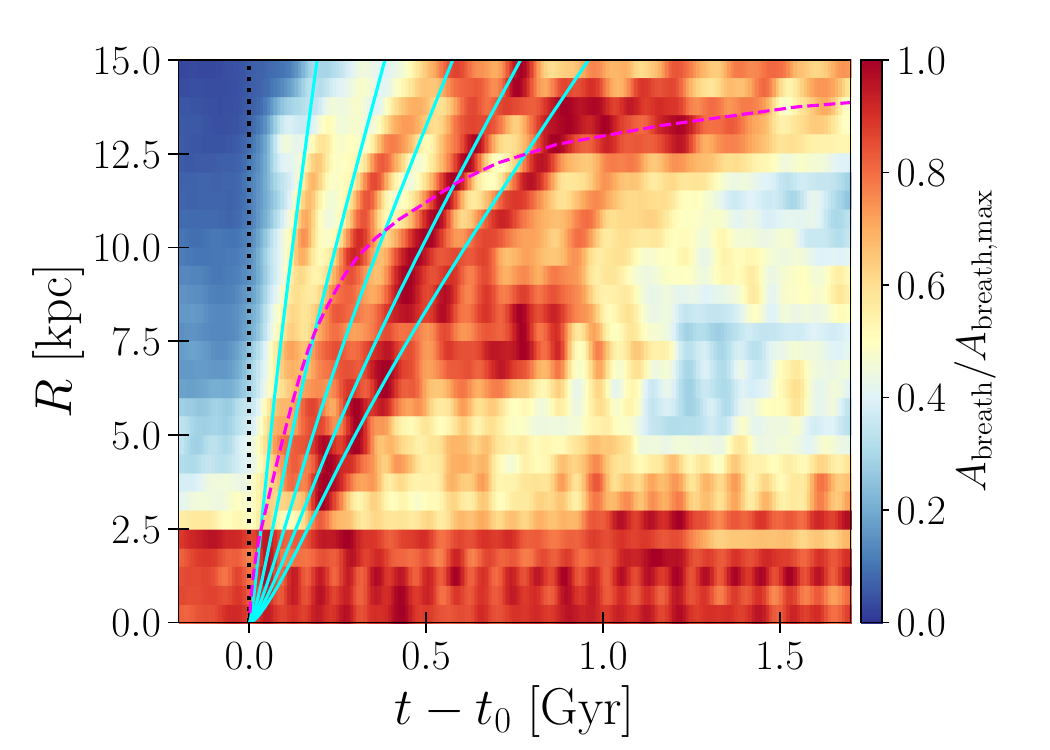}
	\caption{Evolution of the bending mode and the breathing mode in the single-impact model. \textit{First panel:} Bending and breathing amplitude ratio as a faction of time and the galactocentric radius. The horizontal axis indicates the time since the pericentre passage of the dwarf. A vertical dotted line indicates the pericentre time $t-t_0=0$. Dashed curve represents $R/\sigma_R(R)$. \textit{Second panel:} Bending amplitude as a function of time and the galactocentric radius. The amplitude is normalised by the maximum amplitude at each radius. Solid curves represent $(1,3,5,7) \times \pi/\nu(R) - 50 \, \mathrm{Myr}$. \textit{Third panel:} Same as the second panel but for the breathing mode. Solid lines represent $(1,2,3,4)\times \pi/\Omega(R)$.
	The bending mode decays rapidly due to horizontal mixing, while the breathing mode persists longer, driven by spiral arms. This difference of the two modes causes a transition to breathing dominance over time on the timescale of horizontal mixing ($R/\sigma_R$).
	}\label{fig:t_R_single}
\end{figure}

The first panel of Fig.~\ref{fig:t_R_single} shows the ratio of the bending to breathing amplitude as a function of $t-t_0$ and $R$ in the single-impact model, where $t_0\sim0.9$ Gyr is the pericentre passage time of the dwarf.
Shortly after the pericentre passage, the bending mode dominates across the entire disc, but over time, the breathing mode becomes dominant, particularly in the inner regions (Fig.~\ref{fig:t_R_bend_breath}).
The dashed curve represents $R/\sigma_R(R)$, where $\sigma_R(R)$ is the radial velocity dispersion at $R$.
It roughly separates the bending-dominated and breathing-dominated regions.
By analogy with wave damping in horizontally uniform slabs \citep{2022ApJ...935..135B}, we can interpret the timescale $R/\sigma_R(R)$ as being relevant to horizontal mixing.
We describe a mechanism in detail below.

The second panel of Fig.~\ref{fig:t_R_single} shows the bending amplitude, normalised by its maximum value at each radius, as a function of $t$ and $R$.
The solid curves in the panel show $(1, 3, 5, 7) \times \pi / \nu(R) - 50$~Myr, where $\nu(R)$ is the vertical frequency at $R$.
To compute $\nu$, we constructed an axisymmetric potential model based on the parameters for the initial condition described in Section~\ref{sec:initial_condition} using \texttt{Agama} \citep{2019MNRAS.482.1525V}.
The bending amplitude shows a characteristic damping oscillation, with ridges of positive slope closely following the solid lines that represent $2\pi/\nu(R)$.
Over time, the amplitude decays by $\sim 40$--60\%, with the damping timescale corresponding to $R / \sigma_R$.

The damping observed here is consistent with horizontal (lateral)\footnote{In the context of \citet{2022ApJ...935..135B} who analytically studied phase mixing in a horizontally ($x$ and $y$ directions) uniform and vertically ($z$ direction) isothermal slab, the term lateral mixing specifically denotes mixing in the $x$ and $y$ directions. Assuming a Maxwellian velocity distribution and an impulsive perturbation, and neglecting the effects of self-gravity, the decay of the perturbation term in the distribution function due to lateral phase mixing can be expressed as shown in Eq.~\eqref{eq:phase_mixing}. This decay corresponds to the reduction in the oscillation amplitude as well.} phase mixing in a horizontally uniform slab, where the response to an impulsive perturbation decays as
\begin{align}
	 \exp \left( -\frac{\sigma^2 k^2 t^2}{2} \right),
	 \label{eq:phase_mixing}
\end{align}
where $\sigma$ and $k$ are the horizontal velocity dispersion and the horizontal wave number, respectively, and the characteristic timescale of the damping is expressed as $\tau_D = 1/(\sigma k)$ \citep{2022ApJ...935..135B}.
Although our model is more complex than the infinite slab, the damping timescale due to horizontal mixing should be of the same order as that obtained by substituting $\sigma \sim \sigma_R$ and $k \sim 1/R$ into this formula (i.e. $\tau_D = R/\sigma_R$)

Similar damping behaviour is observed in simpler systems. 
Section~5.2 of \citet{2008gady.book.....B} examines the responses of an infinite, homogeneous stellar system with a Maxwellian distribution function.
In this case, the spatial Fourier transform of the polarization function is proportional to Eq.~\eqref{eq:phase_mixing}.
In general, waves in stellar systems are damped by two processes: phase mixing and Landau damping.
Phase mixing is a kinematic process and occurs even in the absence of self-gravity, 
whereas Landau damping arises from self-gravitating effects and occurs more slowly than phase mixing. 
In the inner galaxy, where self-gravity is stronger, the damping timescale is longer than predicted by the $R / \sigma_R$ scaling, as seen in the second panel of Fig.~\ref{fig:t_R_single}.

The third panel of Fig.~\ref{fig:t_R_single} shows the breathing amplitude, normalised by its maximum values at each radius, as a function of $t$ and $R$.
The solid lines represent $(1, 2, 3, 4) \times \pi / \Omega(R)$, where $\Omega(R)$ is the circular frequency at $R$.
The breathing mode takes approximately 1–1.5 circular periods to reach significant amplitudes after the pericentre passage. Once excited, it persists for up to $\sim1$~Gyr, driven by the tidally induced spiral arms responsible for exciting the mode. Unlike the bending mode, the breathing mode does not decay rapidly. As a result, the time evolution of the amplitude ratio between the two modes is primarily governed by the damping of the bending mode. While the bending mode fades, the breathing mode’s persistence reflects the longevity of the spiral arms. This distinction highlights the contrasting roles of phase mixing and spiral-arm dynamics in shaping the vertical oscillations of the disc.

\subsection{Spectral analysis}
\begin{figure*}
	\includegraphics[width=0.33\hsize]{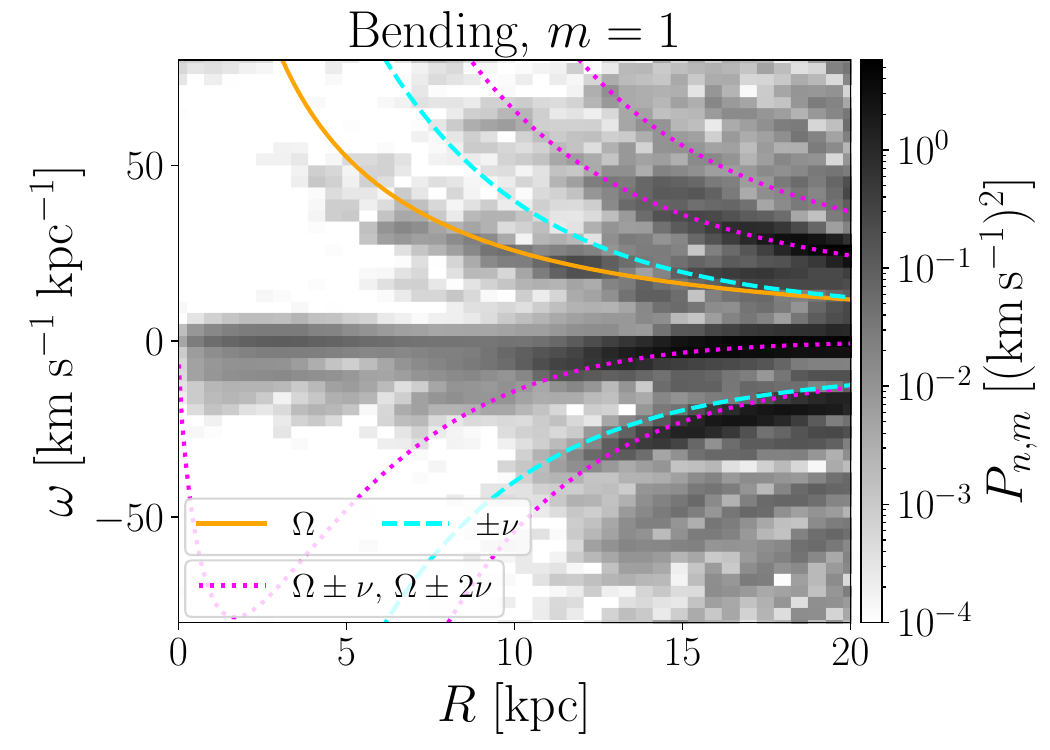}
	\includegraphics[width=0.33\hsize]{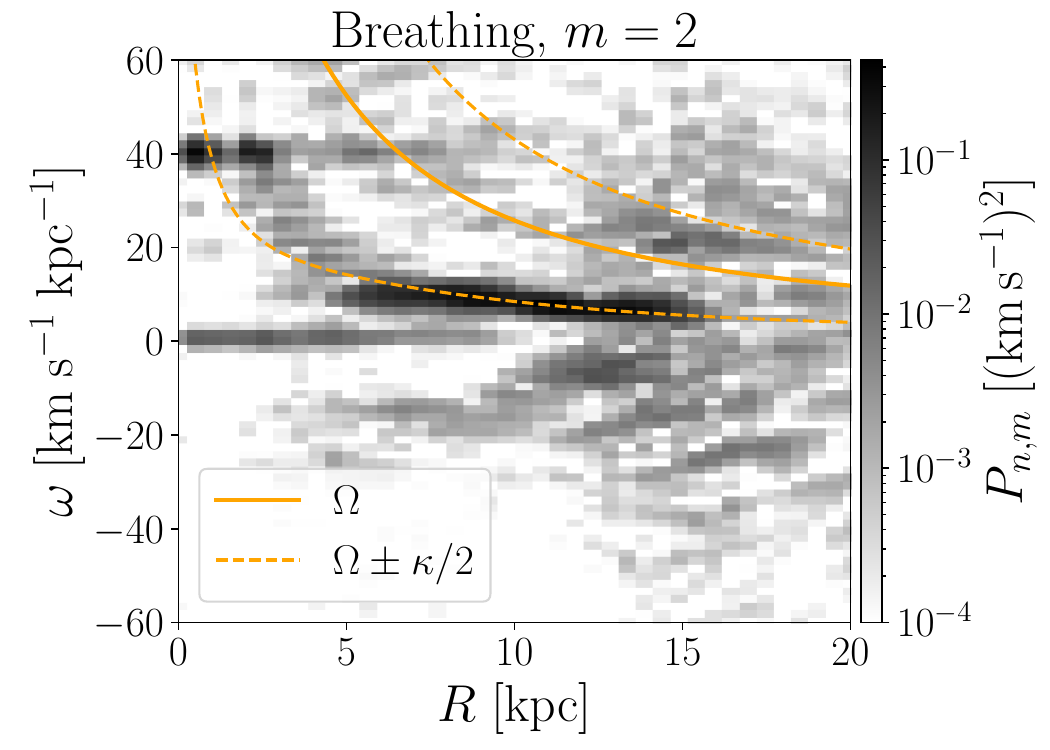}
	\includegraphics[width=0.33\hsize]{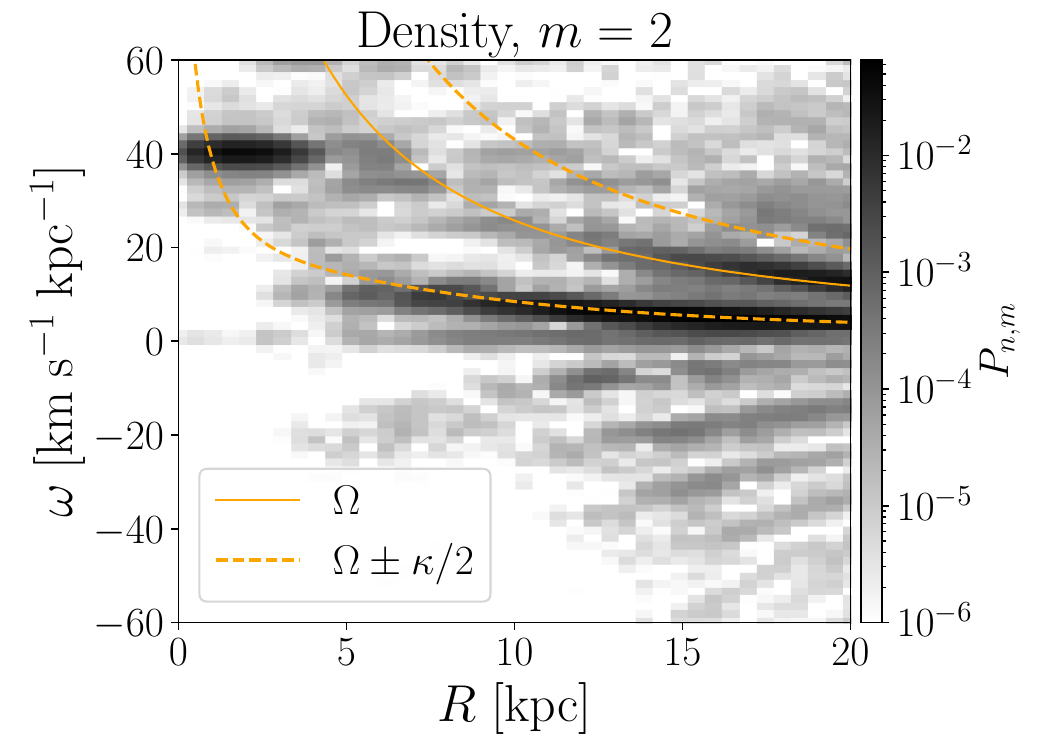}
	\caption{\textit{First panel:} Spectrogram of the $m=1$ bending mode. Lines indicate characteristic frequencies: $\Omega$ (solid), $\pm \nu$ (dashed), $\Omega \pm \nu$, and $\Omega \pm 2 \nu$ (dotted).  \textit{Second panel:} Spectrogram of the $m=2$ breathing mode. Solid and dashed lines indicate $\Omega$ and $\Omega \pm \kappa/2$, respectively. \textit{Third panel:} Spectrogram of the surface  $m=2$ density wave.  Solid and dashed lines indicate $\Omega$ and $\Omega \pm \kappa/2$, respectively. 
		The bending mode shows distinct branches aligned with the resonant curves of $\Omega \pm \nu$ and $\Omega - 2\nu$.
		The breathing mode and the density wave rotate at nearly the same pattern speeds, suggesting that the breathing mode is excited by the bar and the tidally induced arms.
	}\label{fig:spectogram}
\end{figure*}

To investigate the connection between bending and breathing modes and density waves (such as bars and spiral arms), we performed a spectral analysis \citep{1986MNRAS.221..195S,2012MNRAS.426.2089R, 2017MNRAS.472.2751C, 2018MNRAS.480.4244C, 2021MNRAS.508..541P, 2022MNRAS.517L..55K}. This analysis provides insights into the frequencies of these modes across the galactocentric radius by visualising their power spectra in the radius-frequency space.
We followed the method formalised by \citet{2017MNRAS.472.2751C} for the computation of the spectrum.

The temporal discrete Fourier transform of the complex Fourier coefficients $Q_{n, m}(R,t)$ ($n \in$ \{bend, breath, dens\}) was defined as
\begin{align}
	F_{n,m}(R,\omega_k)=\sum_{j=0}^{N-1}Q_{n,m}(R,t_j)w(j)\exp \left( 2\pi i \frac{jk}{N} \right),
\end{align}
where $w(j)$ is a window function.
To suppress spectral leakage, we used the Gaussian function with a width of $N/4$.
The time interval used is between the pericentre passage and the end of the simulation, with  $N = 185$  time steps in this interval.
The frequency $\omega_k$ is given by 
\begin{align}
	\omega_k=\frac{2\pi}{m}\frac{k}{N\Delta t};\quad k=-N/2\ldots N/2,
\end{align}
where $\Delta t=9.78$~Myr is the output cadence of the snapshot.
The power spectrum was defined as
\begin{align}
	P_{n,m}(R,\omega_k)=\frac{1}{N\sum_jw(j)}|F_{n,m}(R,\omega_k)|^2.
\end{align}

The first panel of Fig.~\ref{fig:spectogram} shows the power spectrum $P_{n,m}(R,\omega)$ on $R$-$\omega$ plane (spectrogram) of the $m=1$ bending mode, which is the dominant Fourier component.
Characteristic frequencies, such as $\Omega$, $\pm \nu$, $\Omega \pm \nu$, and $\Omega \pm 2\nu$, are overplotted for comparison.
At $R\gtrsim13$~kpc, three distinct branches are visible: $\Omega \pm \nu$, corresponding to the pattern speed of the kinematic bending wave \citep{2008gady.book.....B, 2018MNRAS.480.4244C}, and $\Omega - 2\nu$, which has also been observed in other simulations \citep{2017MNRAS.472.2751C, 2018MNRAS.480.4244C, 2021MNRAS.508..541P}.\footnote{This branch might follow $-\nu$ rather than $\Omega-2\nu$, as $\Omega \sim \nu$ in the outer galaxy.} Higher-order frequencies, such as $\Omega \pm 3\nu$, are also present but significantly weaker.  In the inner galaxy ($R \sim 5$--13~kpc), a faint branch approximately follows $\Omega$. Interestingly, this branch lies within the forbidden region ($\Omega - \nu < \omega < \Omega + \nu$) according to WKB analysis \citep{2008gady.book.....B, 2018MNRAS.480.4244C}. Its presence suggests that the linear approximation may break down in the inner galaxy, likely due to the stronger influence of self-gravity.

The second and third panels of Fig.~\ref{fig:spectogram} show the spectrograms of the $m=2$ breathing and density components, respectively. Their similarities imply a strong connection between the breathing mode and non-axisymmetric density structures such as spiral arms and bars.
A horizontal line at $\omega \sim 40\,\kmskpc$ at $R\lesssim 4\,\kpc$ corresponds to the bar.
Meanwhile, a branch along $\Omega - \kappa/2$ is visible, aligning with the pattern speed of tidally induced spiral arms \citep{2022A&A...668A..61A}. This matches the kinematic density wave pattern speed \citep{2008gady.book.....B}.

In the outer galaxy (\( R \gtrsim 12 \) kpc), the spectrogram of the density mode shows a branch that follows $\Omega$, consistent with dynamic spiral arms that corotate with disc stars \citep{2013ApJ...763...46B}.
These dynamic arms differ from tidally induced spiral arms, which rotate more slowly relative to the disc. The correspondence between the $m=2$ breathing mode and spiral arms supports the hypothesis that spiral structures play a significant role in driving the breathing mode.

\section{Phase spirals}\label{sec:phase_spiral}
In the previous section, we focused on the global vertical oscillations of the disc, namely the bending and breathing modes, which provide insights into the large-scale impact of the dwarf galaxy. In this section, we shift our focus to the local scale, examining the formation and evolution of phase spirals—vertical phase-space features that are critical for understanding the disc’s response on smaller scales. These structures offer complementary insights into the dynamical processes studied in the global analysis.

\subsection{Maps of $z$-$v_z$ space in the solar neighbourhood}
\begin{figure*}
	\begin{center}
	\includegraphics[width=0.18\hsize]{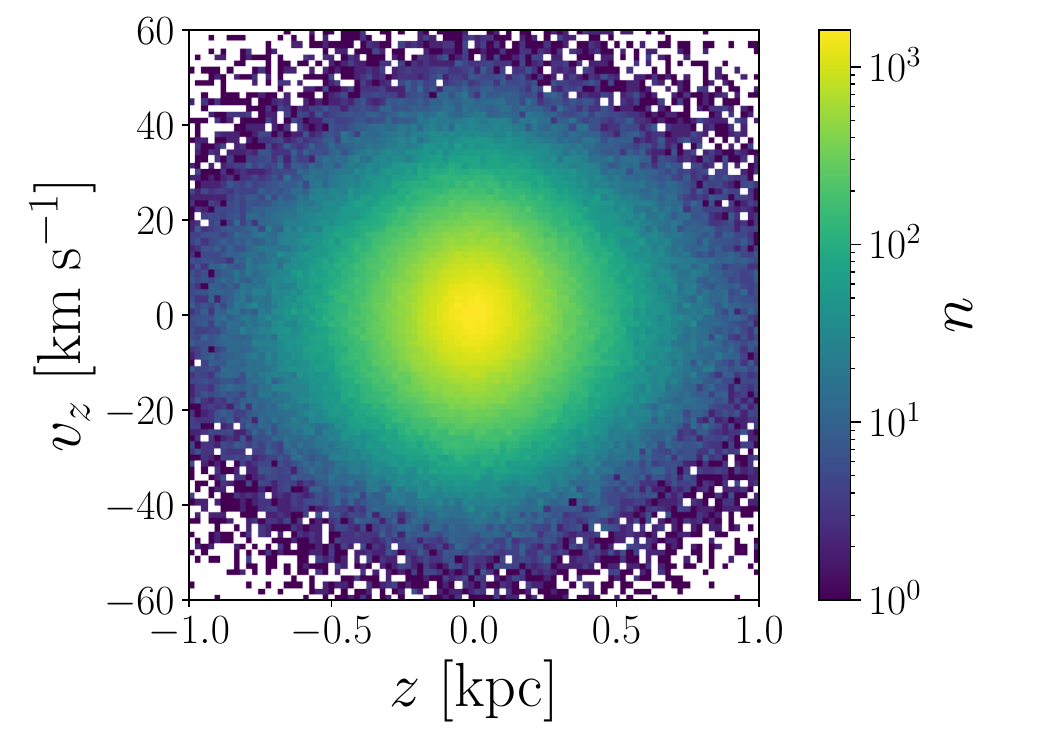}
	\includegraphics[width=0.18\hsize]{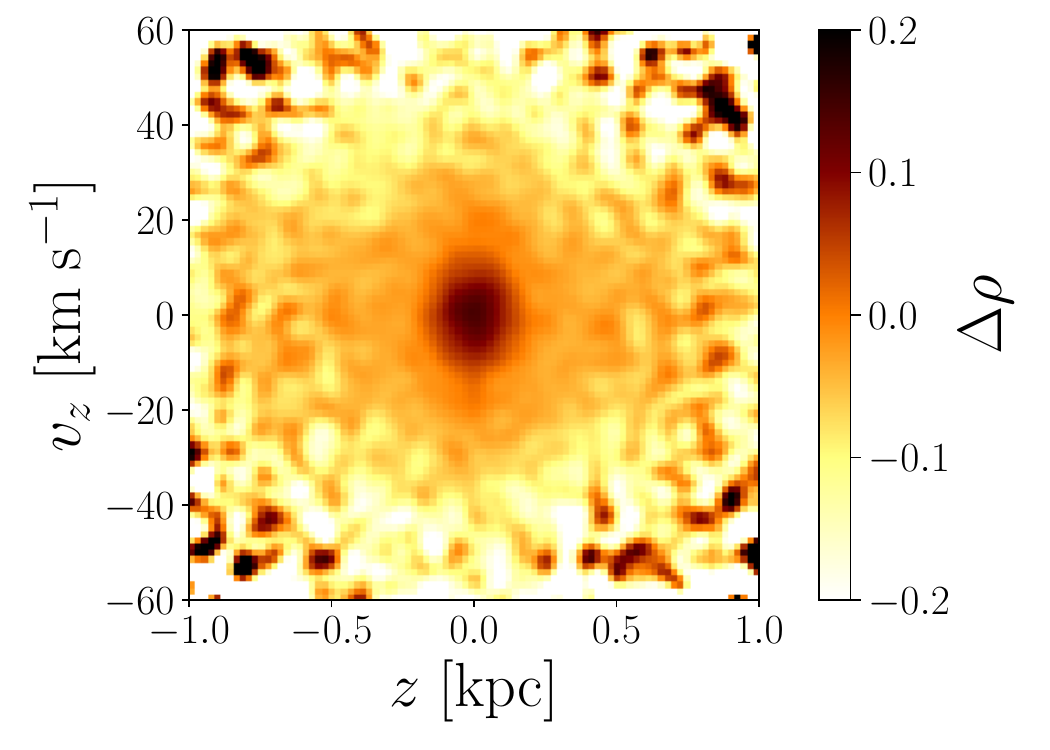}
	\includegraphics[width=0.18\hsize]{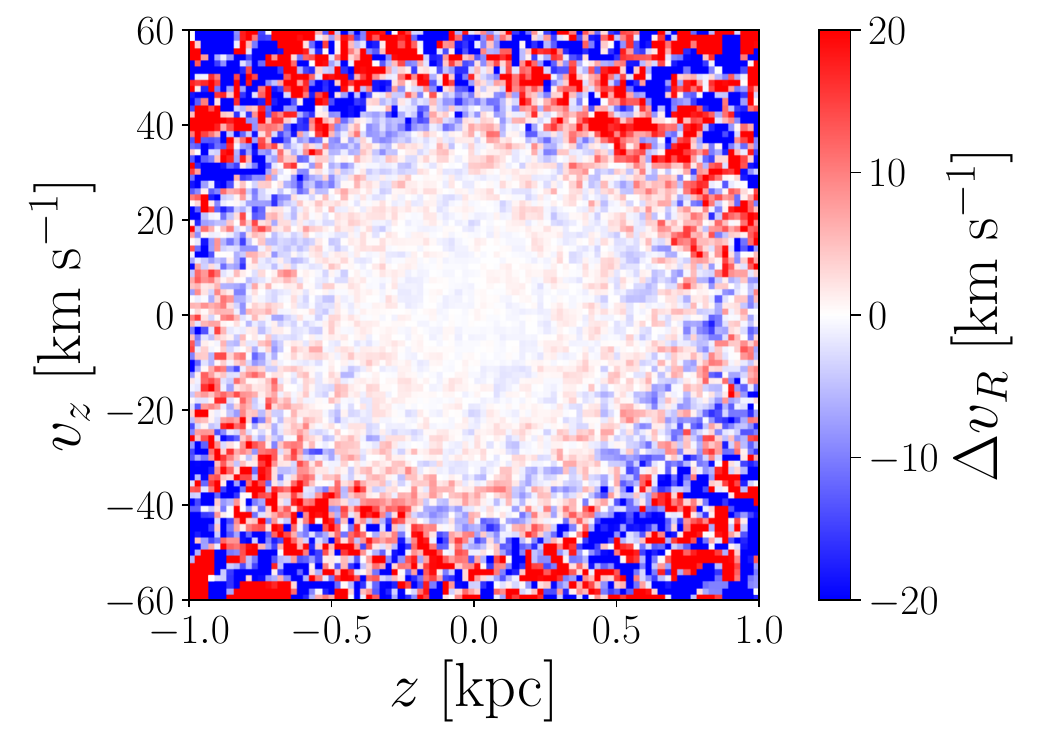}
	\includegraphics[width=0.18\hsize]{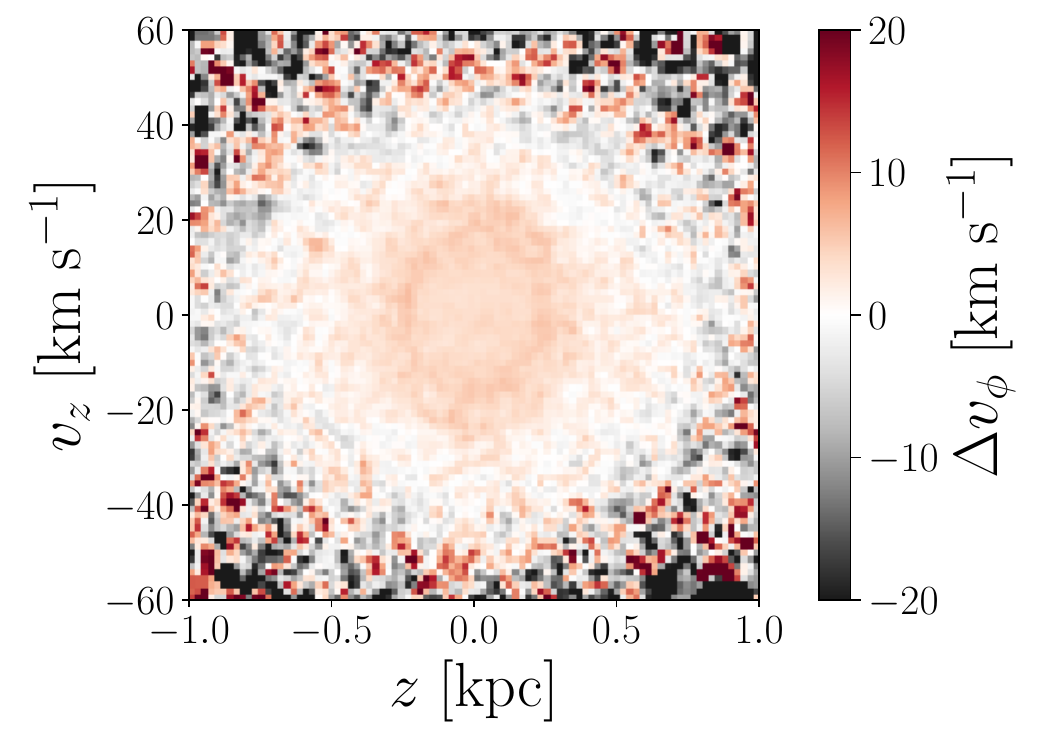}

	\includegraphics[width=0.18\hsize]{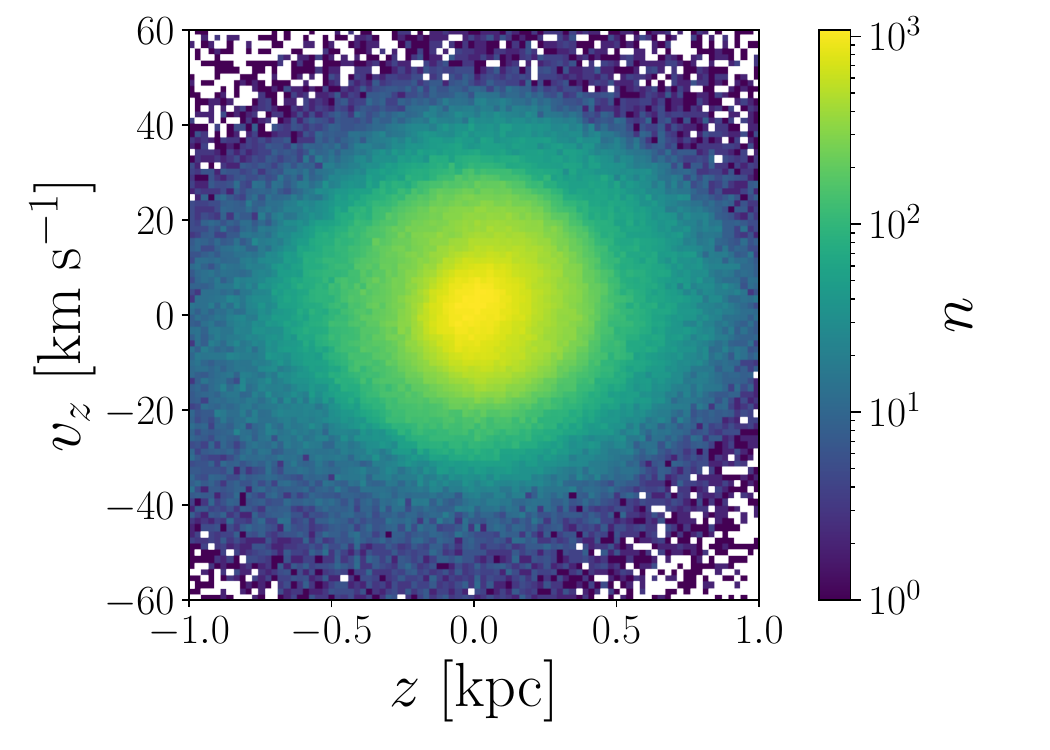}
	\includegraphics[width=0.18\hsize]{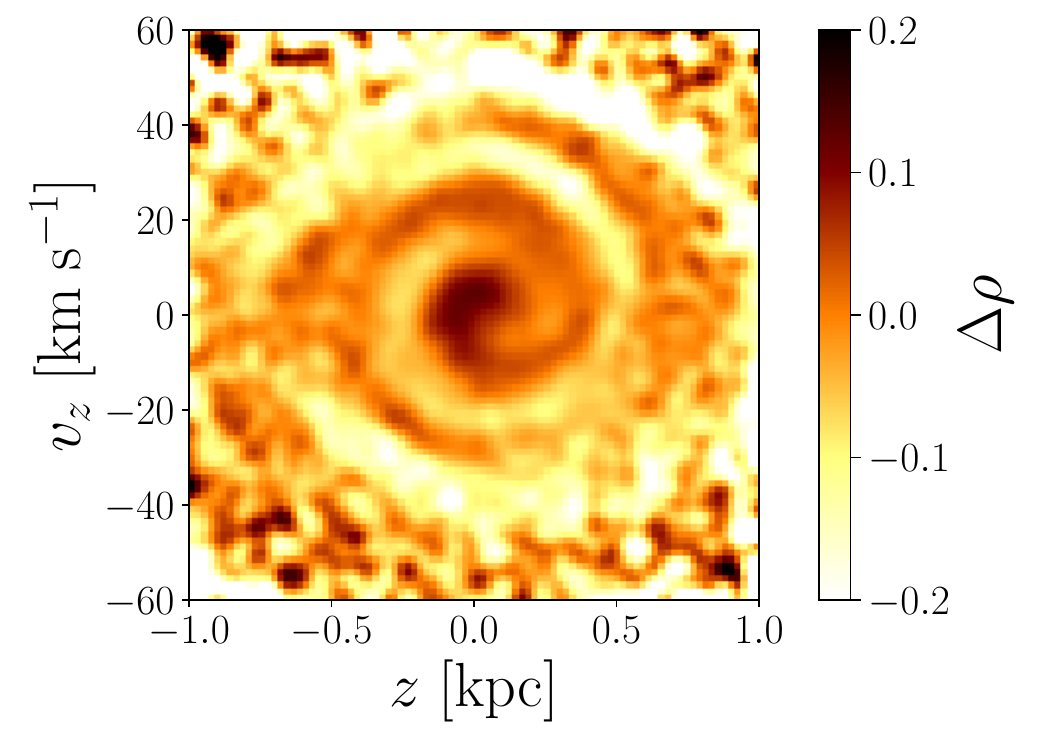}
	\includegraphics[width=0.18\hsize]{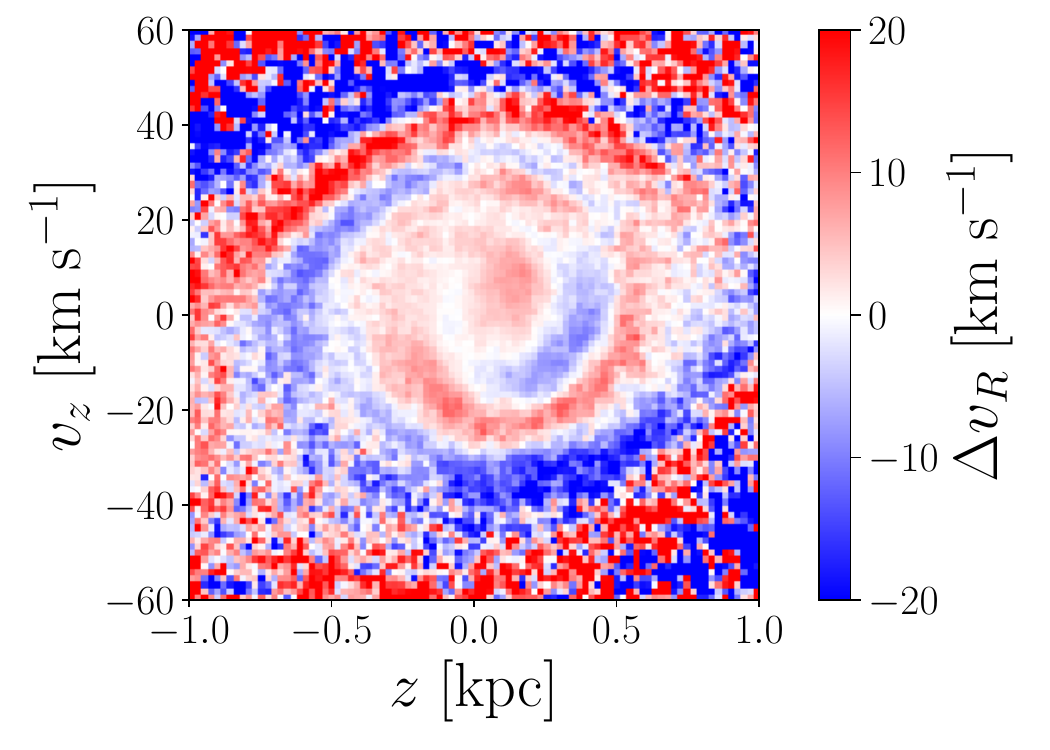}
	\includegraphics[width=0.18\hsize]{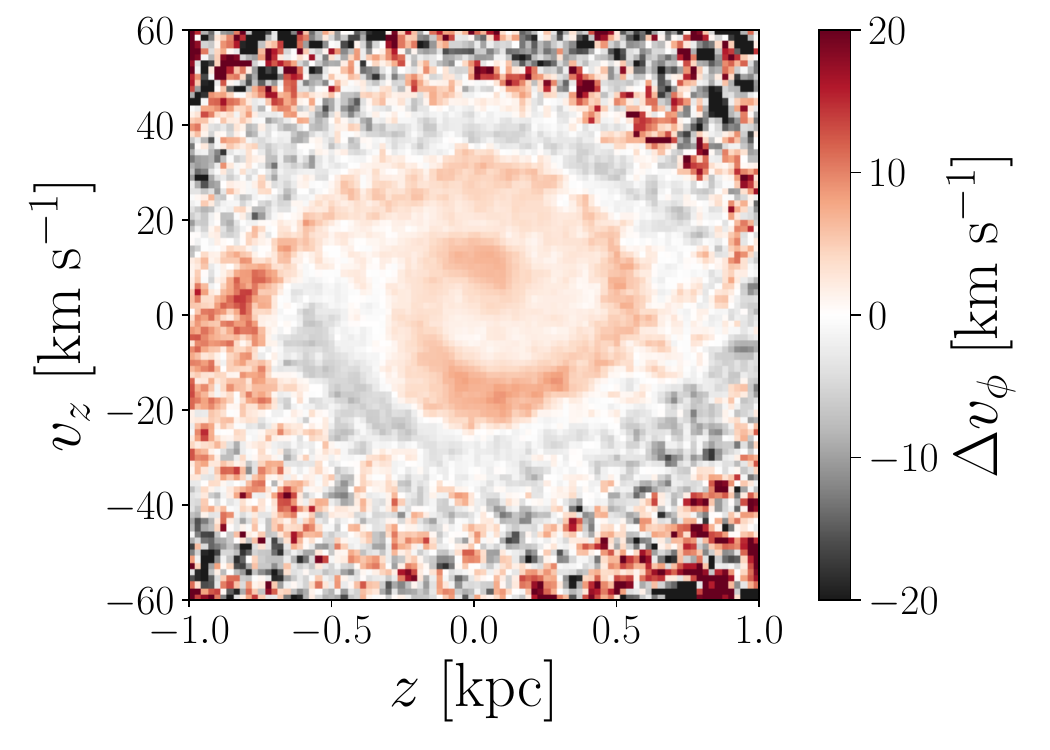}

	\includegraphics[width=0.18\hsize]{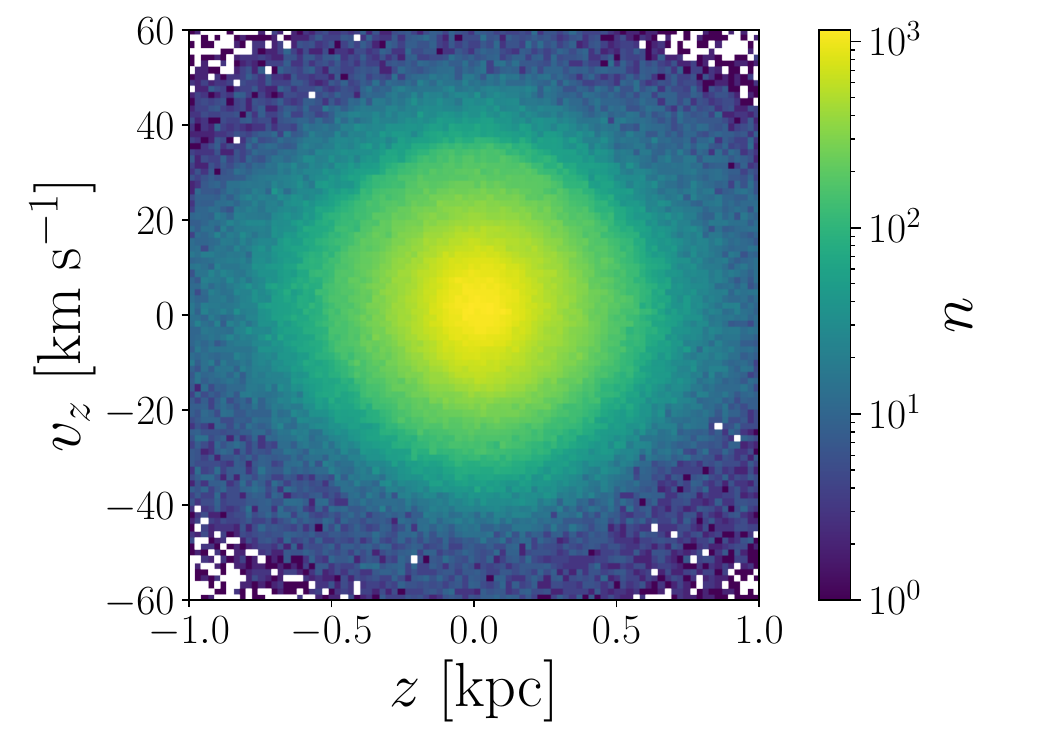}
	\includegraphics[width=0.18\hsize]{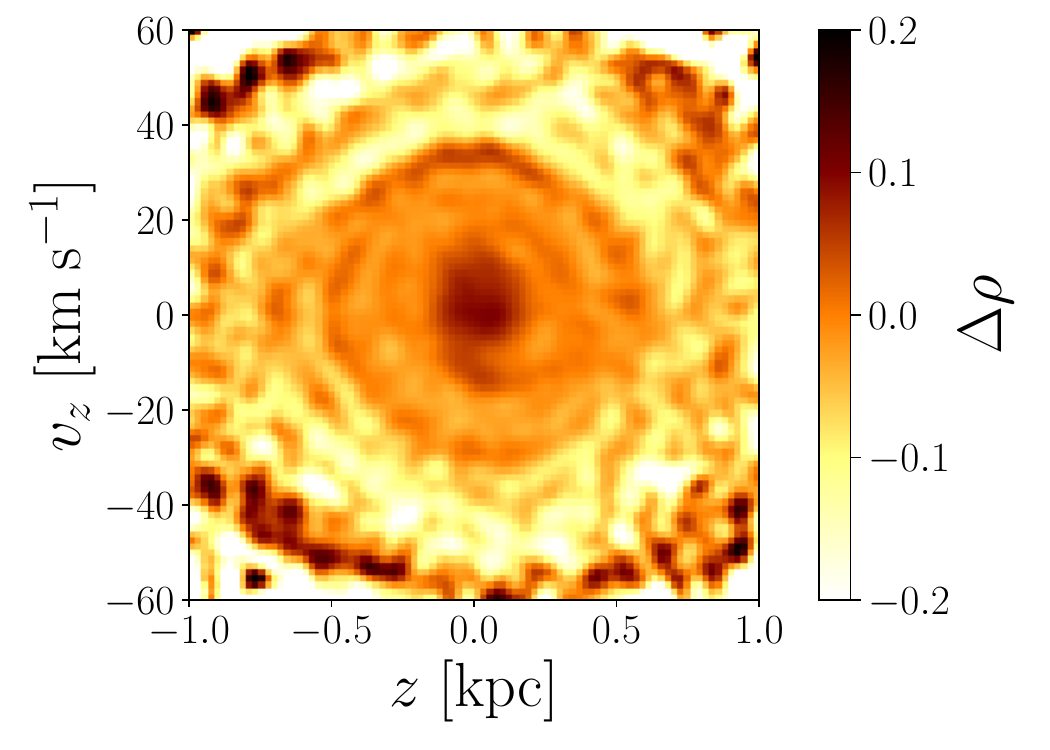}
	\includegraphics[width=0.18\hsize]{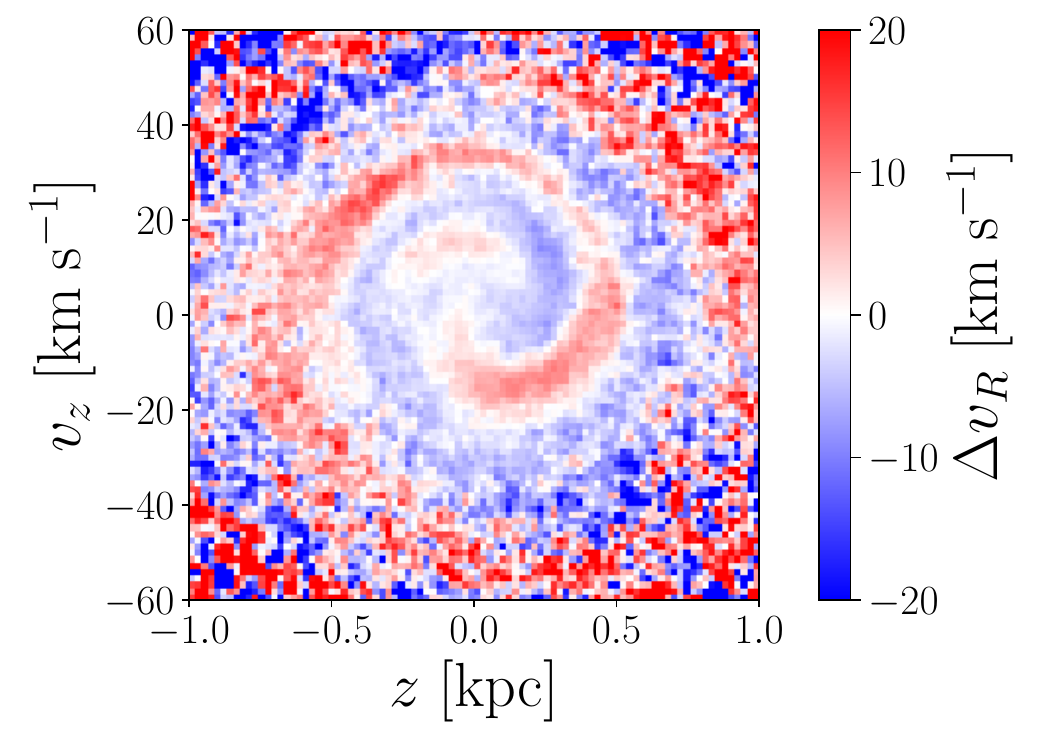}
	\includegraphics[width=0.18\hsize]{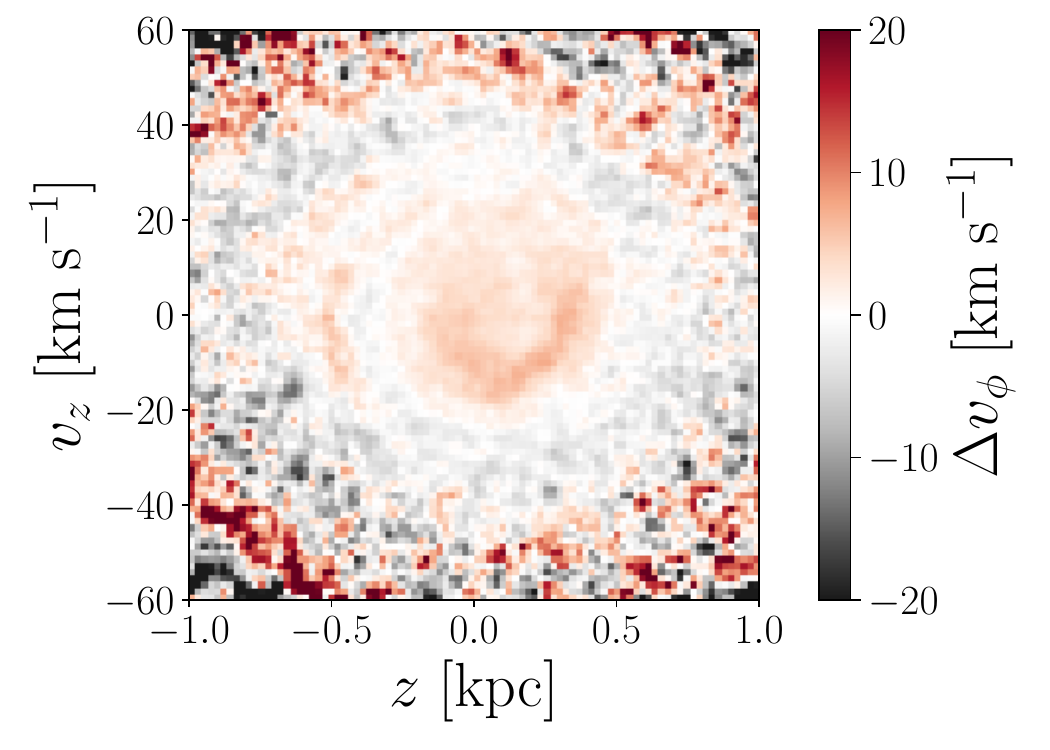}

	\includegraphics[width=0.18\hsize]{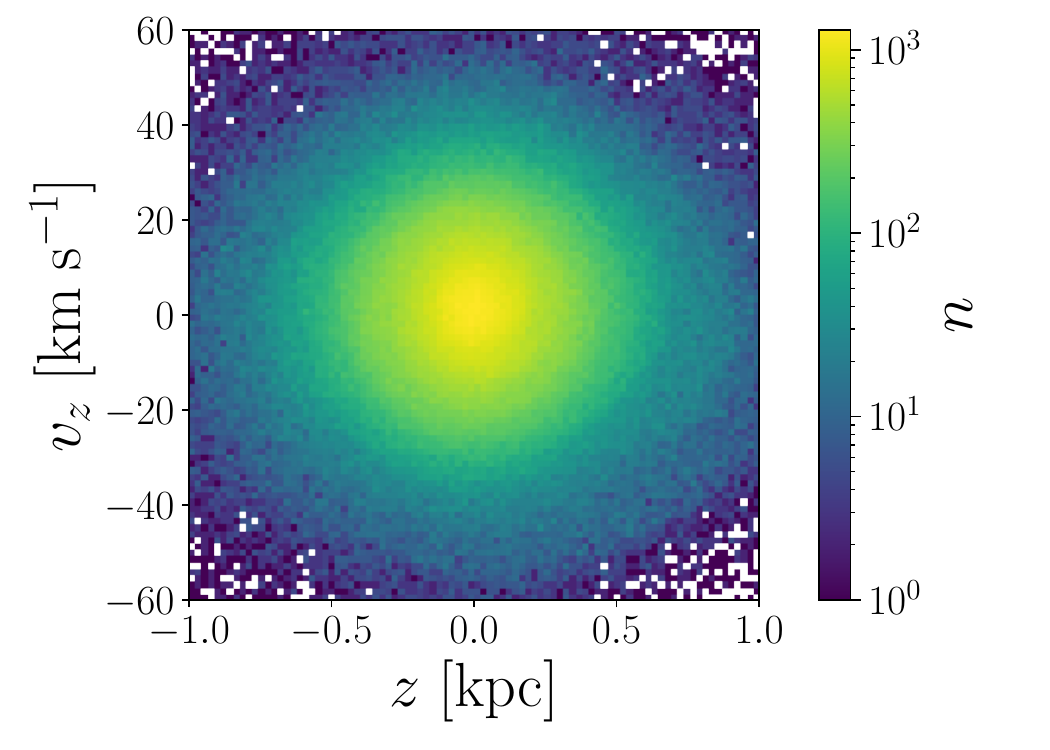}
	\includegraphics[width=0.18\hsize]{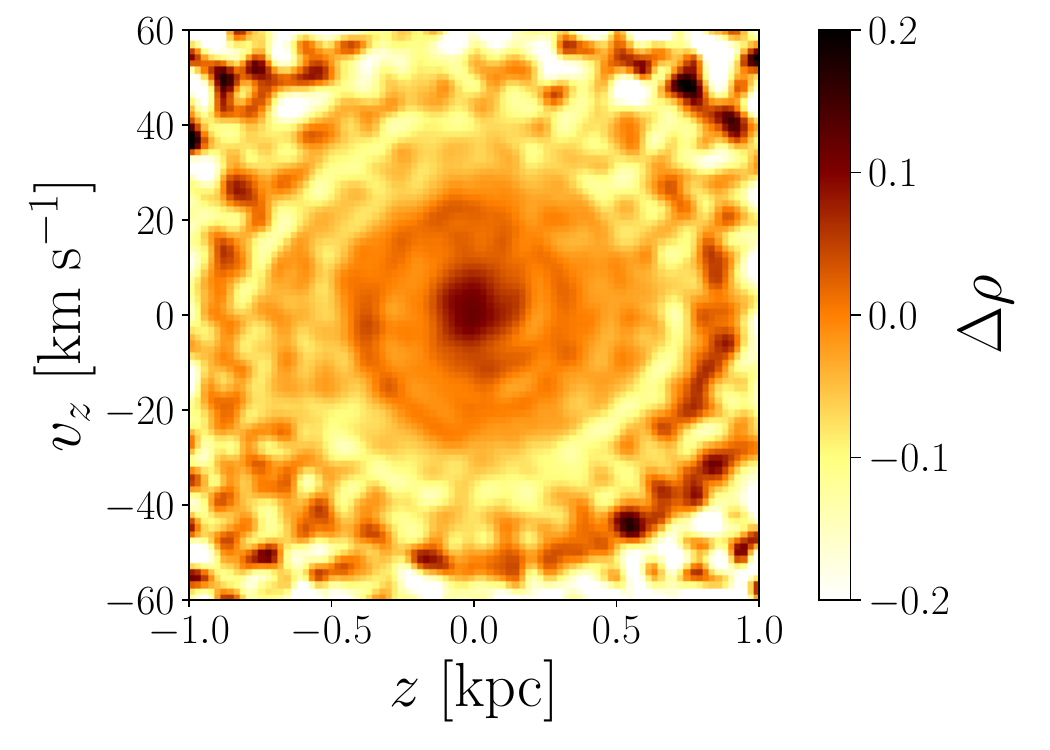}
	\includegraphics[width=0.18\hsize]{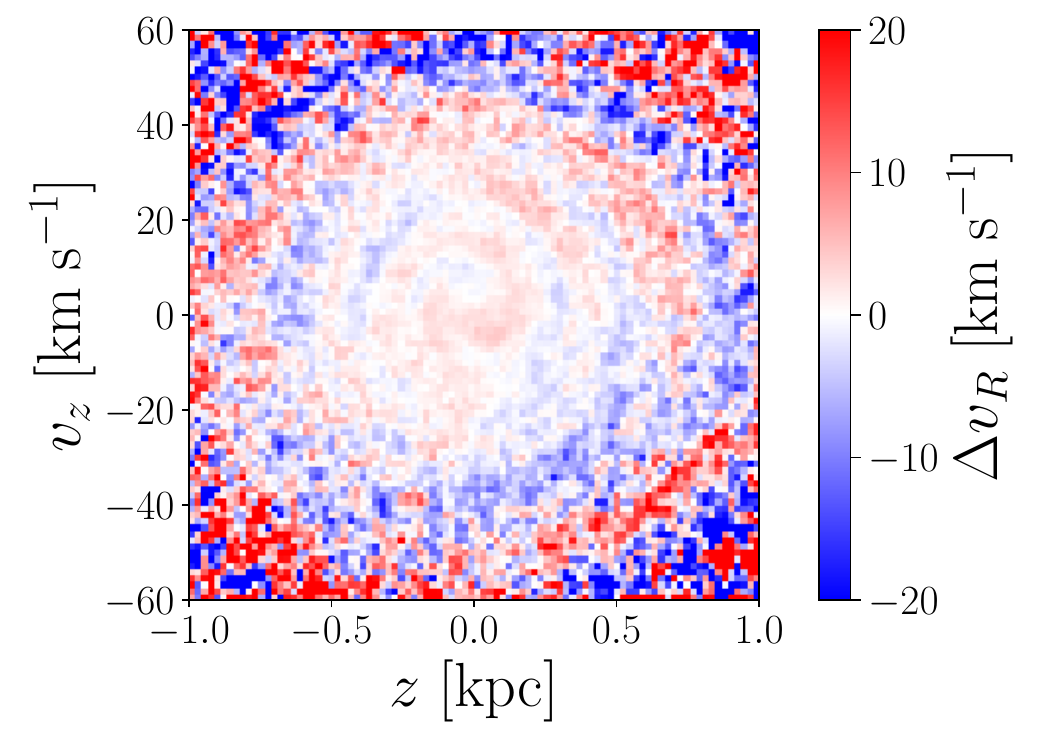}
	\includegraphics[width=0.18\hsize]{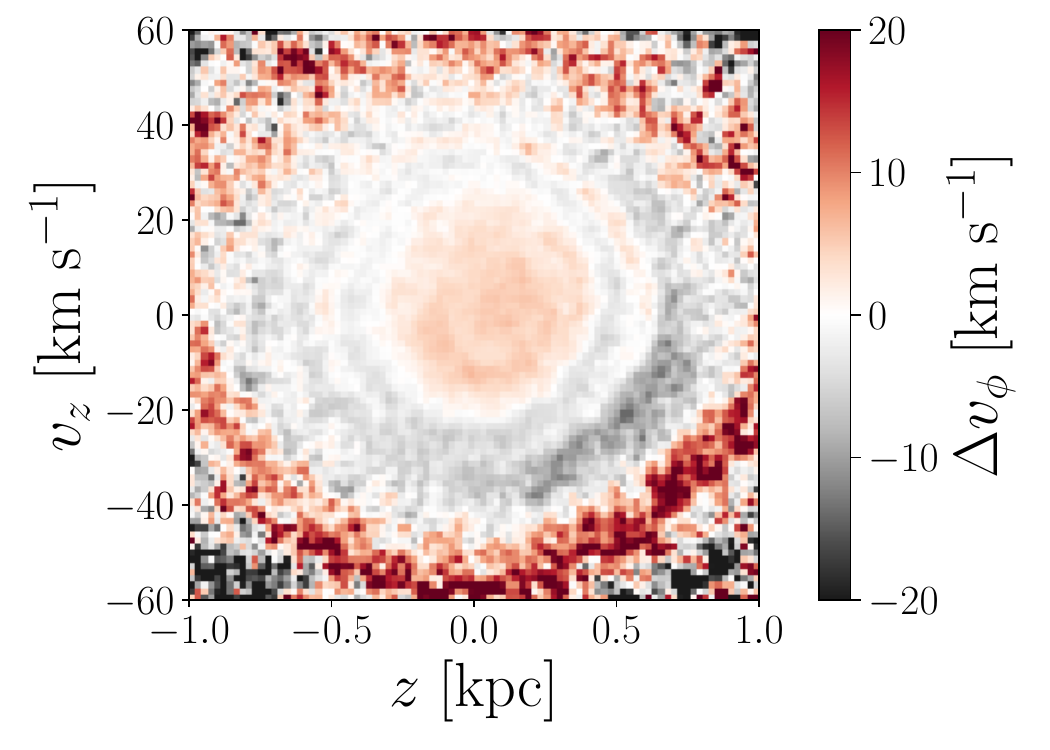}
	\caption{Maps of $z$-$v_z$ of the particles in the `solar neighbourhood'. The panels are colour-coded by number density (\textit{left column}), density contrast (\textit{middle left column}), radial velocity  (\textit{middle right column}), and azimuthal velocity (\textit{right column}).
		The maps in the first row are at $t=0$~Gyr. Second, third, and fourth rows correspond to 0.6~Gyr after the first pericentre passage, 0.5~Gyr after the second pericentre passage, and 0.5~Gyr after the third pericentre passage, respectively.
		Prominent one-arm phase spirals emerge in the solar neighbourhood after the first pericentre passage, becoming fainter after subsequent pericentre passages which have weaker impacts on the galactic disc.
	}\label{fig:phase_spirals_SN}
	\end{center}
\end{figure*}

In Fig.~\ref{fig:phase_spirals_SN}, we show the $z$-$v_z$ maps for the particles within 1~kpc from $(x,y,z)=(-8.277, 0, 0)~\kpc$, which corresponds to the solar position in the MW \citep{2022A&A...657L..12G}.
These maps highlight vertical phase-space structures and are divided into $90\times90$ bins, each with a size of $0.022~\kpc \times 1.33\,\kms$.

The left and middle-left panels are colour-coded by the particle count ($n$) and the density contrast ($\Delta \rho$), respectively. The density contrast is defined as $\Delta \rho(z, v_z) = (n*G_{1.5})(z, v_z)/(n*G_3)(z, v_z)$, where $n*G_{\sigma}$ represents the particle count convolved with the Gaussian filter with a standard deviation of $\sigma$ pixels. 
This visualisation method, used in previous studies \citep[e.g. ][]{2019MNRAS.485.3134L, 2023A&A...673A.115A}, enhances the visibility of phase spiral patterns. The middle-right and right panels are colour-coded by the deviations in mean radial velocity ($\Delta v_R = \bar{v}_R - \bar{v}_R * G_{10}$) and azimuthal velocity ($\Delta v_{\phi} = -\bar{v}_{\phi} + \bar{v}_{\phi} * G_{10}$) from the Gaussian-smoothed values, respectively.

The first row of Fig.~\ref{fig:phase_spirals_SN} shows the $z$-$v_z$ maps at $t=0$~Gyr, before the dwarf's pericentre passage. No asymmetric patterns are evident, indicating an initially symmetric phase-space distribution. 
While \citet{2019MNRAS.486.1167B} identifies a quadrupole pattern in maps colour-coded by $v_R$ due to the tilting of the velocity ellipsoid, this feature is removed here by subtracting the Gaussian-smoothed radial velocity.

In the second row, corresponding to $t = 1.5$Gyr (0.6 Gyr after the first pericentre passage), prominent one-arm phase spirals are visible in maps colour-coded by $\Delta \rho$, $\Delta v_R$, and $\Delta v_{\phi}$. 
This one-arm feature is qualitatively consistent with the phase spiral observed in the solar neighbourhood \citep{2018Natur.561..360A, 2023A&A...673A.115A}.
The phase spiral in the maps colour-coded by radial and azimuthal velocities arises from the coupling between vertical and planar motion \citep{2018Natur.561..360A, 2018MNRAS.481.1501B, 2019MNRAS.484.1050D, 2021MNRAS.504.3168B}.
The vertical orbital frequency depends not only on the vertical oscillation amplitude but also on the angular momentum ($L_z=Rv_{\phi}$).
As a result, phase spirals are particularly pronounced in $v_{\phi}$-coded maps, where angular momentum variations are highlighted. In contrast, $v_R$-coded spirals originate from the interplay of vertical and planar waves triggered by satellite perturbations \citep{2023A&A...673A.115A}.

The third and fourth rows show the $z$-$v_z$ maps at 0.5~Gyr after the second and third pericentre passages, respectively. Phase spirals remain visible, though they are notably fainter than those following the first pericentre passage. In the third row, spiral structures are more distinct in maps colour-coded by $\Delta v_R$ and $\Delta v_{\phi}$ than in those colour-coded by $\Delta \rho$. By contrast, the fourth row reveals faint spirals across all panels, suggesting significantly reduced perturbations.
As shown in Fig.~\ref{fig:dwarf_pos_mass}, the dwarf loses approximately 50\% of its mass during each pericentre passage. 
This substantial mass loss weakens the perturbations, resulting in insufficient excitation of phase spirals during the third pericentre passage.

\subsection{Spatial variation of the phase spirals}\label{sec:phase_spiral_spatial}
\begin{figure}
	\begin{center}
		\includegraphics[width=\hsize]{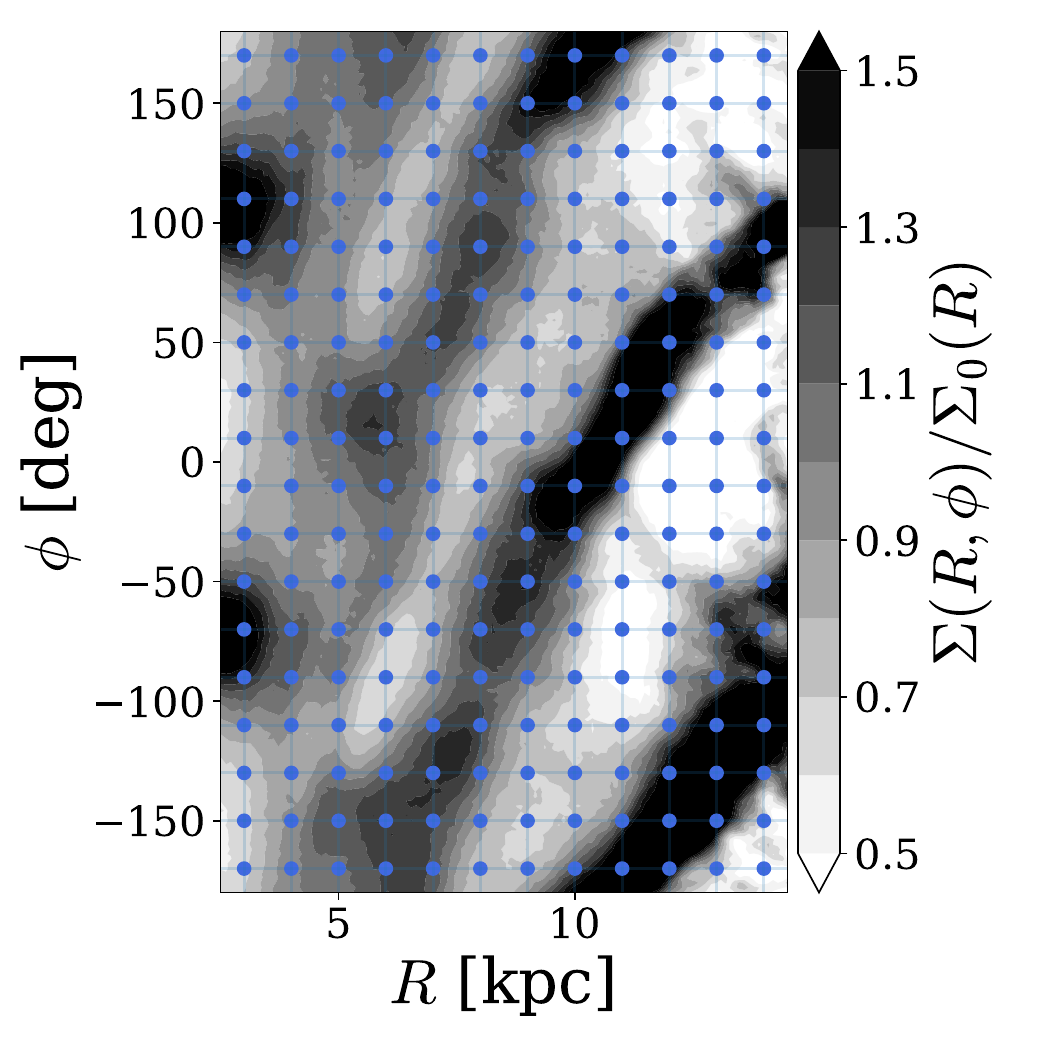}
		\caption{Grid in the $R$-$\phi$ space. The grid size is $1\,\kpc \times 20^{\circ}$. The contour map shows the normalised surface density at $t=1.47$ Gyr.}\label{fig:R_phi_grid}
	\end{center}
\end{figure}
\begin{figure}
	\begin{center}
		\includegraphics[width=\hsize]{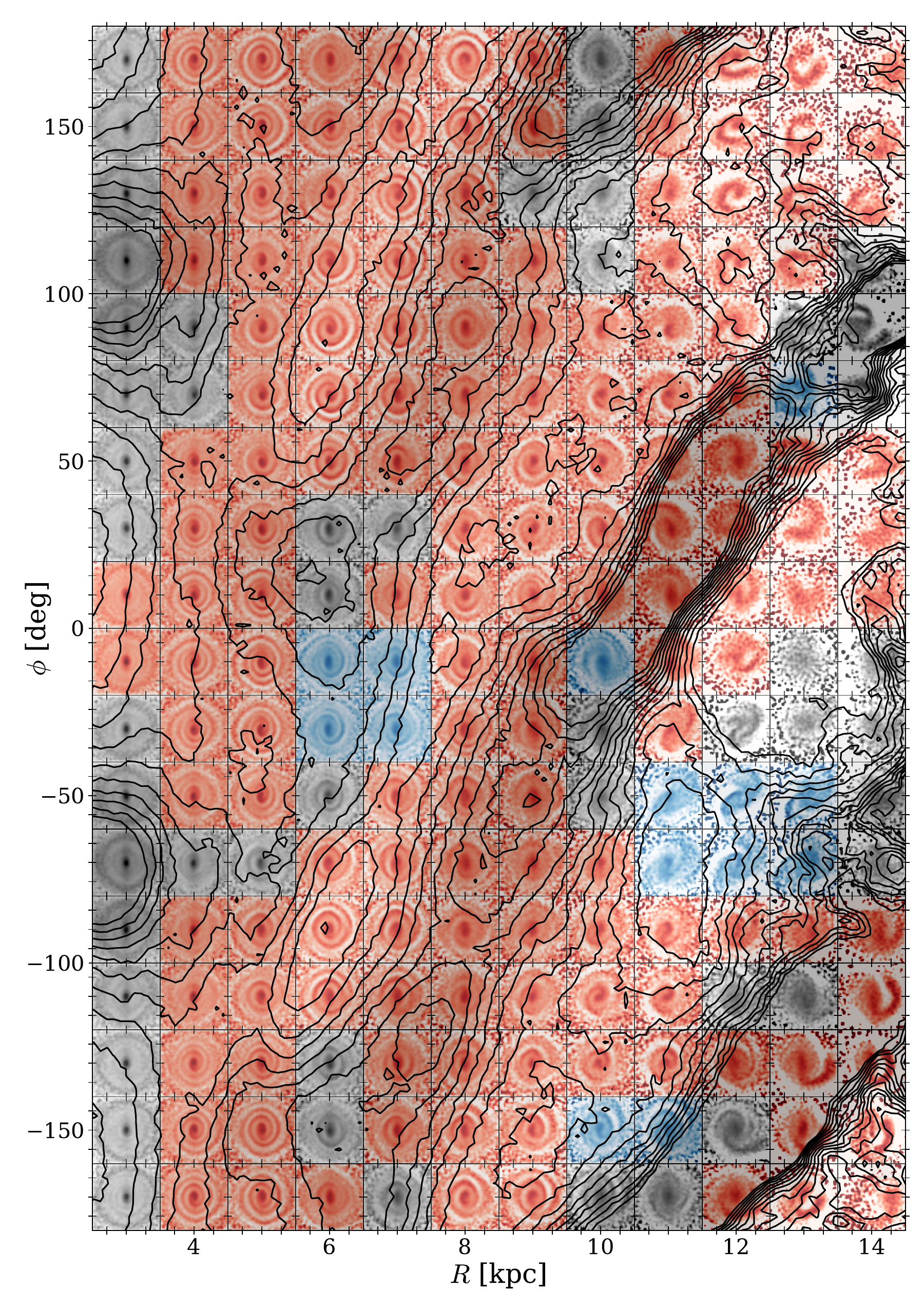}
		\caption{Variation of $R$-$\phi$ of the phase spiral in the snapshot at $t=1.47$ Gyr. Each panel shows $z/h_z$-$v_z/\sigma_z$ map for a set of particles grouped by $R$ and $\phi$. The maps are colour-coded by the density contrast, $\Delta \rho$. Panels exhibiting one-arm phase spirals are painted in red tone, while those with two-arm phase spirals are painted in blue tone. Panels where the phase spiral shapes cannot be classified as either one-arm or two-arm are painted in grey tones. The surface density contour of Fig.~\ref{fig:R_phi_grid} is overlaid.
			While one-arm phase spirals are widely observed, two-arm phase spirals are only found in limited areas.
		}\label{fig:z_vz_R_phi_dens_150}
	\end{center}
\end{figure}

In this section, we examine how the morphology of the phase spiral varies with the galactocentric radius and the azimuth.
Fig.~\ref{fig:R_phi_grid} shows the normalised surface density at $t=1.47$~Gyr on the $R$-$\phi$ plane, where the galaxy rotates in the direction in which $\phi$ decreases (i.e. from top to bottom in the figure).
We divided the $R$-$\phi$ space into a grid with a size of $1~\kpc \times 20^{\circ}$ as shown in the figure and made the vertical phase-space maps for particles within 1~kpc of each grid point.

Fig.~\ref{fig:z_vz_R_phi_dens_150} shows the resulting phase-space maps, highlighting spatial variations in phase spirals at $t=1.47$~Gyr.
Each panel corresponds to a grid point in Fig.~\ref{fig:R_phi_grid} and displays a vertical phase-space map colour-coded by the density contrast, $\Delta \rho$. The horizontal and vertical axes are normalised by the standard deviations $h_z$ and $\sigma_z$, respectively, and cover the range $[-3.5, 3.5] \times [-3.5, 3.5]$.
To classify the phase spiral shapes, we performed a visual inspection: red panels indicate one-arm phase spirals, blue panels indicate two-arm phase spirals, and grey panels correspond to cases where the shapes are ambiguous.

Phase spirals or asymmetric distributions appear in most panels.
In the outer galaxy, the structures are highly distorted, with some panels displaying U-shaped or C-shaped patterns rather than spirals.
At $R=3$~kpc, the spiral pattern is unclear due to the low density contrast.
The phase spirals in the inner galaxy are more tightly wound, reflecting the faster vertical winding rate in this region. 
No clear trends in azimuthal variation are evident from Fig.~\ref{fig:z_vz_R_phi_dens_150}. A more quantitative analysis—such as of parameterising features such as amplitude and rotation—is required to explore this further, as suggested by \citet{2023A&A...678A..46A}, but such an analysis is beyond the scope of this paper.

While one-arm phase spirals are widely observed, two-arm phase spirals are only found in specific regions, such as $(R,\phi) = (6\kpc, -30^{\circ})$.
As discussed earlier, tidally induced spiral arms are a likely driver of the breathing mode, which may contribute to the formation of two-arm phase spirals.\footnote{\citet{2023MNRAS.524.6331L} have demonstrated that slowing bars can also generate two-arm phase spirals. The bar in our model slows down at a rate of $\sim 2\,\mathrm{km \, s^{-1} \, kpc^{-1} \, Gyr^{-1}}$, but two-arm phase spirals appear only after the formation of the tidally induced spiral arms, suggesting that these arms are the primary driver of the two-arm phase spirals.}
This connection is particularly plausible in the inner galaxy, where the breathing mode dominates.
However, Fig.\ref{fig:z_vz_R_phi_dens_150} does not show a clear spatial correlation between the spiral arms and the two-arm phase spirals.

\begin{figure}
	\begin{center}
		\includegraphics[width=\hsize]{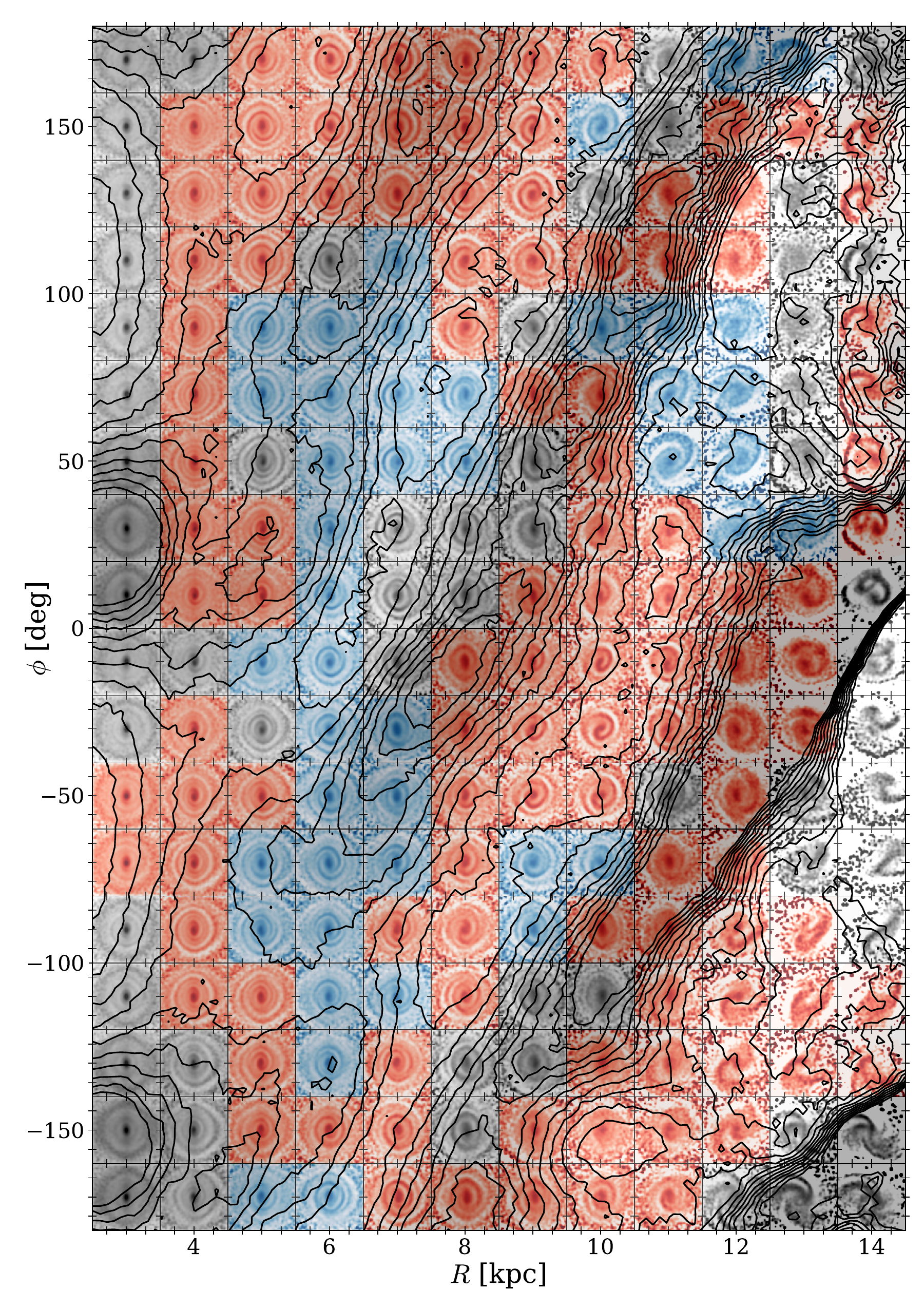}
		\caption{Same as Fig.~\ref{fig:z_vz_R_phi_dens_150} but for the snapshot at $t=1.65$~Gyr.
			Two-arm phase spirals are more widely distributed than in Fig.~\ref{fig:z_vz_R_phi_dens_150}, especially around $R\sim 6$--7.5~kpc, consistent with observations showing two-arm spirals primarily for stars with small guiding radii.
		}\label{fig:z_vz_R_phi_dens_169}
	\end{center}
\end{figure}

Fig.~\ref{fig:z_vz_R_phi_dens_169} shows the similar phase-space maps as Fig.~\ref{fig:z_vz_R_phi_dens_150}, but for the snapshot at $t=1.65$~Gyr. 
Compared to $t = 1.47$Gyr (Fig.~\ref{fig:z_vz_R_phi_dens_150}), two-arm phase spirals are more widely distributed, particularly between $R = 6$~kpc and $7$~kpc.
At $R>8$~kpc, the one-arm phase spirals are more widely observed than the two-arm phase spirals.
This trend is consistent with the findings of \citet{2022MNRAS.516L...7H}, who reported that the two-arm phase spirals are observed only in the samples with small guiding radii.
For further analysis, Appendix\ref{appendix:Rg_thetaphi_phasespirals} provides phase-space maps classified by guiding radius ($\Rg$) and azimuthal angle ($\theta_\phi$) instead of $R$ and $\phi$.

Previous studies \citep{2014MNRAS.440.1971W, 2022ApJ...935..135B, 2023ApJ...952...65B} have shown that satellites can excite the breathing mode if their perturbations are impulsive, meaning that the satellites move sufficiently faster than disc stars.
Analytical calculations by \citet{2023ApJ...952...65B} indicate that the bending mode dominates over the breathing mode in the solar neighbourhood and the inner galaxy; on the contrary, a substantial breathing mode amplitude is expected in the outer galaxy.
Consistent with this prediction, in the $N$-body model by \citet{2021MNRAS.508.1459H}, two-arm phase spirals are observed only in the outer galaxy, and it is suggested that these phase spirals are associated with the breathing mode directly excited by external perturbations from the satellite.
In our model, the two-arm phase spirals in the inner galaxy are likely excited by the spiral arms.
In contrast, those in the outer galaxy, such as those observed around $(R,\phi) = (12\,\kpc, -50^{\circ})$ in Fig.~\ref{fig:z_vz_R_phi_dens_150} and $(R,\phi) = (11\,\kpc, 70^{\circ})$ in Fig.~\ref{fig:z_vz_R_phi_dens_169}, might be directly excited by the satellite impact, consistent with the mechanism proposed in \citet{2021MNRAS.508.1459H}'s model.

Some panels, such as $(R,\phi) = (6\,\kpc, -150^{\circ})$ in Fig.~\ref{fig:z_vz_R_phi_dens_150} and $(9\,\kpc, -130^{\circ})$ in Fig.~\ref{fig:z_vz_R_phi_dens_169}, exhibit ring-like structures rather than spirals. 
These features, absent in \textit{Gaia} data, are of uncertain origin but may be linked to spiral arms, as they are located within arm regions.

\subsection{Time evolution of the phase spiral}\label{sec:phase_spiral_time}
\begin{figure*}
	\begin{center}
		\includegraphics[width=0.49\hsize]{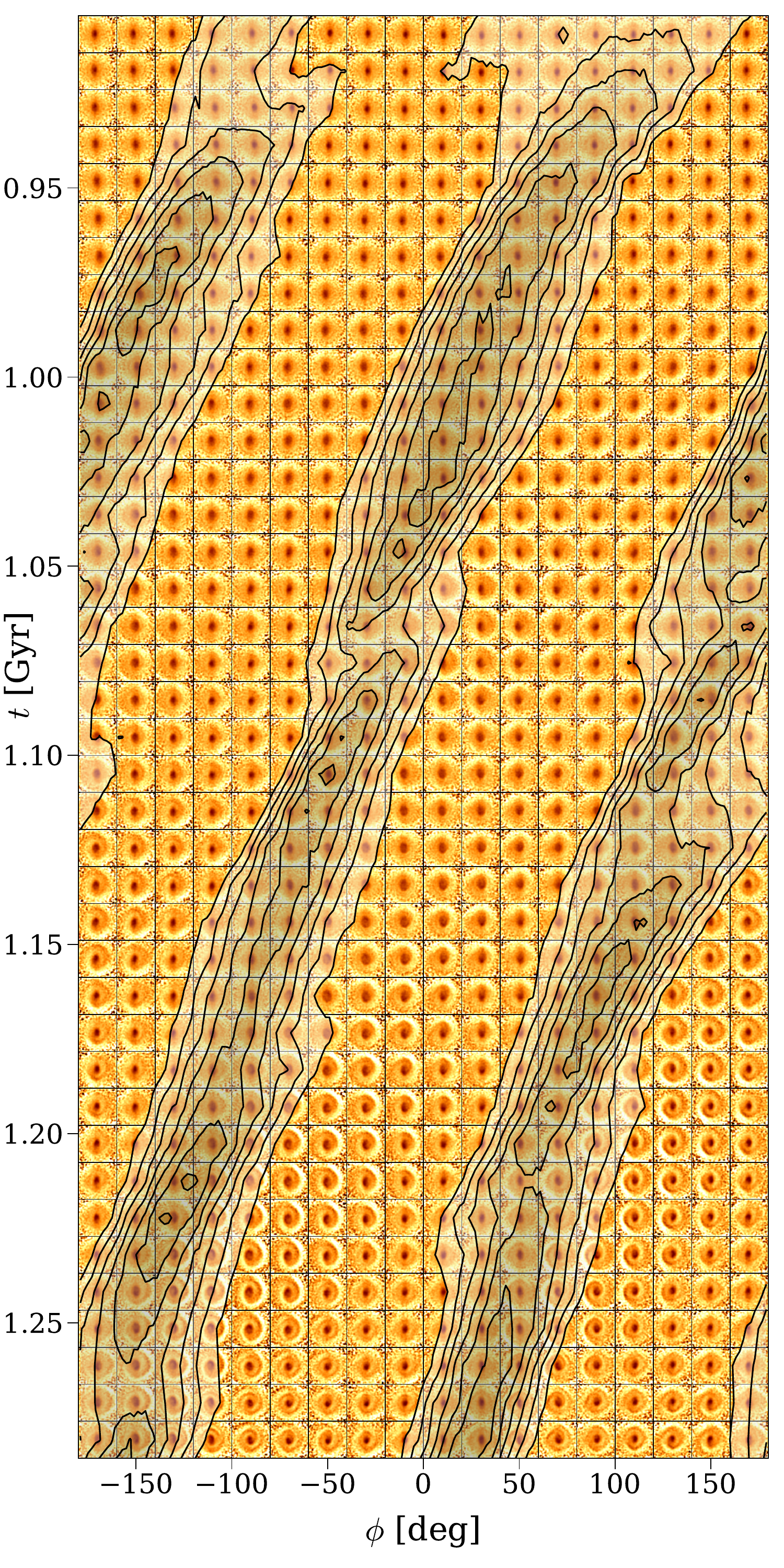}
		\includegraphics[width=0.49\hsize]{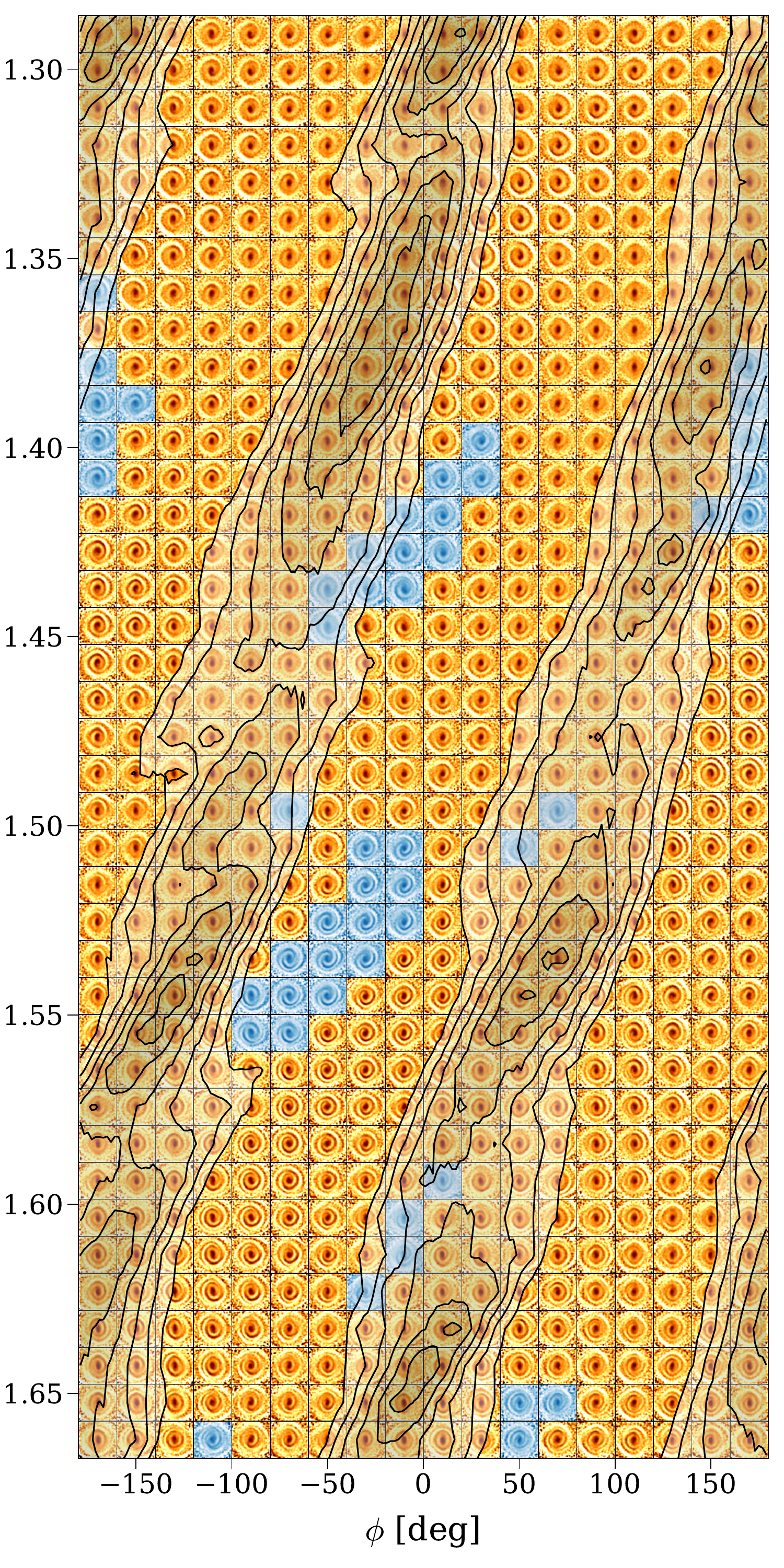}
		\caption{Time evolution of the phase spiral at $R=8$~kpc.  Columns represent azimuths equally spaced by $20^{\circ}$. Rows correspond to time, from $t=0.9$~Gyr to $1.28$~Gyr on the left block, and $t=1.29$~Gyr to $1.67$~Gyr on the right block.  Each panel shows $z/h_z$-$v_z/\sigma_z$ maps colour-coded by the density contrast, $\Delta \rho$. Panels showing two-arm phase spirals are painted in blue tone. Contours indicate the normalised surface density, $\Sigma(R=8\,\kpc, \phi, t)/\Sigma_0(R=8\,\kpc, t)$. Arm regions ($\Sigma/\Sigma_0>1$) are shaded.
		One-arm spirals emerge first, followed by intermittent appearances of two-arm spirals around 200--250 Myr after the bending-to-breathing mode transition.
		}\label{fig:z_vz_time_8}
	\end{center}
\end{figure*}

Finally, we examined the time evolution of the phase spirals. 
For all snapshots from $t=0.9$~Gyr to $1.67$ Gyr, we created $z/h_z$-$v_z/\sigma_z$ maps at eighteen different positions at $R=8$~kpc, equally spaced by $20^{\circ}$ in azimuth.
Fig.~\ref{fig:z_vz_time_8} illustrates these results, with rows representing time and columns representing azimuth.
This type of visualisation was introduced by \citet{2021MNRAS.504.3168B}, who referred to it as a `chronogram' in their paper.
Each panel shows the phase-space map colour-coded by the density contrast, $\Delta \rho$. 
The overlaid contours represent the normalised surface density, $\Sigma(R=8\,\kpc, \phi, t)/\Sigma_0(R=8\,\kpc, t)$, and are spaced by 0.1 between 1 and 1.8. Regions where $\Sigma/\Sigma_0>1$ are shaded to indicate arm regions.

At $t \lesssim 1.15$~Gyr, phase spirals are not yet visible, but asymmetric density distributions are evident in the $z/h_z$-$v_z/\sigma_z$ space. Around $\phi \sim -180^\circ$, phase spirals begin to emerge at $t \sim 1.15$~Gyr ($\sim250$~Myr after the first pericentre passage) and subsequently appear at other azimuths. By $t \sim 1.3$~Gyr, phase spirals are visible throughout the entire ring at $R=8$~kpc, though their prominence varies with azimuth. As expected, the phase spirals wind up over time, reflecting the underlying dynamics.

Two-arm phase spirals, highlighted in blue in Fig.~\ref{fig:z_vz_time_8}, first appear at $t \sim 1.35$~Gyr. These structures are limited to specific azimuthal ranges and shift positions following the galaxy’s rotation. They tend to occur more frequently in inter-arm regions. For example, a two-arm phase spiral emerges at $\phi = 30^\circ$ at $t \sim 1.4$~Gyr, moves leftward (in the direction of rotation), and becomes faint after entering the arm region at $t \sim 1.44$~Gyr. Another example begins at $\phi = -10^\circ$ at $t \sim 1.5$~Gyr and disappears around $\phi = -100^\circ$ at $t \sim 1.55$~Gyr after entering an arm region. After entering the arm region, these two-arm spirals transform into ring-like structures, as seen at $\phi \sim -150^\circ$ at $t \sim 1.57$~Gyr. This transformation suggests that the breathing perturbation from spiral arms deforms phase spirals into ring structures. 

\begin{figure*}
	\begin{center}
		\includegraphics[width=0.49\hsize]{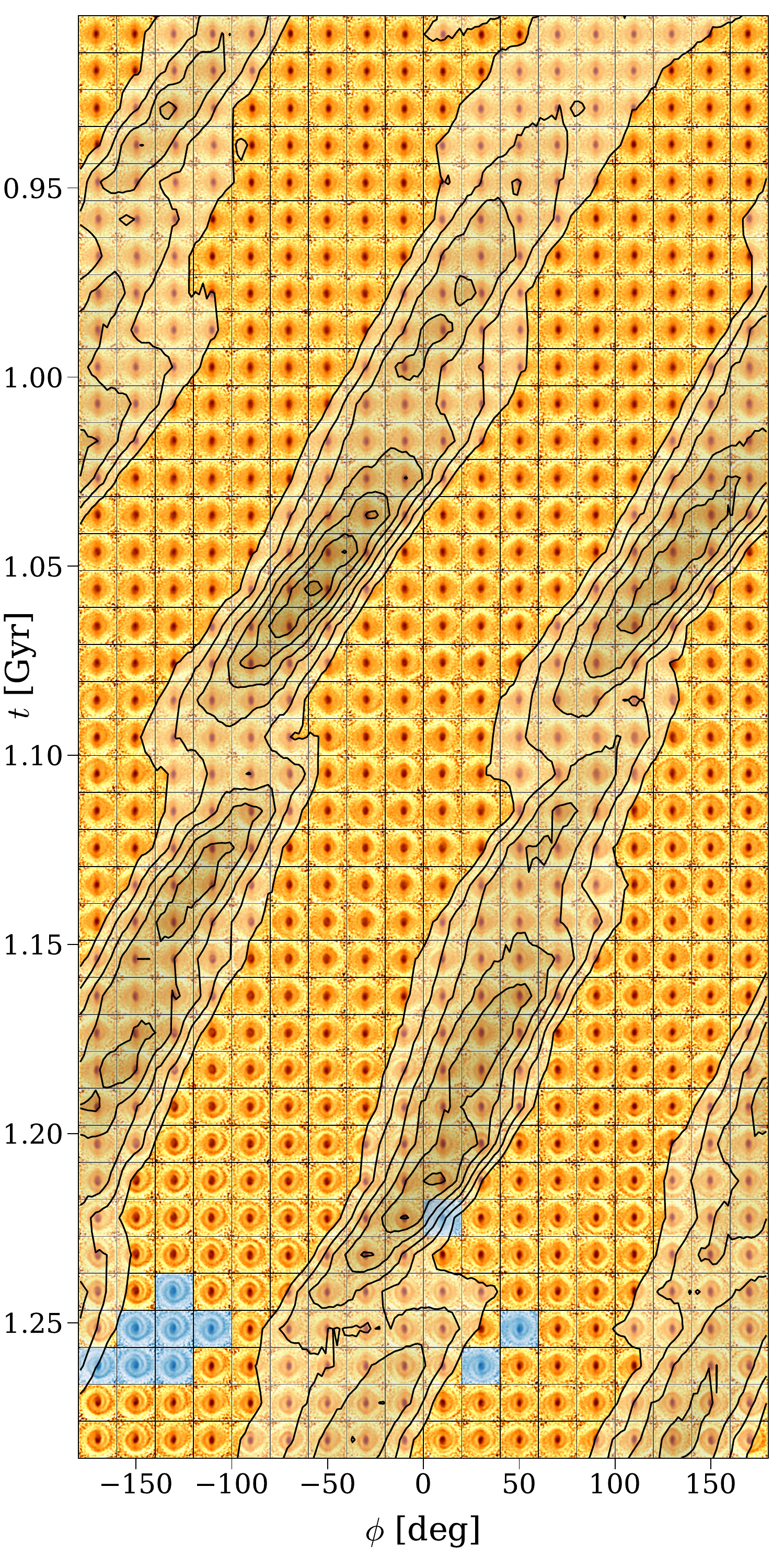}
		\includegraphics[width=0.49\hsize]{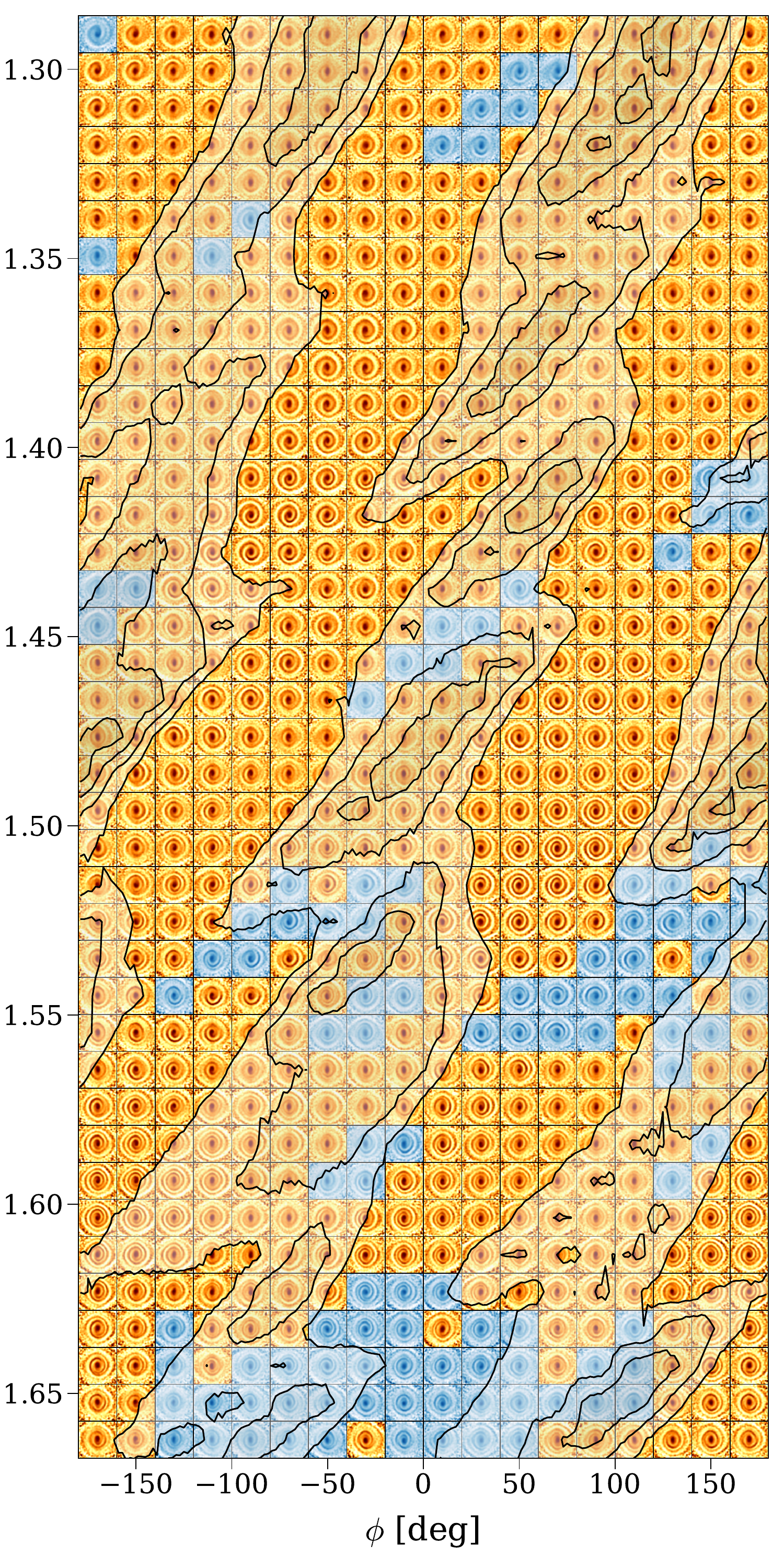}
		\caption{Same as Fig.~\ref{fig:z_vz_time_8} but at $R=6$~kpc.
			Two-arm phase spirals emerge earlier at $R=6$~kpc than at $R=8$~kpc, reflecting the faster transition to breathing mode dominance in the inner galaxy.
		}\label{fig:z_vz_time_6}
	\end{center}
\end{figure*}

Fig.~\ref{fig:z_vz_time_6} shows the time evolution of phase spirals at $R=6$~kpc. 
Phase spirals appear at $t \sim 1.15$~Gyr and become increasingly tightly wound, consistent with observations at $R=8$~kpc. However, two-arm phase spirals emerge earlier at $t \sim 1.25$~Gyr, reflecting the faster transition from bending mode to breathing mode dominance in the inner galaxy. Unlike at $R=8$~kpc, two-arm phase spirals at $R=6$~kpc are not confined to specific azimuths. At earlier times ($t \lesssim 1.5$~Gyr), the patterns resemble those at $R=8$~kpc. After $t \sim 1.5$~Gyr, particularly around $t \sim 1.65$~Gyr, two-arm phase spirals become more widespread, likely due to weakening spiral arms. Between $t = 1.0$~Gyr and $1.6$~Gyr, the spiral arm density at $R = 6$~kpc decreases by $\sim 50$\%, blurring the boundaries between arm and inter-arm regions. Additionally, the bar may influence phase spirals at $R=6$~kpc, as this radius is near the bar’s corotation radius.

At both $R=6$~kpc and $R=8$~kpc, two-arm phase spirals appear intermittently rather than continuously after their initial emergence.
This temporal irregularity is likely due to the oscillation in spiral strength and the breathing amplitude (see Fig.~\ref{fig:bend_breath_dens_amp}).
Although predicting the exact timing and location of two-arm spirals is challenging, their first appearance correlates with the transition to breathing mode dominance.

Comparing Figs.~\ref{fig:z_vz_time_8}, \ref{fig:z_vz_time_6}, and the third panel of Fig.~\ref{fig:t_R_bend_breath}, we find that the two-arm phase spiral emerges $\sim200$~Myr after the dominant mode transition at both $R=6$~kpc and 8~kpc.
This trend holds for other radii between $R=5$~kpc and $R=10$~kpc, where two-arm spirals consistently appear $200$–$250$~Myr after the mode transition.
The delay in the emergence of the two-arm phase spiral relative to the dominant mode transition occurs because the bending and breathing amplitudes reflect the current states of these oscillation modes, while the phase spirals correspond to their histories.

In Appendix~\ref{appendix:vR_vphi_phasespirals}, we present the time evolution of the vertical phase space maps colour-coded by $\Delta v_R$ and $\Delta v_{\phi}$.
The phase spirals labelled by $\Delta v_R$ and $\Delta v_{\phi}$ emerge earlier than those labelled by $\Delta \rho$, and their morphologies (one-arm or two-arm) do not always coincide.

\section{Discussions}\label{sec:discussion}
\subsection{Dating the perturbation to the MW disc}
The phase spiral contains information about the timing of the perturbation event, as it winds up over time at a fixed rate determined by the galactic potential.
Several previous studies \citep[e.g.][]{2018Natur.561..360A, 2023A&A...673A.115A,2023MNRAS.521.5917F, 2023ApJ...955...74D} estimated the dynamical age of the one-arm phase spiral to be $\sim200$–900 Myr by `unwinding' the spiral under assumed galactic potentials.
However, these analyses often neglect the disc’s self-gravity, which slows the winding rate of phase spirals \citep{2023MNRAS.522..477W} and could lead to an underestimation of their dynamical age.

An alternative approach was proposed by \citet{2022A&A...668A..61A}, who linked the perturbation time to tidally induced spiral arms.
In our simulation, tidal forces from a satellite induce spiral arms in the disc that rotate at $\Omega - \kappa/2$.
These arms create a global velocity pattern that manifests as a wave in $v_R$ as a function of $L_z$ when observed within a narrow azimuthal range.
The wavelength of this wave corresponds to the time since the formation of the arms, as the wave shortens over time due to the winding of the arms.
\citet{2019MNRAS.490.5414F} detected a similar $L_z$-$v_R$ wave in the \textit{Gaia} data, and \citet{2022A&A...668A..61A} identified two distinct wavelengths corresponding to the arm formation times of $< 0.6$~Gyr and 0.8--2.1~Gyr through Fourier analysis. 
Although the origin of these waves remains debated \citep{2024ApJ...975..292C}, this method offers a complementary avenue for estimating perturbation times.

Our results suggest a new method for estimating the perturbation time using the relative strengths of the bending and breathing modes. In our simulations, we observed a transition from bending-dominated to breathing-dominated vertical oscillations. This transition progresses outward from the inner to the outer galaxy, with the timescale determined by $R/\sigma_R$. While it is challenging to determine azimuthally averaged amplitude ratios from observational data as in Section~\ref{sec:bend_breath_ana}, local observations of phase spirals can provide constraints. For example, in Section~\ref{sec:phase_spiral_time}, we showed that two-arm phase spirals emerge $\sim200$–250 Myr after the bending-to-breathing transition. The detection of two-arm phase spirals in \textit{Gaia} data implies that part of the MW disc transitioned to a breathing-dominated state at least $\sim200$ Myr ago.

The two-arm phase spiral is predominantly observed for stars with small guiding radii, roughly $R_g\lesssim 7$~kpc \citep{2022MNRAS.516L...7H, 2024A&A...690A..15A}.
The bending-to-breathing transition takes $R/\sigma_R\sim200$~Myr at $R=7$~kpc, where $\sigma_R\sim35$~\kms  \citep{2016ARA&A..54..529B, 2018A&A...616A..11G}.
This suggests that the MW disc was perturbed more than $\sim400$ Myr ago ($200$ Myr for the transition plus $200$ Myr for the emergence of two-arm phase spirals).
This estimate is consistent with the previous studies \citep{2018Natur.561..360A, 2022A&A...668A..61A, 2023A&A...673A.115A,  2021MNRAS.503.1586L, 2023MNRAS.520.3329L, 2023MNRAS.521.5917F, 2023ApJ...955...74D}.

\subsection{Sgr mass problem}
Previous studies \citep[e.g.][]{2018MNRAS.481.1501B, 2022ApJ...927..131B} suggest that satellites must be heavier than  $\sim 3\times10^{10}M_{\odot}$ to excite phase spirals.
This required mass is several orders of magnitude greater than the present-day mass of Sgr, estimated at $\sim5 \times 10^8~M_{\odot}$ \citep{2020MNRAS.497.4162V}.
In our dwarf model, the satellite initially has a total mass of $5 \times 10^{10}M_{\odot}$ and loses 50\% of its mass during each orbital loop, reducing its final mass to match the present-day Sgr.
However, the phase spirals produced in the final epoch of the simulation are much fainter than those observed in the \textit{Gaia} data.

This discrepancy between the theoretically required mass and the observationally estimated mass may be resolved by properly accounting for Sgr’s mass loss history.
\citet{2021MNRAS.504.3168B} suggest that the Sgr could have excited the phase spiral during earlier pericentre passages if it had experienced mass loss of 0.5--1 dex per orbit loop. 
Such high mass loss rates are dynamically plausible.
\citet{2000MNRAS.314..468J}, using a semi-analytic model and $N$-body simulations, demonstrated that a wide range of orbital histories can reproduce the present-day position and velocity of the Sgr. 
An extreme scenario involves Sgr initially having a mass of $\sim 10^{11}M_{\odot}$ and being located at $r \sim 250$~kpc from the MW centre.
By fine-tuning the initial conditions of the Sgr model, it might be possible to simultaneously reproduce the phase spiral and the present-day state of Sgr. Such adjustments could help resolve the tension between the mass required to excite the phase spiral and Sgr’s currently observed mass.

\subsection{Comparison with an isolated galaxy model}
\begin{figure}
	\begin{center}
		\includegraphics[width=\hsize]{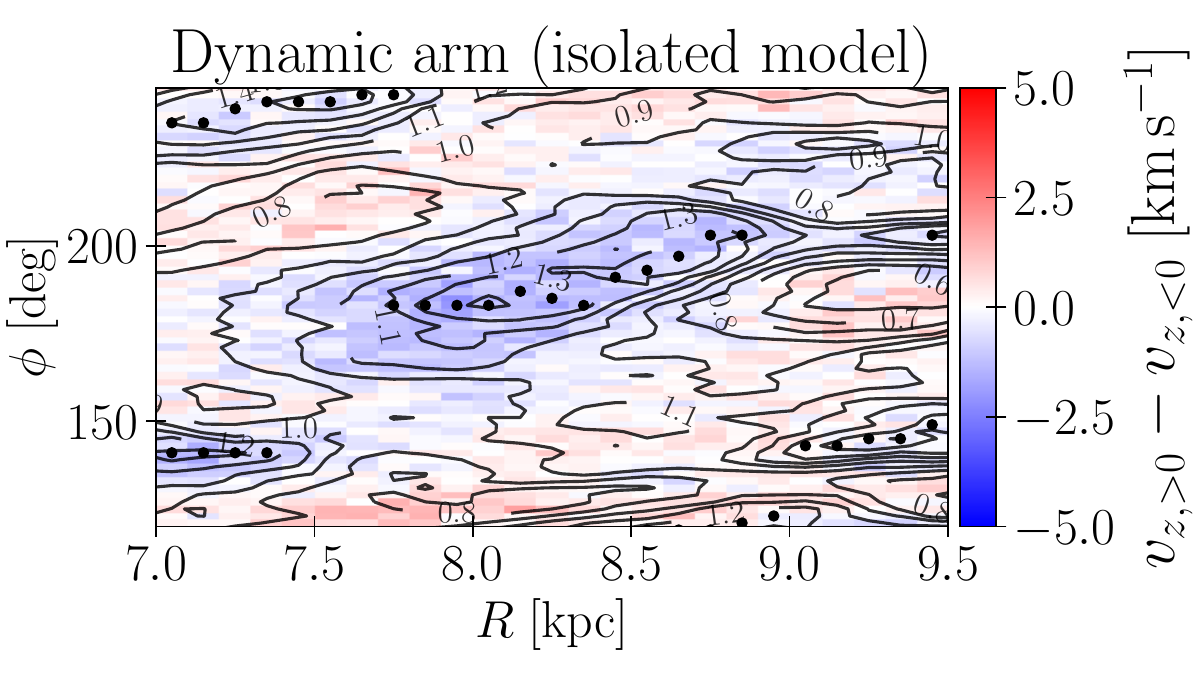}
		\includegraphics[width=\hsize]{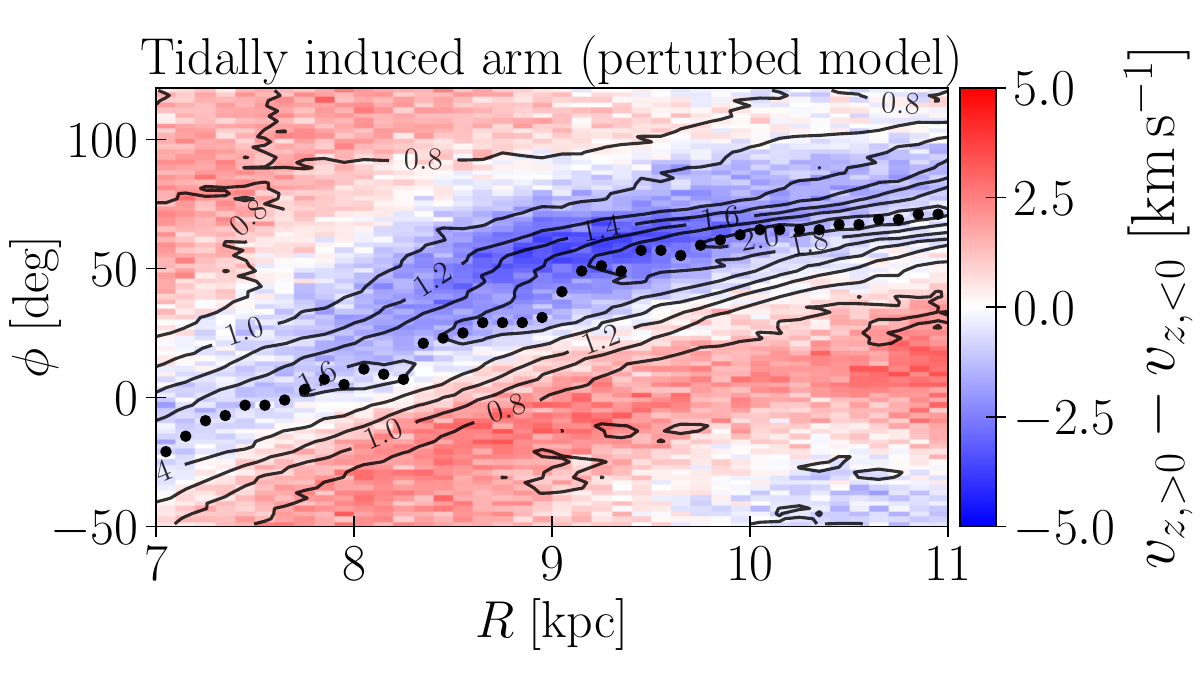}
		\caption{Breathing signature around spiral arms. \textit{Upper panel:} Breathing signature in the isolated model. Colour map shows the breathing velocity, $v_{z,>0} - v_{z,>0}$. Contours indicate the normalised density, $\Sigma(R,\phi)/\Sigma_0(R)$. Black dots indicate the density peaks. \textit{Lower panel:} Same as the upper panel but in the perturbed model.
			In the perturbed model (tidally induced arms), the breathing velocity minimum is displaced from the density peak to the trailing side, unlike in the isolated model (dynamic arms) where they are aligned.
		}\label{fig:breath_R_phi}
	\end{center}
\end{figure}
\begin{figure}
	\begin{center}
		\includegraphics[width=0.8\hsize]{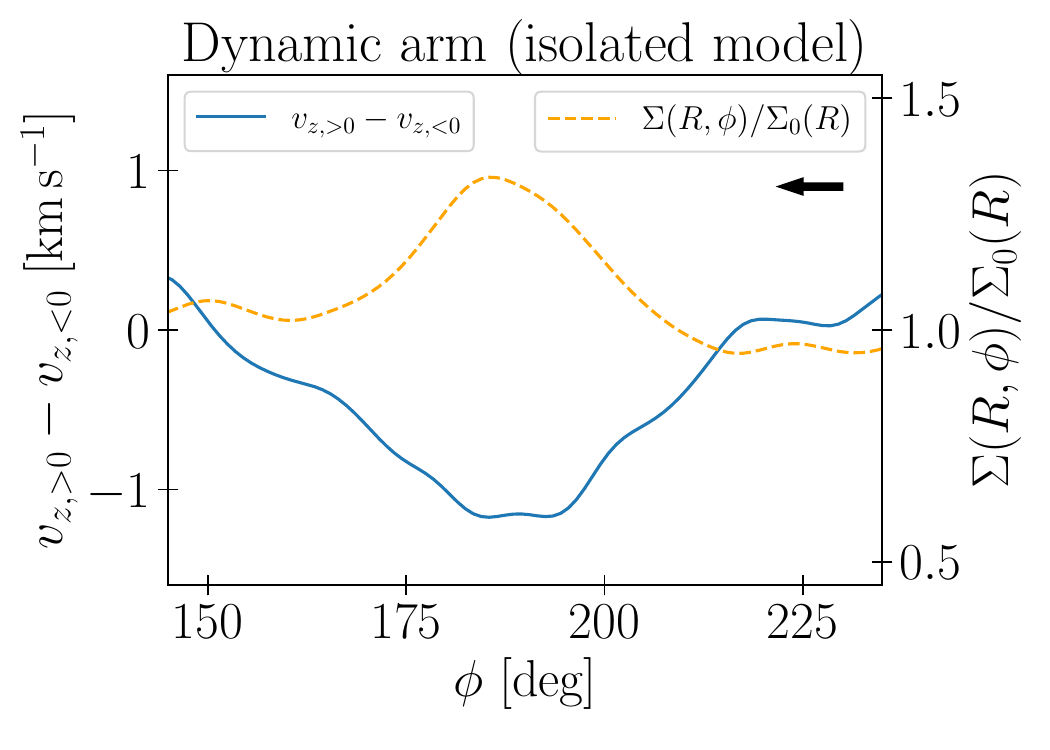}
		\includegraphics[width=0.8\hsize]{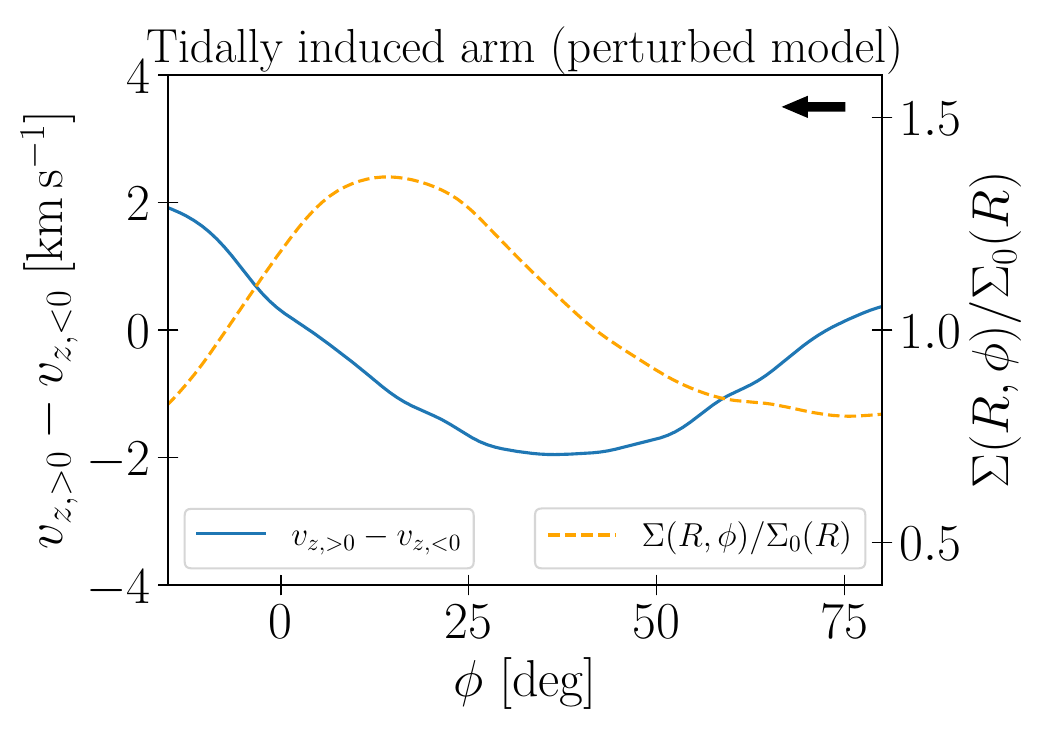}
		\caption{Breathing velocity and normalised density as functions of the azimuth. \textit{Upper panel:} Isolated model. The solid line and dashed line represent the breathing velocity and the normalised density, respectively,  averaged over $R$ between $R=8$~kpc and 8.5~kpc. The arrow indicates the direction of the galaxy rotation. \textit{Lower panel:} Same as the upper panel but in the perturbed model.
			This plot quantifies the displacement in Fig.~\ref{fig:breath_R_phi}, showing that the breathing velocity minimum is shifted to the trailing side relative to the density peak in the perturbed model with tidally induced arms. This contrasts with the alignment seen in the isolated model with dynamic arms.
		}\label{fig:breath_dens}
	\end{center}
\end{figure}

Breathing modes are not exclusive to perturbed galaxies; isolated galaxies can also exhibit breathing modes if they possess spiral arms or bars. \citet{2024MNRAS.529L...7A} investigate the relation between the breathing mode and the evolution of dynamic spiral arms in an isolated galaxy model.
In this section, we compare the characteristics of breathing modes in an isolated galaxy with those in a perturbed galaxy, focusing on their spatial alignment and amplitude.

The upper panel of Fig.~\ref{fig:breath_R_phi} shows the breathing velocity, $v_{z,>0} - v_{z,<0}$, around a spiral arm accompanying strong breathing mode in the isolated model of \citet{2019MNRAS.482.1983F}.
In this frame, the galaxy rotates from top to bottom.
The contours represent the normalised surface density, $\Sigma(R,\phi)/\Sigma_0(R)$, and the black dots indicate its peaks, detected using the \texttt{signal.find\_peaks} routine in \texttt{scipy} package \citep{2020NatMe..17..261V}.
The compressing breathing mode (indicated by the blue colour) is observed within the spiral arms, with no systematic displacement between the compressing area and the density peaks.

The lower panel of Fig.~\ref{fig:breath_R_phi} shows the same plot for the perturbed model.
In this case, the breathing signature is observed globally along the tidally induced spiral arm.
The compressing mode is stronger on the trailing side of the arm than on the leading side, and the expanding mode is more prominent in the inter-arm regions.

This displacement is more evident in Fig.~\ref{fig:breath_dens}, where the breathing velocity and the normalised surface density are shown as functions of $\phi$, averaged over $R$ between $R = 8$~kpc and $R = 8.5$~kpc.
In the isolated model (upper panel), the minimum of the breathing velocity almost coincides with the maximum of the density at $\phi \sim 190^{\circ}$.
In contrast, in the perturbed model (lower panel), the density peaks are located at $\phi \sim 20^{\circ}$, while the minimum breathing velocity is shifted to $\phi \sim 35^{\circ}$.

The difference in the spatial alignment of the breathing modes between the two models can be attributed to the distinct pattern speeds of dynamic arms versus tidally induced arms.
For density-wave-like arms with pattern speeds slower than the galaxy’s rotation, compressing breathing modes and expanding breathing modes appear on the trailing and leading sides of the arms, respectively \citep{2014MNRAS.443L...1D, 2014MNRAS.440.2564F, 2016MNRAS.457.2569M}.
This scenario aligns with the tidally induced arms in our perturbed model, where the breathing velocity minimum is shifted to the trailing side.
Conversely, dynamic arms, which corotate with disc stars at every radius, show no displacement between arm locations and breathing modes \citep{2013ApJ...763...46B}.

\begin{figure}
	\begin{center}
	\includegraphics[width=0.49\hsize]{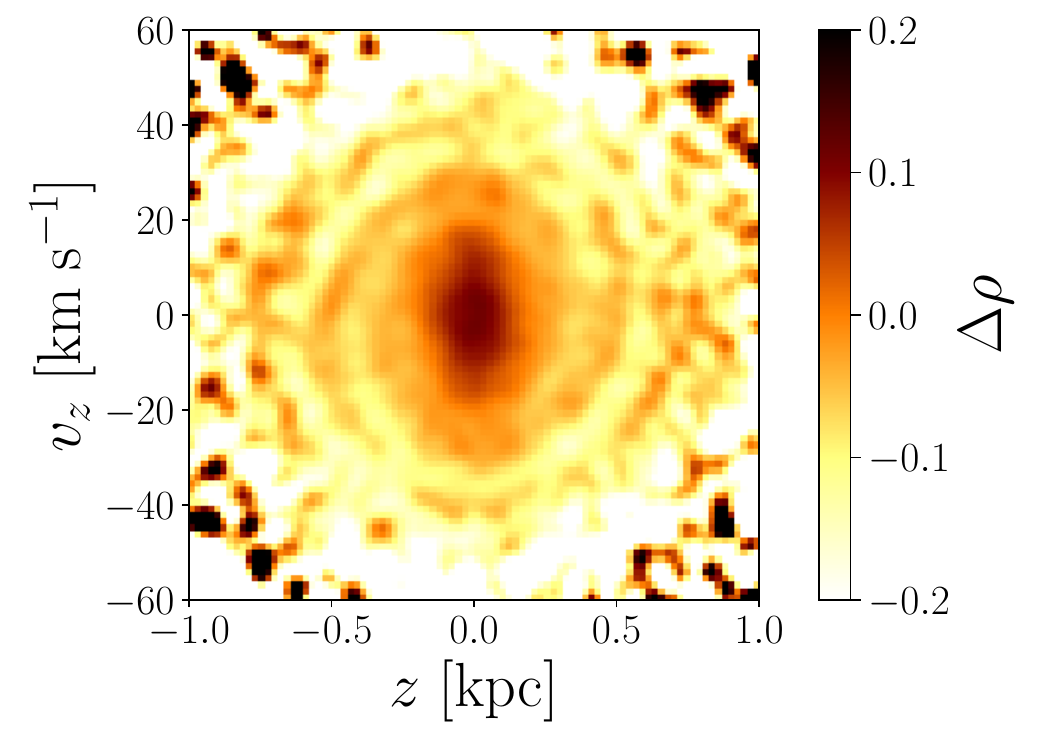}
	\includegraphics[width=0.49\hsize]{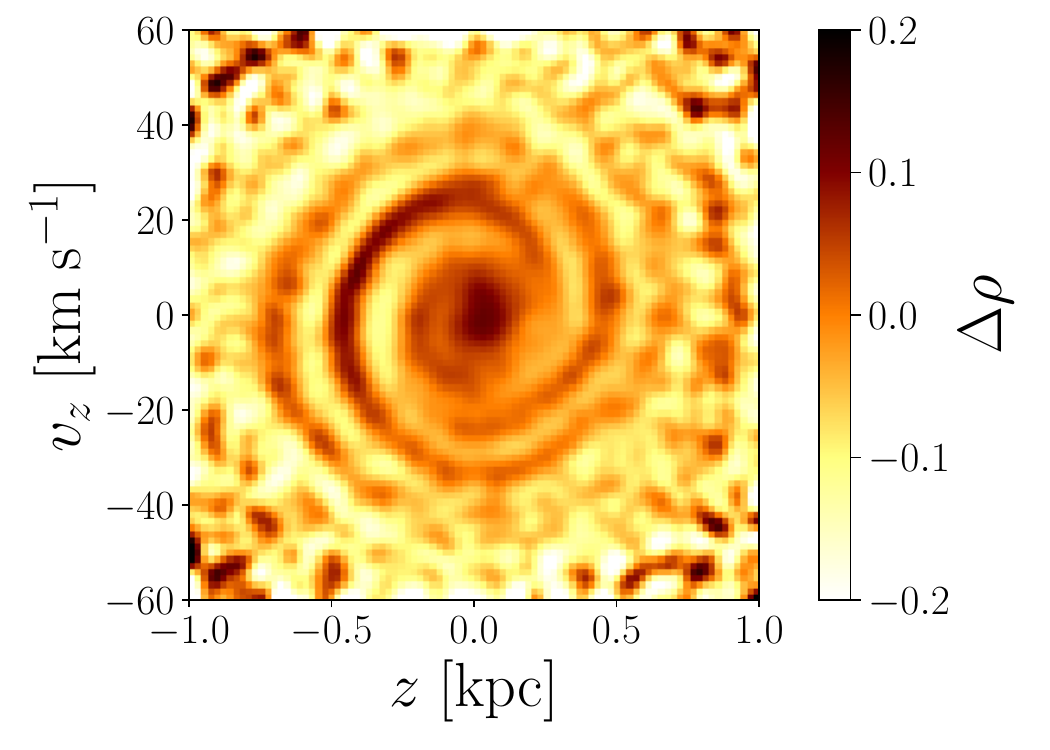}
	\caption{Two-arm phase spiral in the isolated model (\textit{left}) and in the perturbed model (\textit{right}).
	The two-arm phase spiral is fainter in the isolated model than in the  perturbed model.
	}\label{fig:phase_spiral_isolated}
\end{center}
\end{figure}

The amplitude of the breathing mode also differs between the models. In the perturbed model, the amplitude is higher than in the isolated model, as tidally induced arms are stronger and longer-lived than dynamic arms. This difference in amplitude affects the morphology of the phase spiral. Fig.~\ref{fig:phase_spiral_isolated} compares the two-arm phase spirals in the isolated and perturbed models, showing that the phase spiral in the isolated model is fainter.

\cite{2024MNRAS.529L...7A} conclude that the Local arm aligns with the dynamic arm model because there is no systematic displacement between the compressing breathing mode and the arm location.
They also discussed the growth and disruption of the other arms based on the tentative signals of the breathing mode around them, assuming the dynamic arm model.
However, they do not rule out the tidally induced arm model, as the spatial correlations between the arms and the breathing mode are less clear compared to the Local arm due to the observational uncertainties.
The detection of the prominent two-arm phase spiral suggests that the MW has strong spiral arms, even if not tidally induced arms.

\subsection{Model limitations}
Our high-resolution cold disc model captures the spatial and temporal evolution of vertical oscillations in the MW disc in detail.
However, several important components that could influence the disc's dynamical response are not included.
While the MW disc is composed of multiple components, specifically the thin disc, thick disc, and gas disc, our model only has a single component corresponding to the thin disc.
Observational estimates suggest that the thick and gas discs each account for only $\sim$10--20\% of the total disc mass  \citep{2016ARA&A..54..529B, 2016PASJ...68....5N, 2017MNRAS.465...76M}. 
Since the thin disc makes the dominant contribution to the total disc potential, it is reasonable to assume that the thin disc dominates the overall dynamical response to external perturbations.
Nonetheless, we discuss below the potential influence of the thick and gas discs based on insights from recent studies.

\citet{2024ApJ...977..252S} performed $N$-body simulations of warm and cold stellar discs perturbed by the Sgr and the Large Magellanic Cloud (LMC). 
Their models exhibit only minor differences in the resulting bending amplitudes, implying that the inclusion of a thick disc (i.e. a dynamically hotter component) does not significantly alter the global bending motion.
However, the two models show notable difference in the $v_R$ fields, which are related to the tidally induced arms.
This suggests that the thick disc may modulate the amplitude of breathing modes, which are excited by the spiral arms.

The impact of the gas disc is less well understood. 
\citet{2022MNRAS.515.5951T} study the evolution of vertical corrugations in a combined stellar+gas disc using hybrid $N$-body and hydrodynamical simulations. 
They demonstrate that, although stellar and gas corrugations are initially excited in phase, the gas component is damped more quickly due to dissipative processes.
Based on their results, if the interaction between the gas disc and the stellar disc is not negligible, gas disc may influence the evolution of bending modes in the stellar component. 
We also expect that self-gravity is more important in stellar+gas discs than in pure stellar discs, as the gas helps keep the disc dynamically cold.
Therefore, high-resolution $N$-body and hydrodynamical simulations are important to understand influences of self-gravity on the disc oscillation.
Additionally, if we can include star formations in such simulations, they will allow us to test the possible correlation between the Sgr impact and the star formation activity in the MW disc suggested by \citet{2020NatAs...4..965R}.

Our study focuses on the perturbation of the Sgr on the vertical structure of the MW disc.
However, it is important to recognise that the MW is also interacting with the LMC, whose gravitational influence cannot be neglected.
Some recent studies \citep[e.g.][]{2018MNRAS.481..286L, 2024ApJ...977..252S} performed $N$-body simulations of the MW-Sgr-LMC system to investigate the combined impact of these two satellites.
One of the notable features seen in their models is the disc warp similar to the warp in the MW disc.
Although our Sgr-only model also produces a warp, its orientation in the present-day snapshot ($t = 1.78$~Gyr) does not align with that of the observed MW. 
In our model, the outer disc exhibits a vertical displacement of $z<0$ at $y>0$ and $z>0$ at $y<0$, whereas the observed MW warp displays the opposite pattern.
This discrepancy suggests that the LMC plays a crucial role in shaping the actual Galactic warp.

The outer disc appears to be strongly perturbed by the combined influences of Sgr and the LMC, while the inner disc is thought to be primarily affected by Sgr.
The LMC is currently located at a Galactocentric distance of $\sim 50$~kpc, having recently passed its pericentre.
The disc is only beginning to respond to the LMC's perturbation from the outer part.
There has not yet been sufficient time for vertical features such as corrugations or phase spirals to fully develop \citep{2019MNRAS.485.3134L}.
Therefore, our Sgr-only model remains valuable for studying the disc dynamics at inner and intermediate radii, where the Sgr’s influence dominates.
As an example, we used the ratio $R/\sigma_R$ to trace the spatial transition from bending-dominated to breathing-dominated regimes. This diagnostic should remain valid for $R\lesssim 15$~kpc, even in the presence of the LMC.
At larger radii, however, models incorporating both Sgr and the LMC are necessary for a more accurate representation.

In addition to its direct perturbation, the LMC may also influence the orbit of the Sgr dwarf, thereby altering its interaction history with the MW.
\citet{2021MNRAS.501.2279V}  demonstrate that including the LMC in $N$-body simulations of the MW-Sgr interaction modifies Sgr’s orbits, particularly its eccentricity. This affects the pericentre distances and potentially alters the strength of Sgr’s impact during previous pericentre passages.

In summary, Sgr appears to be the primary perturber of the MW disc, at least in the solar neighbourhood, and the response of the thin disc plays a key role in determining the overall dynamical behaviour.
However, a comprehensive understanding of the MW’s response to external perturbations requires simulations of multi-component discs that include both Sgr and the LMC, and we plan to perform such simulations in future work.
They will help us reconstruct the history of perturbation events encoded in the bending and breathing signatures in the MW disc.

\section{Conclusions}\label{sec:conclusion}
In this study, we used high-resolution $N$-body simulations to investigate the interaction between the MW and the Sgr dwarf galaxy. This interaction generates vertical oscillations in the MW disc, including bending and breathing modes, as well as phase spirals associated with these oscillations. Our findings highlight the interplay between direct satellite perturbations and indirect effects mediated by tidally induced spiral arms, shaping the vertical structure of the MW disc.
Our main results are summarised as follows:

\begin{enumerate}
	\item The satellite's perturbation directly excites the bending mode across a wide radius in the galactic disc. Simultaneously, it induces the spiral arms within the disc.
	\item The tidally induced spiral arms subsequently excite the breathing mode in the disc. As such, the breathing mode is the indirect result of the satellite interaction.
	\item Initially, the bending mode dominates, but the dominant mode transitions to the breathing mode over time.
    This transition originates from the different timescales of the two modes: the bending mode decays rapidly mainly due to horizontal mixing, while the breathing mode persists, sustained by the spiral arms. The transition occurs faster in the inner galaxy than in the outer regions.
	\item The bending mode is dominated by its $m=1$ Fourier component, with its spectrogram displaying distinct branches aligned with the resonant curves of $\Omega \pm \nu$ and $\Omega - 2\nu$. In contrast, the dominant Fourier component for the breathing mode is $m=2$, which rotates at $\sim 40 , \kmskpc$ in the inner galaxy and $\Omega - \kappa/2$ in the outer galaxy, corresponding to the pattern speeds of the bar and the tidally induced spiral arms, respectively.
	\item The simulation successfully reproduces the one-arm phase spiral observed in the solar neighbourhood. Additionally, it reveals two-arm phase spirals, particularly in the inner galaxy, linked to the breathing mode excited by spiral arms. The simulations also indicate that the two-arm phase spiral emerges approximately 200-250 Myr after the transition to the breathing-dominated state.
	\item The detection of the two-arm phase spiral in the \textit{Gaia} data suggests that parts of the MW disc transitioned to a breathing-dominated state at least $\sim200$~Myr ago.
    Considering the bending-to-breathing transition timescale, we estimate that the MW disc was perturbed more than $\sim400$~Myr ago.
\end{enumerate}

These results provide insights into the MW disc's response to external perturbations, particularly from the Sgr dwarf galaxy, and highlight the importance of both direct and indirect effects in shaping the disc's vertical structure.
We plan to extend this work by incorporating the LMC and a multi-component disc model to further investigate the MW's response to external perturbations and to reconstruct the history of perturbation events encoded in the MW disc.

\begin{acknowledgements}
We thank the anonymous referee for the useful comments.
We appreciate discussions with the members of the \textit{Gaia}-UB group. 
This research used computational resources of Pegasus and Cygnus provided by Multidisciplinary Cooperative Research Program in Center for Computational Sciences, University of Tsukuba.
Part of the numerical computations were carried out on HA-PACS at Tsukuba University, Piz Daint at the Swiss National Supercomputing Centre and Little Green Machine II.
In this research work, we used the ``mdx: a platform for the data-driven future''.
TA was supported by Japan Society for the Promotion of Science (JSPS) Research Fellowships for Young Scientists and JSPS Overseas Challenge Program for Young Researchers.
This work was supported by Grant-in-Aid for JSPS Fellows Number JP22J11943.
JB acknowledges support from JSPS under Grant Numbers 21K03633 and 21H00054.
We acknowledge the grants PID2021-125451NA-I00 and CNS2022-135232 funded by MICIU/AEI/10.13039/501100011033 and by ``ERDF A way of making Europe'', by the ``European Union'' and by the ``European Union Next Generation EU/PRTR''.

This work made use of the following software packages: \texttt{Bonsai} \citep{2012JCoPh.231.2825B, 2014hpcn.conf...54B}, \texttt{Galactics} \citep{1995MNRAS.277.1341K, 2005ApJ...631..838W, 2008ApJ...679.1239W}, \texttt{astropy} \citep{2013A&A...558A..33A, 2018AJ....156..123A, 2022ApJ...935..167A}, \texttt{Jupyter} \citep{2007CSE.....9c..21P, 2016ppap.book...87K}, \texttt{matplotlib} \citep{2007CSE.....9...90H}, \texttt{numpy} \citep{numpy}, \texttt{pandas} \citep{mckinney-proc-scipy-2010, pandas_10537285}, \texttt{scipy} \citep{2020NatMe..17..261V, scipy_10155614} and \texttt{Agama} \citep{2019MNRAS.482.1525V}.
This research has made use of NASA's Astrophysics Data System.Software citation information aggregated using \texttt{\href{https://www.tomwagg.com/software-citation-station/}{The Software Citation Station}} \citep{2024arXiv240604405W, software-citation-station-zenodo}.
\end{acknowledgements}

\bibliographystyle{aa}
\bibliography{extracted}

\begin{appendix}
\FloatBarrier
\section{Additional material about the phase spiral}\label{appendix:phase_spiral}
\subsection{Dependence of $R_g$ and $\theta_{\phi}$ of the phase spiral}\label{appendix:Rg_thetaphi_phasespirals}
We created vertical phase-space maps by dividing the local sample based on the guiding radius ($\Rg$) and azimuthal angle ($\theta_{\phi}$) in a similar way to how \citet{2021MNRAS.508.1459H, 2022MNRAS.516L...7H} did.
This sample division allowed us to estimate the global phase mixing state from the local sample \citep{2021ApJ...911..107L, 2022ApJ...928...80G}.
We first constructed the potential model from the $N$-body snapshot using \texttt{Agama} \citep{2019MNRAS.482.1525V}.
We applied multipole expansion to the DM halo and the classical bulge and azimuthal harmonic expansion to the disc. 
We converted the angular momentum, $L_z$, to the guiding radius, $\Rg$, by interpolating the $R$-$L_z$ relation obtained from the potential model. 
The azimuthal angle was computed with St\"{a}ckel fudge \citep{2012MNRAS.426.1324B} implemented in \texttt{Agama}.
It is important to note that the approximation of axisymmetric potentials may lead to significant biases in the estimation of angle-action variables when the galaxies have strong spiral arms \citep{2025MNRAS.537.1620D}.

\begin{figure}
	\begin{center}
		\includegraphics[width=0.8\hsize]{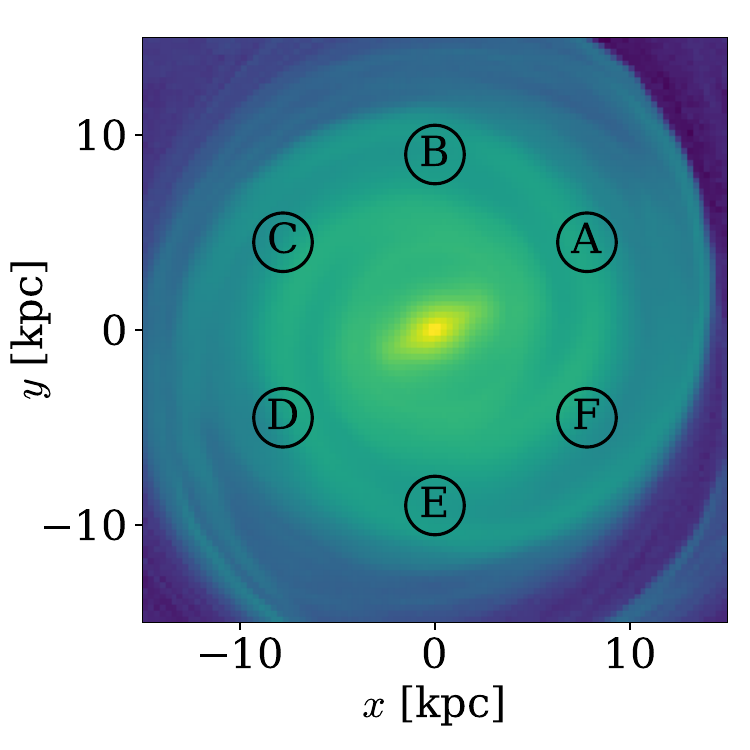}
		\caption{Face-on density map at $t=1.65$~Gyr. Six circles indicate regions A--F.}\label{fig:local_areas}
	\end{center}
\end{figure}

\begin{figure}
	\begin{center}
		\includegraphics[width=\hsize]{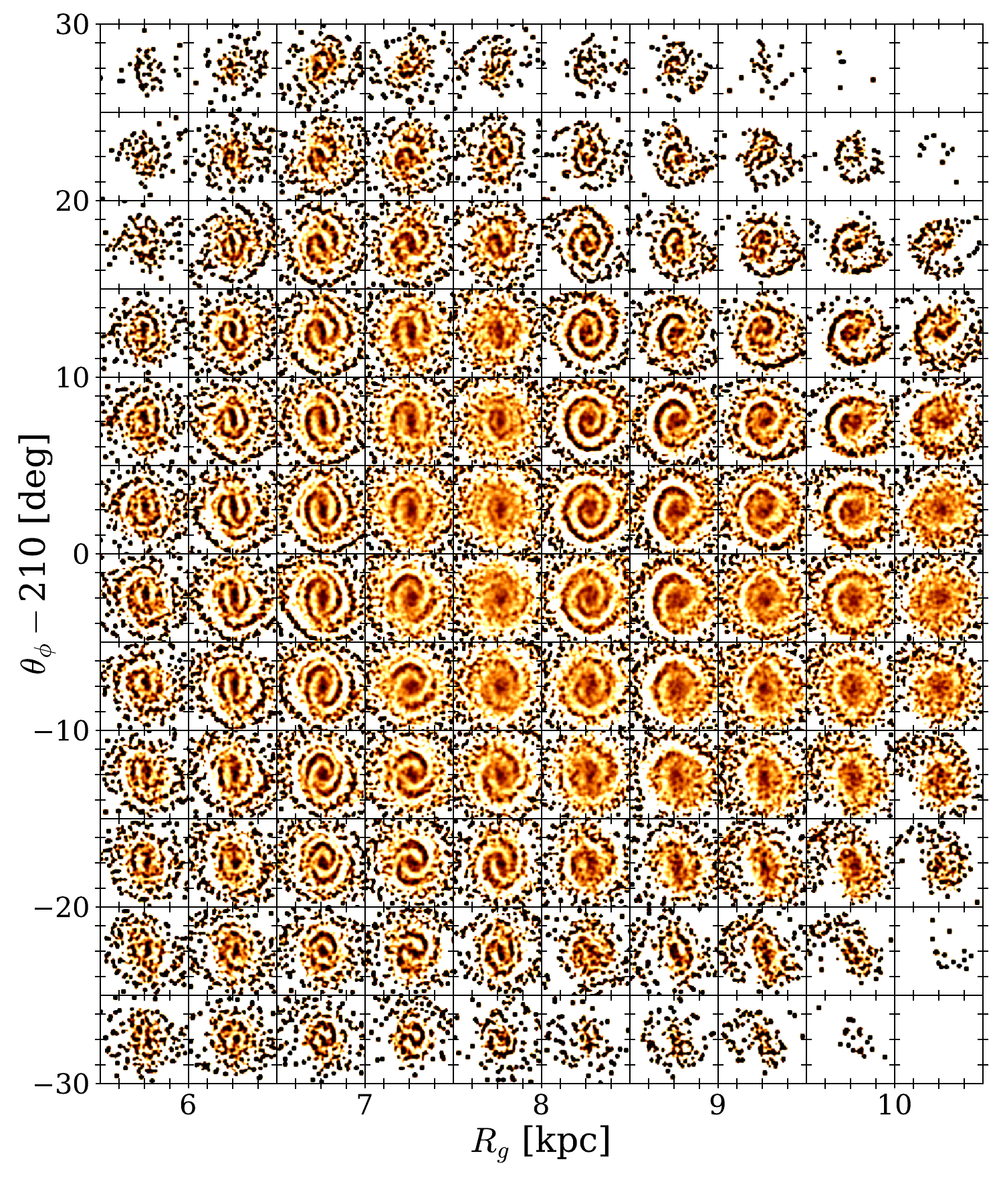}
		\caption{Dependence of $\Rg$-$\theta_{\phi}$ of the phase spiral in Region D. Each panel shows the $z/h_z$-$v_z/\sigma_z$ map colour-coded by $\Delta \rho$. The galaxy rotates from top to bottom in this frame.}\label{fig:z_vz_Rg_theta_D}
	\end{center}
\end{figure}

\begin{figure}
	\begin{center}
		\includegraphics[width=\hsize]{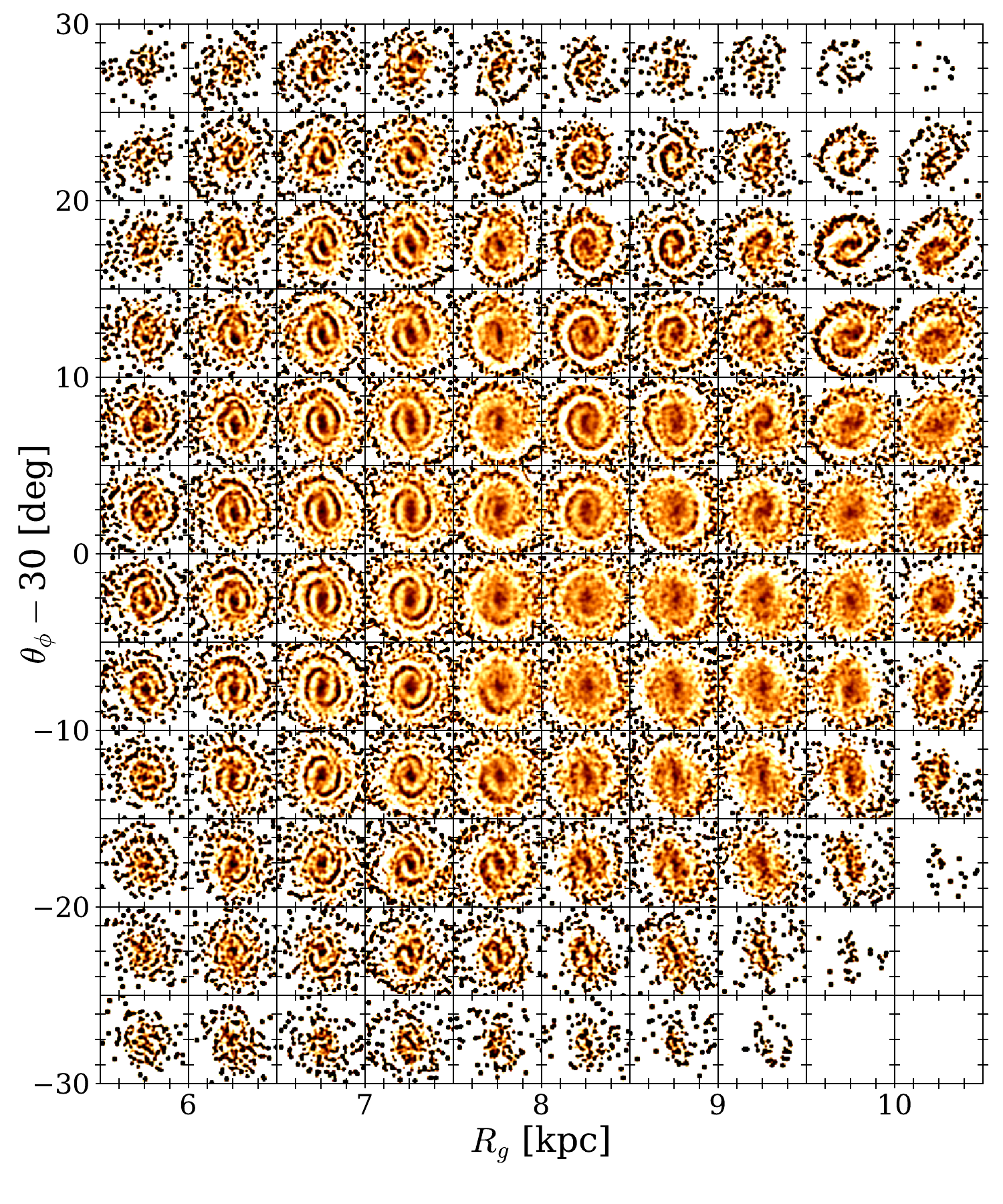}
		\caption{Same as Fig.~\ref{fig:z_vz_Rg_theta_D} but for Region A.}\label{fig:z_vz_Rg_theta_A}
	\end{center}
\end{figure}

\begin{figure}
	\begin{center}
		\includegraphics[width=\hsize]{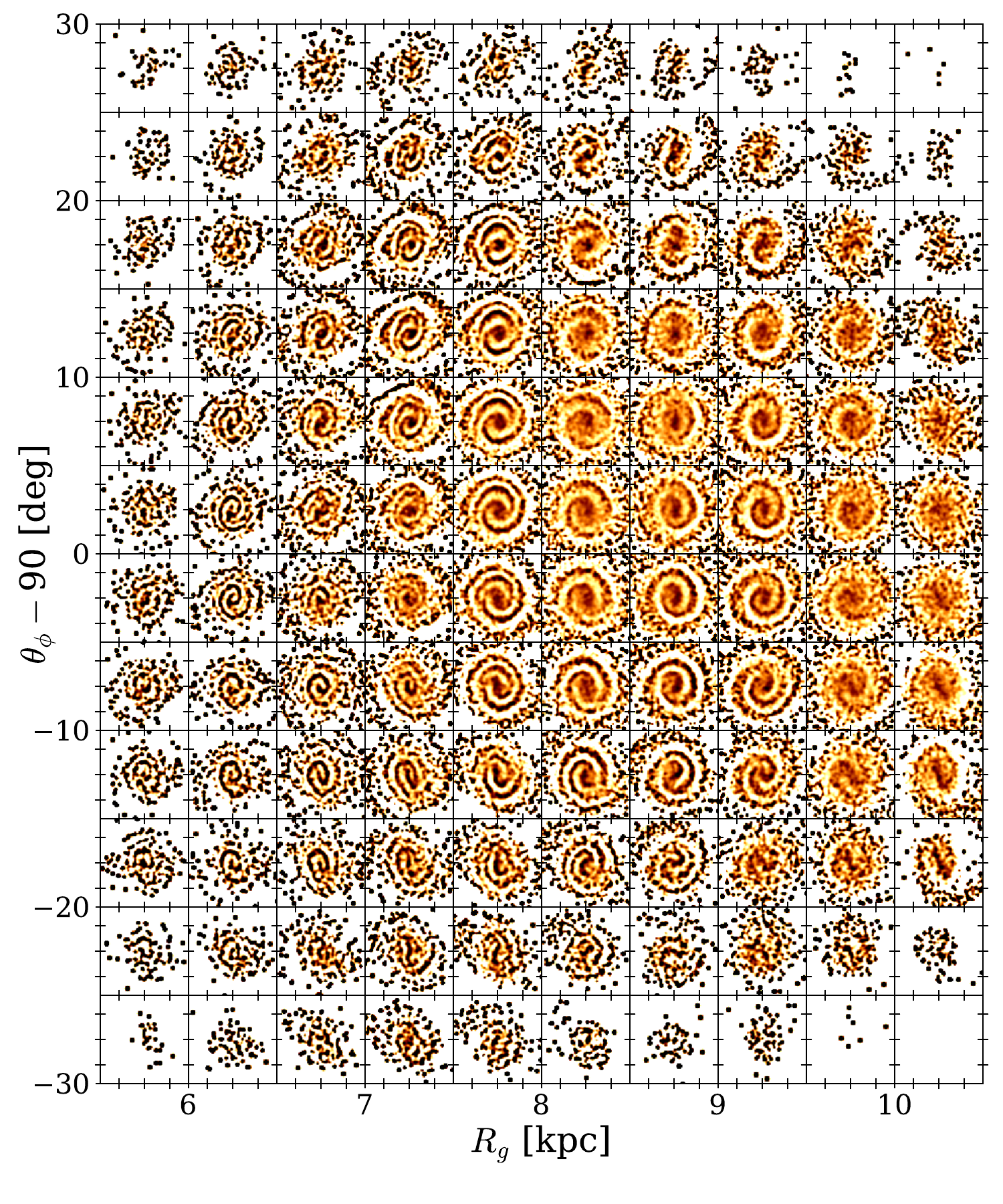}
		\caption{Same as Fig.~\ref{fig:z_vz_Rg_theta_D} but for Region B.}\label{fig:z_vz_Rg_theta_B}
	\end{center}
\end{figure}

\begin{figure}
	\begin{center}
		\includegraphics[width=\hsize]{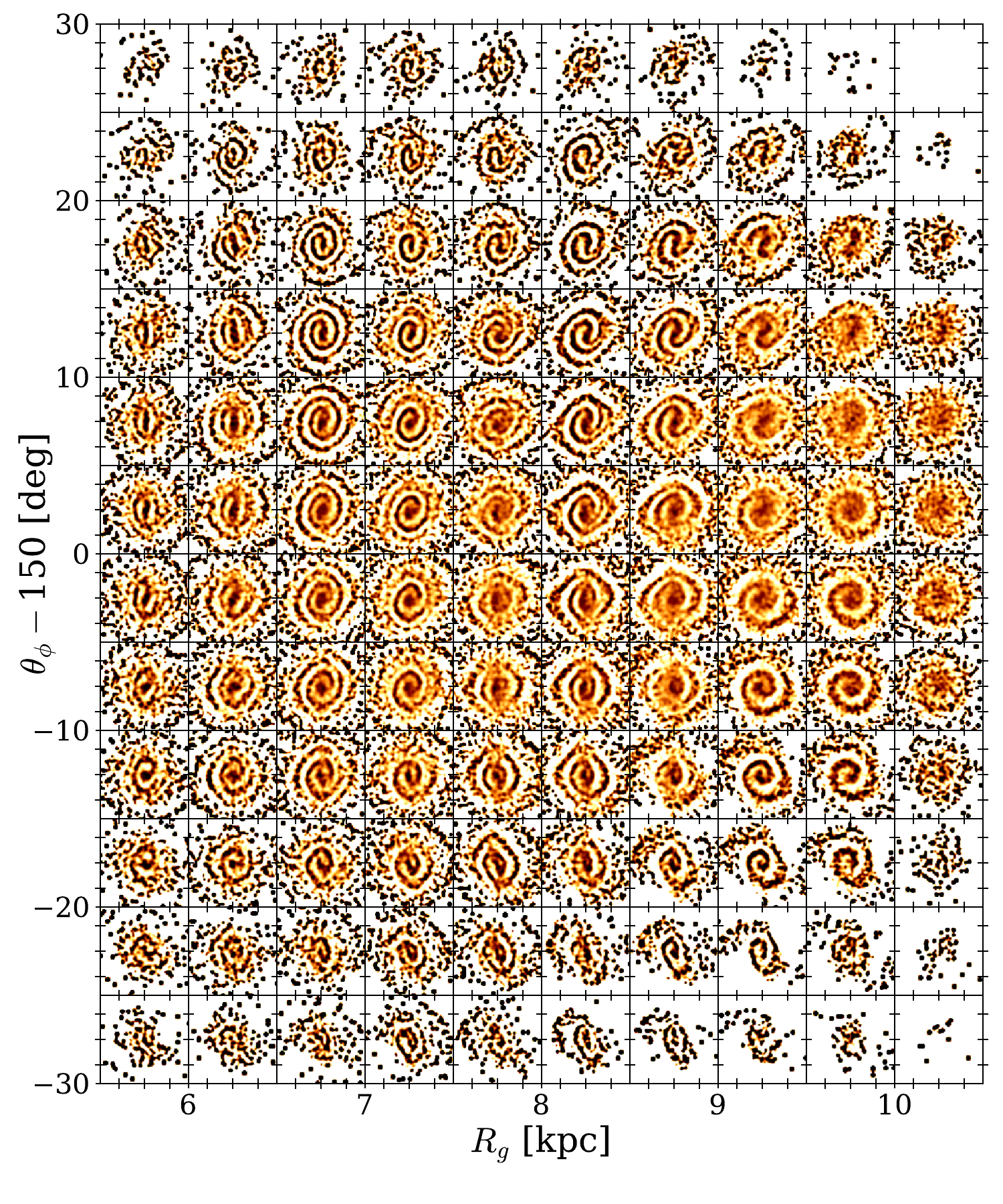}
		\caption{Same as Fig.~\ref{fig:z_vz_Rg_theta_D} but for Region C.}\label{fig:z_vz_Rg_theta_C}
	\end{center}
\end{figure}

\begin{figure}
	\begin{center}
		\includegraphics[width=\hsize]{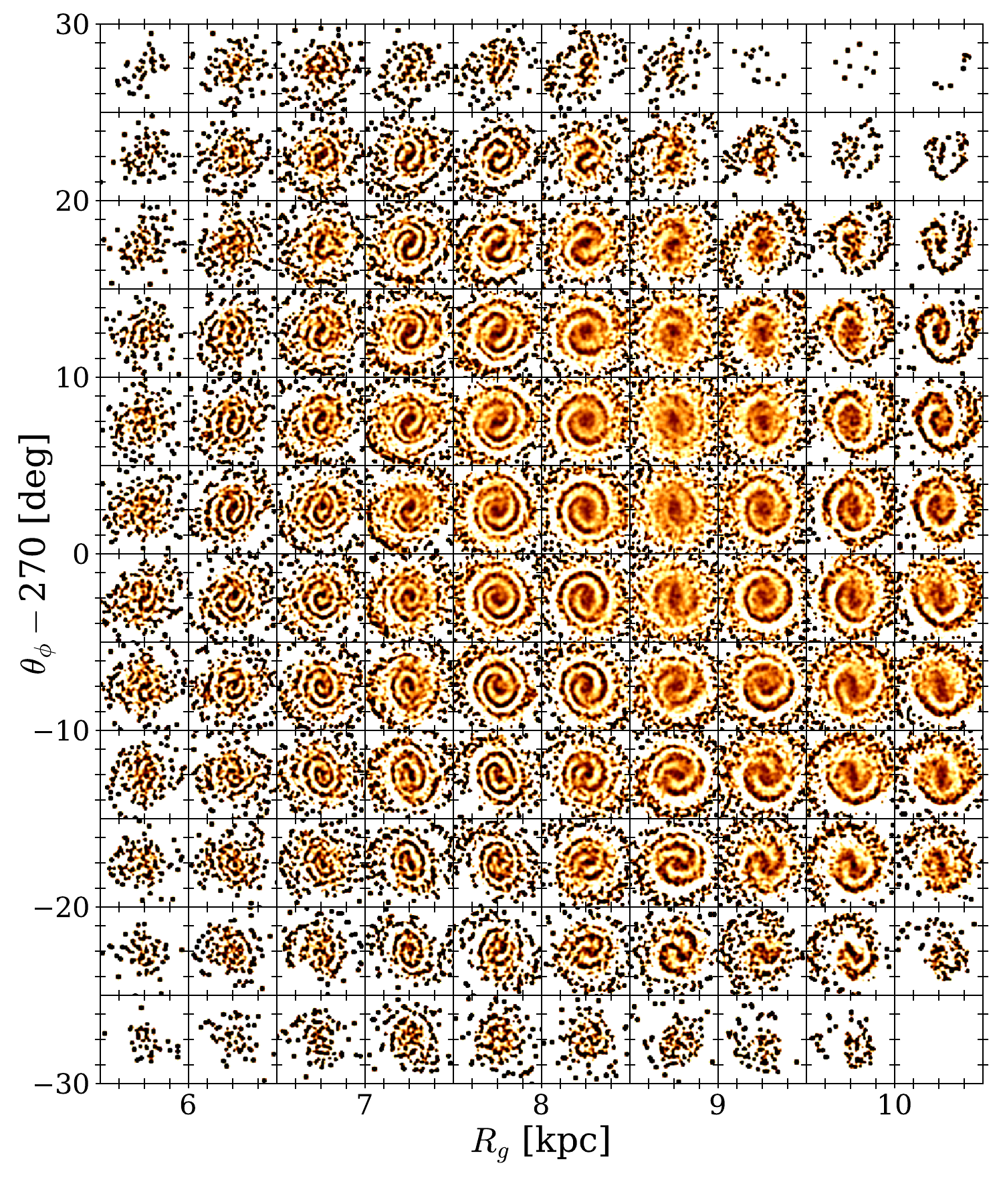}
		\caption{Same as Fig.~\ref{fig:z_vz_Rg_theta_D} but for Region E.}\label{fig:z_vz_Rg_theta_E}
	\end{center}
\end{figure}

\begin{figure}
	\begin{center}
		\includegraphics[width=\hsize]{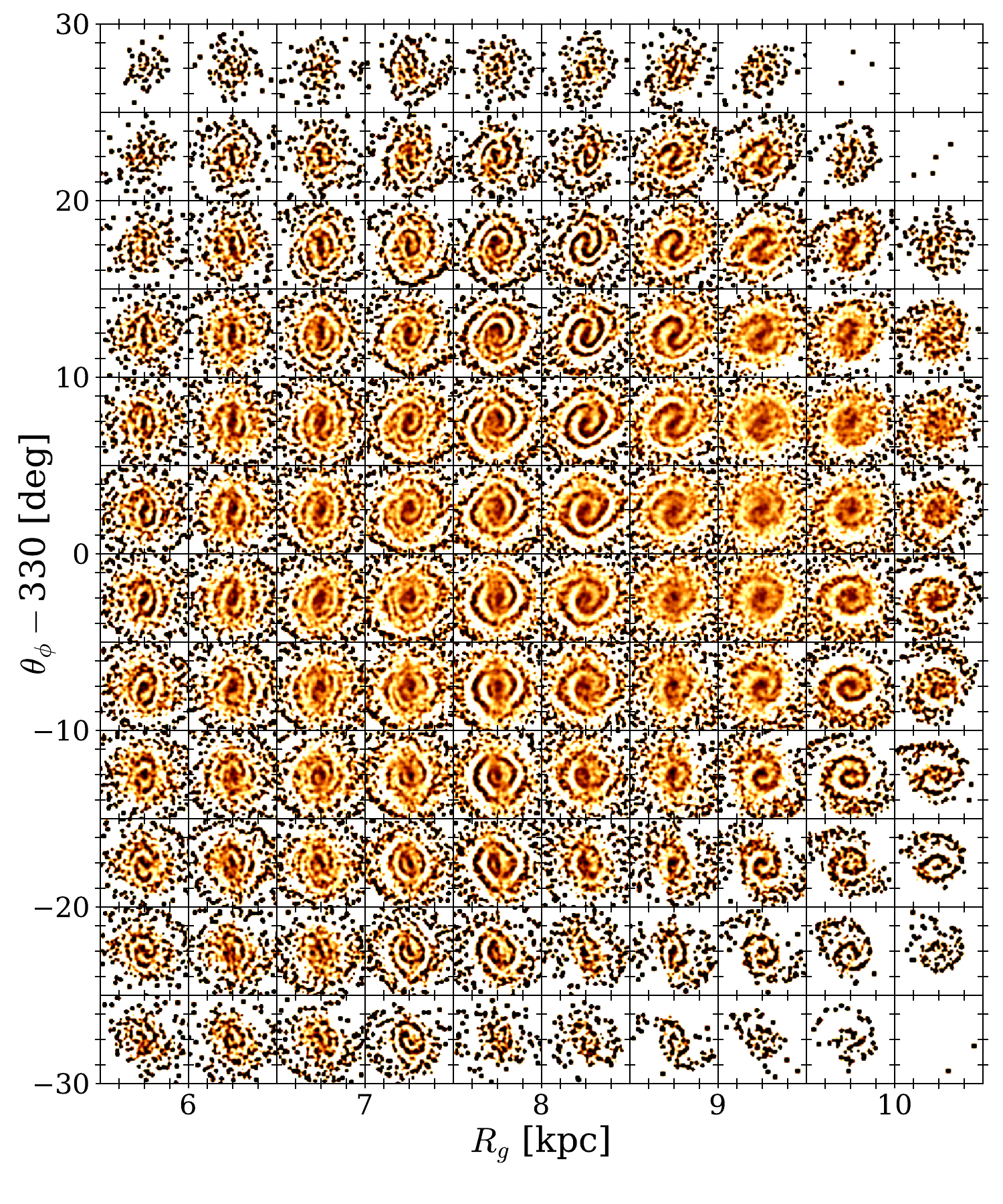}
		\caption{Same as Fig.~\ref{fig:z_vz_Rg_theta_D} but for Region F.}\label{fig:z_vz_Rg_theta_F}
	\end{center}
\end{figure}

Fig.~\ref{fig:local_areas} shows the face-on map of the galactic disc at $t=1.65$~Gyr, where six circles indicate the local regions labelled A--F.
Each circle is centred at $R=9$~kpc and has a radius of 1.5~kpc.
Fig.~\ref{fig:z_vz_Rg_theta_D} shows the $R_g$-$\theta_{\phi}$ dependence of the phase spiral in Region D.
Each panel shows the $z/h_z$-$v_z/\sigma_z$ map colour-coded by $\Delta \rho$.
Two-arm phase spirals are observed in the small $\Rg$ (inner galaxy) bins, while one-arm phase spirals appear in the large $\Rg$ (outer galaxy) bins. This spatial variation in spiral morphology consistent with what is observed in the \textit{Gaia} data \citep{2022MNRAS.516L...7H}.
Figs.~\ref{fig:z_vz_Rg_theta_A}--\ref{fig:z_vz_Rg_theta_F} show the same plots for the other regions.
Whereas  Region A (Fig.~\ref{fig:z_vz_Rg_theta_A}) shows a similar trend, the two-arm and one-arm phase spirals are not clearly separated by $\Rg$ in the  Regions B, C, E and F.
In these regions, the two-arm phase spirals tend to be observed in the intermediate $\Rg$ range around $\sim7$--9~kpc rather than the inner galaxy.

\FloatBarrier

\subsection{Time evolution of phase spirals label by $v_R$ and $v_{\phi}$}\label{appendix:vR_vphi_phasespirals}
\begin{figure*}[t]
	\begin{center}
		\includegraphics[width=0.49\hsize]{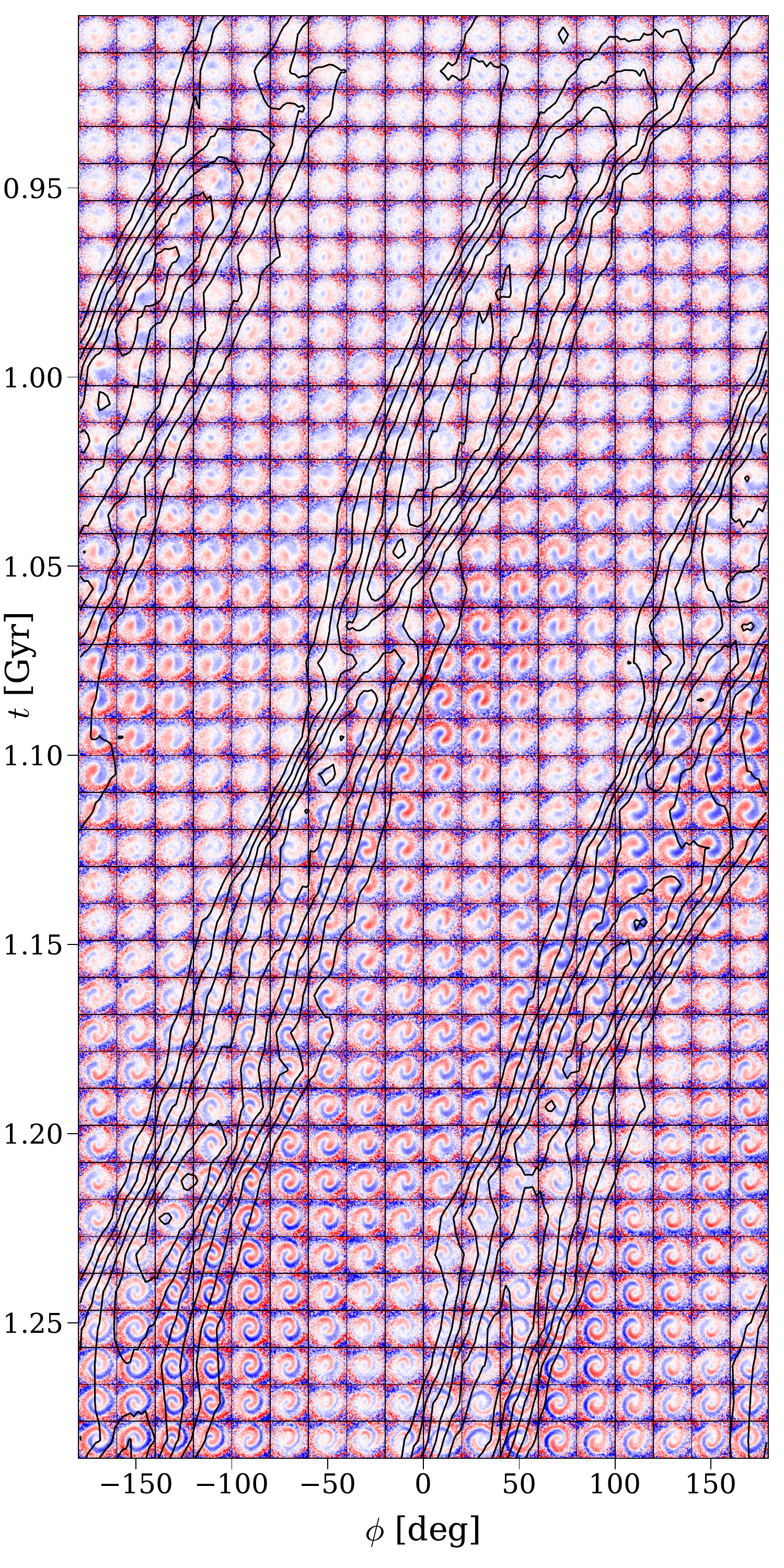}
		\includegraphics[width=0.49\hsize]{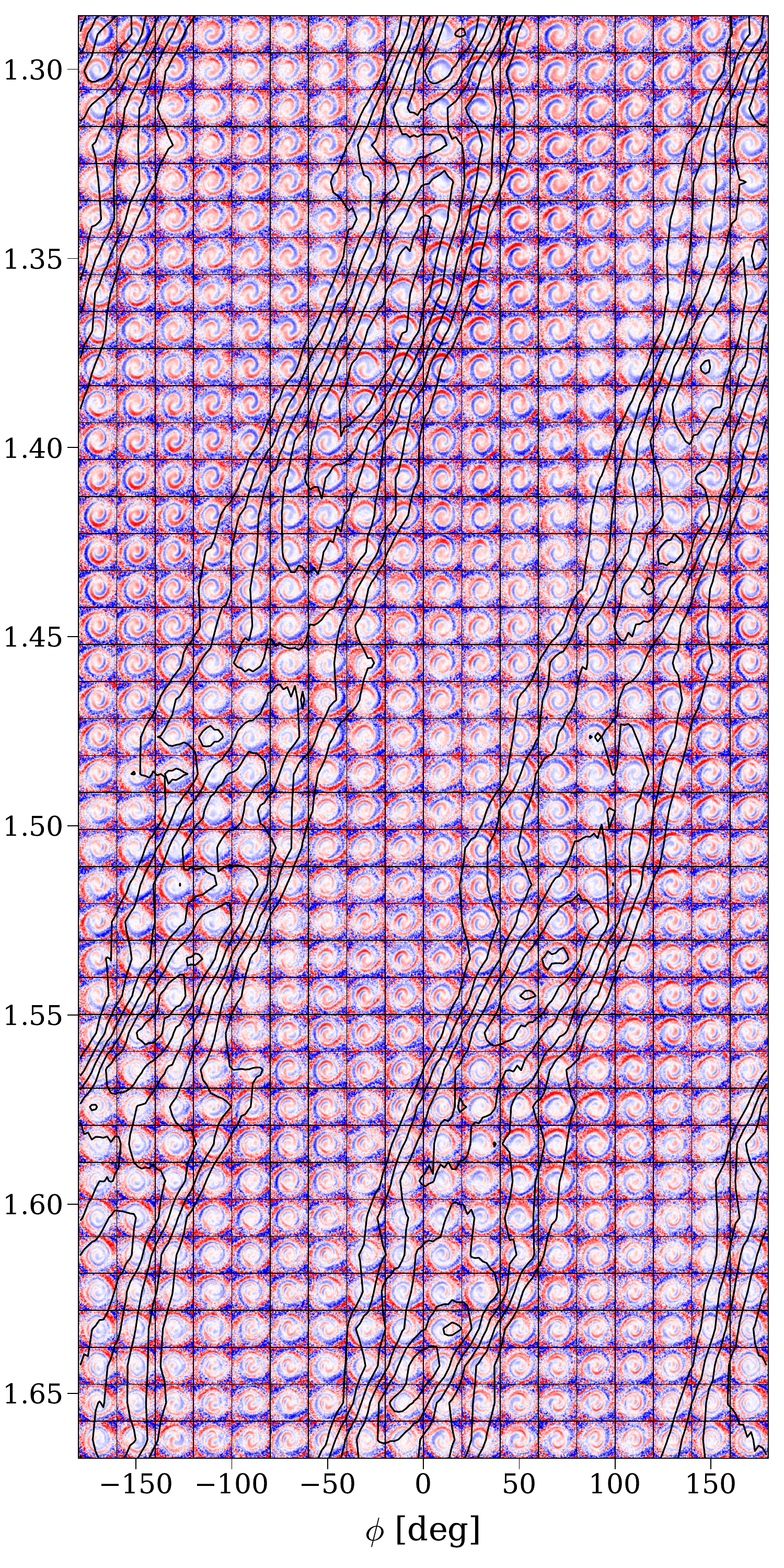}
		\caption{Time evolution of the phase spiral at $R = 8$~kpc. The $z/h_z$-$v_z/\sigma_z$ maps are shown in the same way as Fig.~\ref{fig:z_vz_time_8} but colour-coded by $\Delta v_R$ instead of $\Delta \rho$. }\label{fig:z_vz_time_vR_8}
	\end{center}
\end{figure*}

\begin{figure*}
	\begin{center}
		\includegraphics[width=0.49\hsize]{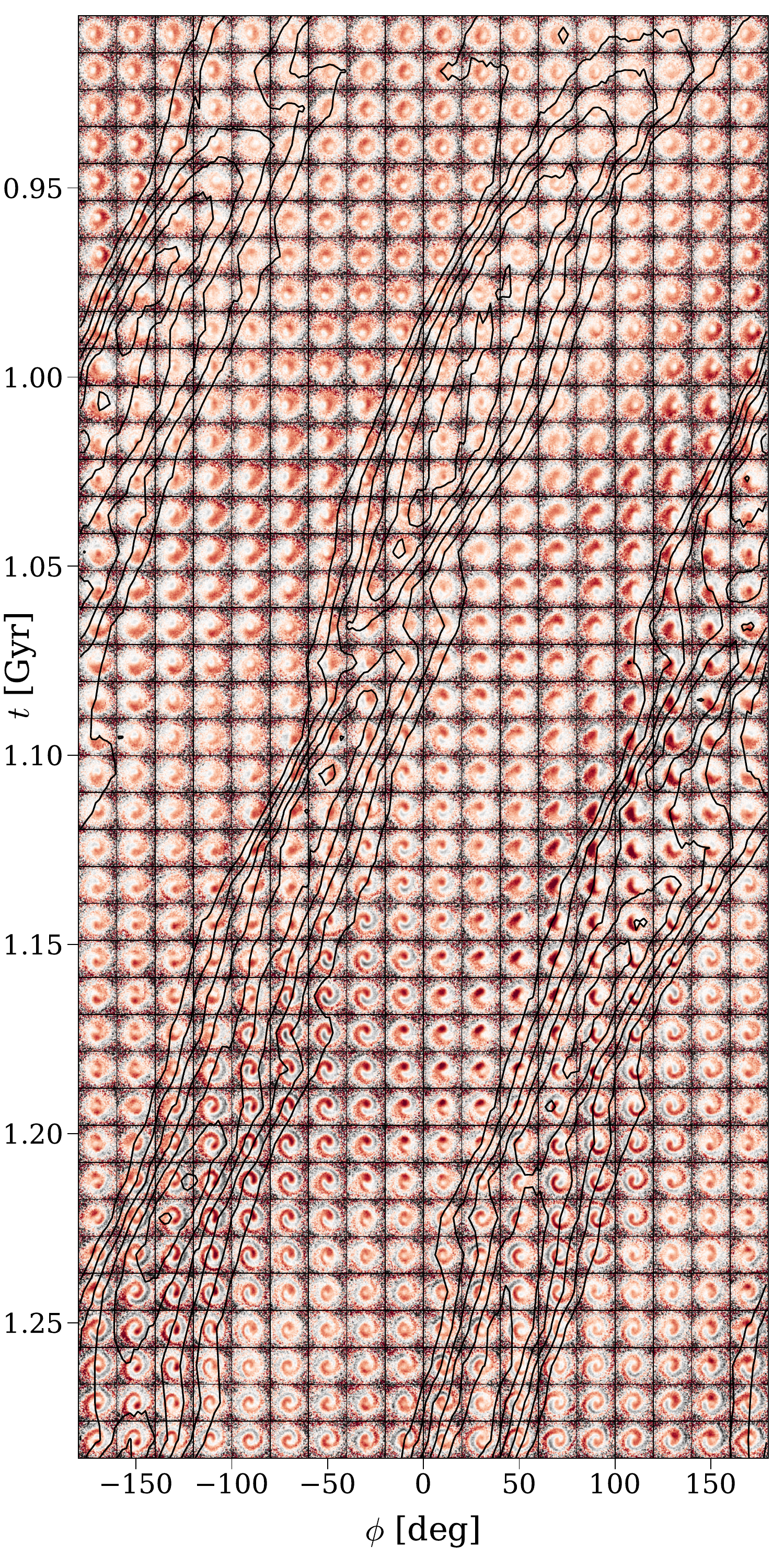}
		\includegraphics[width=0.49\hsize]{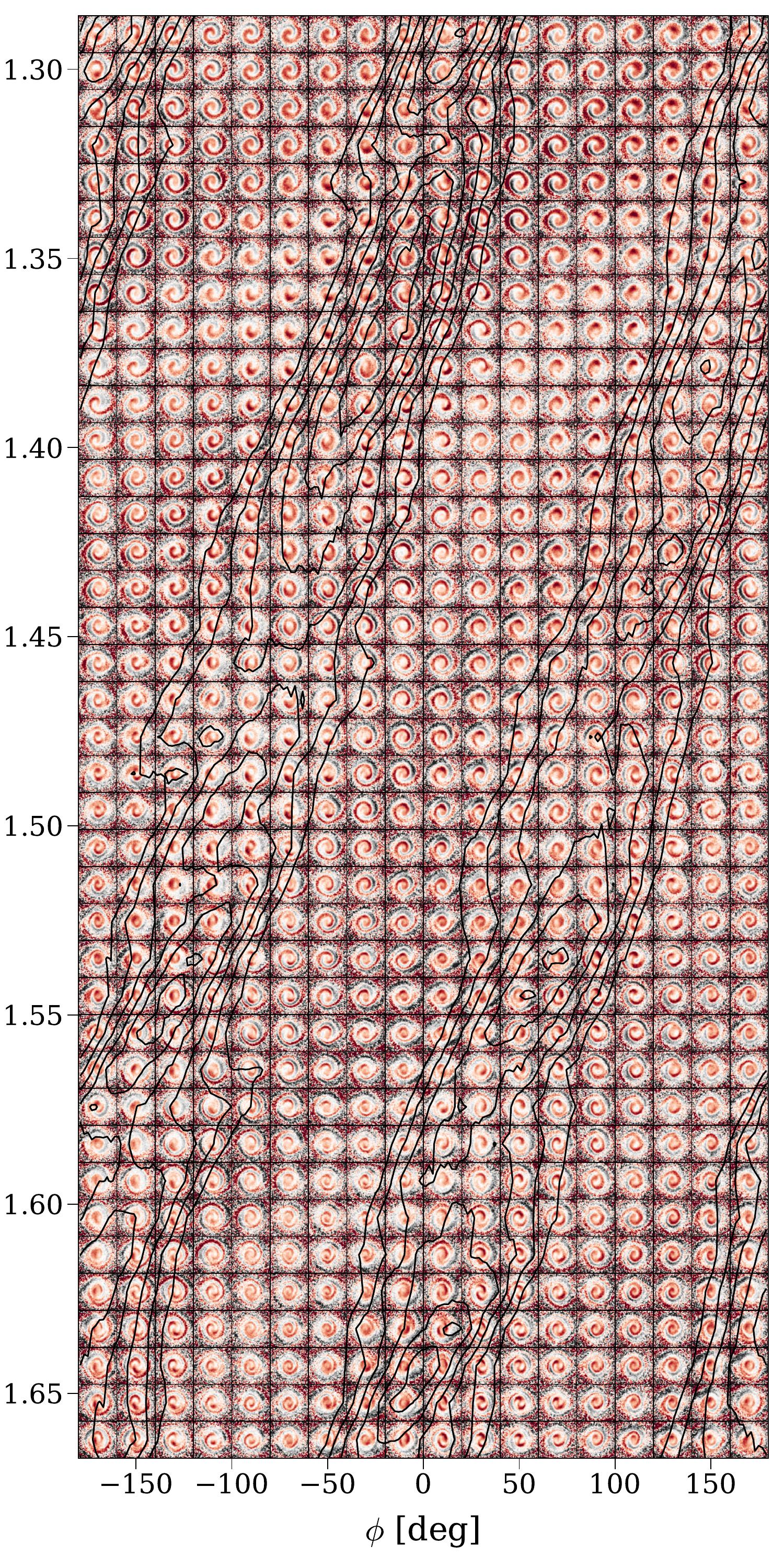}
		\caption{Same as Fig.~\ref{fig:z_vz_time_vR_8} but colour-coded by $\Delta v_{\phi}$.}\label{fig:z_vz_time_vphi_8}
	\end{center}
\end{figure*}

Fig.~\ref{fig:z_vz_time_vR_8} and Fig.~\ref{fig:z_vz_time_vphi_8} show the time evolution of the phase spiral at $R=8$~kpc.
They are visualised similarly to  Fig.~\ref{fig:z_vz_time_8} but are colour-coded by $\Delta v_R$ and $\Delta v_{\phi}$, respectively.
One-arm phase-spirals start to emerge around $t\sim1.05$~Gyr in the maps colour-coded by $\Delta v_R$ (Fig.~\ref{fig:z_vz_time_vR_8}) and $\Delta v_{\phi}$ (Fig.~\ref{fig:z_vz_time_vphi_8}), whereas they emerge around $t\sim1.15$~Gyr in the maps colour-coded by $\Delta \rho$ (Fig.~\ref{fig:z_vz_time_8}).
Morphologies (i.e. one-arm or two-arm) of the phase spirals labelled by $\Delta \rho$, $\Delta v_R$, and $\Delta v_{\phi}$ do not always coincide, suggesting a complex vertical phase mixing process coupled with planar motion.

\end{appendix}
\end{document}